\documentclass[acmsmall]{acmart}
\usepackage{booktabs} % For formal tables
\usepackage[ruled]{algorithm2e} % For algorithms

\usepackage{graphicx}
\usepackage{subfig}
\usepackage{caption}

\usepackage{verbatim} % For comment
\usepackage{multirow}
\usepackage{threeparttable}
\usepackage{color}
\usepackage[T1]{fontenc}
\usepackage[utf8]{inputenc}
\usepackage{amsmath}

\newcommand{\rebuttal}[1]{\textcolor{black}{#1}}

\AtBeginDocument{%
  \providecommand\BibTeX{{%
    \normalfont B\kern-0.5em{\scshape i\kern-0.25em b}\kern-0.8em\TeX}}}

\setcopyright{acmlicensed}
\acmJournal{TOIS}
\acmYear{2021} \acmVolume{1} \acmNumber{1} \acmArticle{1}
\acmMonth{1} 
\acmPrice{15.00}\acmDOI{10.1145/3464301}

\begin{document}

\title{What and How long: Prediction of Mobile App Engagement}


\author{Yuan Tian}
\email{yuan.tian@nottingham.ac.uk}
\affiliation{%
  \institution{University of Nottingham}
  \streetaddress{Wollaton Road}
  \city{Nottingham}
  \postcode{NG8 1BB}
  \country{United Kingdom}
}

\author{Ke Zhou}
\email{ke.zhou@nottingham.ac.uk}
\affiliation{%
  \institution{University of Nottingham}
  \streetaddress{Wollaton Road}
  \city{Nottingham}
  \postcode{NG8 1BB}
  \country{United Kingdom}
}

\author{Dan Pelleg}
\email{pellegd@acm.org}
\affiliation{%
  \institution{Yahoo Research}
  \city{Haifa}
  \country{Israel}}

\renewcommand{\shortauthors}{Tian, et al.}

\begin{abstract}
User engagement is crucial to the long-term success of a mobile app. Several metrics, such as dwell time, have been used for measuring user engagement. However, how to effectively predict user engagement in the context of mobile apps is still an open research question. For example, do the mobile usage contexts (e.g.,~time of day) in which users access mobile apps impact their dwell time? Answers to such questions could help mobile operating system and publishers to optimize advertising and service placement. In this paper, we first conduct an empirical study for assessing how user characteristics, temporal features, and the short/long-term contexts contribute to gains in predicting users' app dwell time on the population level. The comprehensive analysis is conducted on large app usage logs collected through a mobile advertising company. The dataset covers more than 12K anonymous users and 1.3 million log events. Based on the analysis, we further investigate a novel mobile app engagement prediction problem --  can we predict simultaneously what app the user will use next and how long he/she will stay on that app? We propose several strategies for this joint prediction problem and demonstrate that our model can improve the performance significantly when compared with the state-of-the-art baselines. Our work can help mobile system developers in designing a better and more engagement-aware mobile app user experience.
\end{abstract}

\begin{CCSXML}
<ccs2012>
   <concept>
       <concept_id>10002951.10003227.10003245</concept_id>
       <concept_desc>Information systems~Mobile information processing systems</concept_desc>
       <concept_significance>500</concept_significance>
       </concept>
   <concept>
       <concept_id>10003120.10003138.10003140</concept_id>
       <concept_desc>Human-centered computing~Ubiquitous and mobile computing systems and tools</concept_desc>
       <concept_significance>500</concept_significance>
       </concept>
 </ccs2012>
\end{CCSXML}

\ccsdesc[500]{Information systems~Mobile information processing systems}
\ccsdesc[500]{Human-centered computing~Ubiquitous and mobile computing systems and tools}

\keywords{mobile apps,  user engagement, app usage, dwell time, next app prediction, app engagement prediction,  demographics, behavior modeling, user modeling}

\maketitle

\section{Introduction}

\begin{figure}
	 \centering
	   \includegraphics[width = 0.7\textwidth]{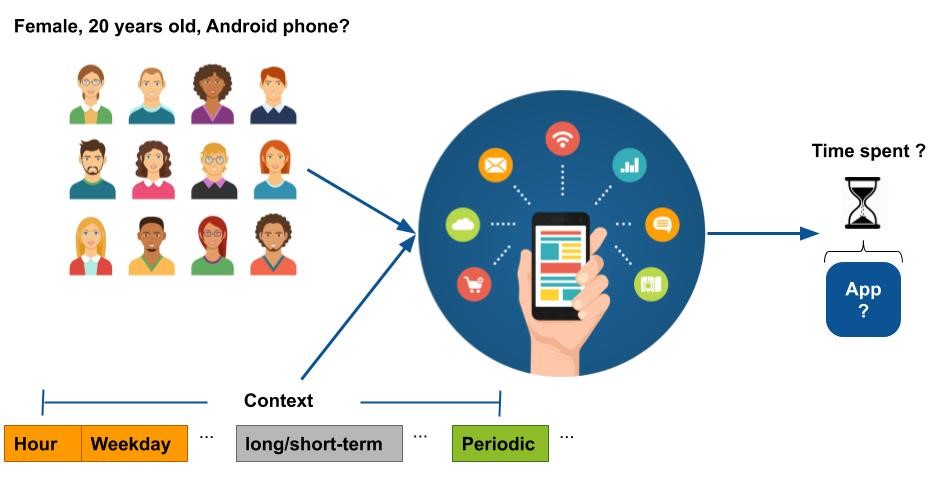}
	   \caption{Next app and app dwell time prediction.}
	   \label{overview_image}
\end{figure}

Mobile devices have become an increasingly ubiquitous part of our everyday life. People access mobile apps to fulfil their information needs or accomplish their daily tasks. Users are generally engaged with an app when they appreciate the mobile app content to which they have given their attention. Properly monitoring user engagement is one of the key ingredients of success to improve user experience and retention. One way user engagement has been measured at a large scale is by tracking how long users spend with content, e.g., the time spent on a webpage. Studies on user engagement in the contexts of desktop-based systems \cite{van2007realism} and websites \cite{diaz2014predicting, yom2013measuring} have shown that simple metrics such as dwell time are meaningful and robust in modelling user engagement. More importantly, these studies have shown that with an awareness of engagement, users' experience with a system can be substantially improved which in turn leads to user growth, user retention, and increasing revenue streams.

However, even many past works in mobile computing have investigated how individuals download, install and use different apps on their mobile devices \cite{zhang2012nihao, pan2011composite, shin2012understanding, huang2012predicting}, 
to our knowledge, few studies have examined how to effectively predict how long users would stay with a mobile app. Real-world mobile app usage behaviour is a complex phenomenon driven by a number of competing factors. Game apps, in general, have a higher probability to be used for a long period, whereas weather app is, not surprisingly, shorter. Intuitively, there are certain times in a day when a user might be more likely to engage with certain mobile apps: John might be more likely to engage with game apps for a longer time at night after work. User characteristics could also make a difference: female users may spend longer time with shopping apps than male users \cite{quinlan2003just}. Despite that the averaged total time spent on different app categories were reported in \cite{bohmer2011falling, xu2011identifying, li2015characterizing}, there is little research that comprehensively analyzes the dwell time of mobile apps. This motivates our first research question \textbf{(RQ1)}:
\begin{itemize}
\item \textit{What are the factors (user characteristics and contexts) that influence the dwell time a user spends on an app?}
\end{itemize}

As far as we know, there have been many works conducted for modelling users' behaviour on choosing one particular app under different contexts. However, they do not describe how users engage with that app. Attention is a scarce resource in the modern world. For instance, a user may become immersed in the video watching or quickly abandon it – the distinction of which will be clear if we know how much time the user spent interacting with this app given different contexts. Next app prediction is characterised as the willingness to use an app, whereas engagement is the usage pattern after accessing the app. Though most researchers have focused on measuring which app user will use \cite{liao2013feature, liao2013mining, baeza2015predicting, xu2020predicting}, the engagement of apps is still not well understood. Hence, we consider which app user will use and the engagement level of how long the user will stay with this app an aggregated measure of users' app usage behaviour. Furthermore, app engagement (dwell time) is much dependent on the app content itself. As we mentioned above, checking the weather app is always shorter than playing games. Therefore, it is  meaningless to predict how long a user will stay regardless of which app user is engaging. Given the inter-dependency between an app and app dwell time, we aim to predict the next app as well as app engagement (dwell time) as shown in Figure~\ref{overview_image}. Therefore our second research question \textbf{(RQ2)} is:
\begin{itemize}
    \item  \textit{Can we simultaneously predict which app user will use next and how long the user will stay on this app?}
\end{itemize}
Answers to these questions have a profound impact on the success of an app, as engagement awareness can radically improve users' experience with digital services \cite{revels2010understanding}. 

To answer \textbf{RQ1}, we first demonstrate how different factors affect users' app dwell time by presenting the first population-level analysis of how different contexts affect the app usage duration. The analysis is based on a large-scale mobile app usage dataset with more than 1.3 million logs from over 9K unique apps and 12K users. The data was collected from a mobile advertising company over a period of one week.
Specifically, we consider the influential factors from two aspects: user characteristics (e.g., age, gender, device type, historical preferences) and context (e.g., hour, weekday, last used app, periodic pattern). We find that, for example, users between 20 and 40 years old are more likely to have a shorter dwell time than teenagers and older people. Both the demographics and device type have an influence on how long user stay with mobile apps. Furthermore, we also establish that the app dwell time differs significantly across different time in a day. 
More importantly, we observe that users' app dwell time maintains periodic patterns and follows historical trends. For example, some users spend a similar length of time to regularly check the shopping apps every day (i.e., after every 24 hours). Additionally, users have different historical engagement habits on their app usage duration. For example, some users always prefer staying long on social apps, while others tend to only check for a short while.

%\todo{highlighting main results}

Based on the comprehensive analysis of users' app dwell time, we are able to conduct the study of how users' app dwell time can be inferred from these features. We then set to answer \textbf{RQ2} -- how to predict the next app and how long the user will stay on this app simultaneously? Based on past work on next app prediction \cite{liao2013feature, liao2013mining, baeza2015predicting}, we propose several joint prediction models (sequential, stacking, and boosting) for such app engagement prediction, whereas the dwell time is represented as discrete levels (light, medium, and intensive) defined based on different app categories. Additionally, different from personalized models in prior work \cite{liao2013feature, liao2013mining, baeza2015predicting}, our models leverage the community-behaviour patterns by extracting predictive features from the entire user population.

To summarize, our main contributions are two-folds:
\begin{itemize}
    \item We conduct the first empirical analysis of mobile app engagement based on dwell time with a large-scale data set collected from thousands of users. %We are the first to assess how user characteristics and context features impact the dwell time with a mobile app.
    %\item A generic app category prediction model is proposed to benefit the further engagement prediction problem, which outperforms prior personalized models for such problem.
    \item Our research investigates a novel problem on simultaneously predicting which app user will use and how long the user will stay on that app. An effective boosting based prediction model is proposed, %for solving this research problem, 
    which outperforms 56.8\% over the baseline approaches. 
\end{itemize}
To the best of our knowledge, this is the first empirical study on inferring predictive features at the scale of millions of logs for app dwell time, assessing how user characteristics and context features impact the dwell time with a mobile app. Our proposed predictive models are empirically validated to be effective for this novel task.
%The app user engagement prediction task 

%and we are the first study conducted to predict the next app and its corresponding dwell time simultaneously. 

The rest of the paper is organized as follows. We first review the prior literature in Section 2. Section 3 describes the dataset and presents a descriptive analysis of the app usage logs. Section 4 introduces all the features we extracted from our dataset and provides insights on how these features could impact app dwell time prediction. Section 5 presents our methodology for predicting the next app and app dwell time simultaneously, and we propose three joint prediction strategies to solve this problem. We then show the experimental results of our proposed models and analyze the effectiveness of different models in Section 6. We discuss the implications and potential limitations of our work in Section 7 and conclude in Section 8.

\section{Related Work}
\label{Sec:Related_Work}
Our work is related to prior literature on user engagement analysis (\S\ref{user_engagement}) and next app/engagement prediction (\S\ref{app_usage_engagement_prediction}).

\subsection{User Engagements}
\label{user_engagement}
\subsubsection{\textbf{Engagement Measurements}}
Approaches to measuring user engagement of online services can be divided into three main groups: (a)~self-reported engagement, (b)~cognitive engagement, and (c)~online behaviour metrics \cite{lehmann2012models}. In the first group (a), questionnaires and interviews are used to elicit user engagement attributes or to create user reports and to evaluate engagement \cite{kim2013study}. The second type of approach (b) uses task-based methods and physiological measures to evaluate the cognitive engagement (e.g., facial expressions, vocal tone, and heart rate) using tools such as eye-tracking \cite{ehmke2007identifying}, heart rate monitoring, brain readings from headset \cite{mathur2016engagement}, swipe on the screen \cite{nelissen2018swipe} and mouse tracking \cite{huang2011no}. However, these two types of methods have known drawbacks, e.g.,~reliance on user subjectivity of the self-reported engagement, and only be able to measure a small number of user interactions of the cognitive engagement. 

The third type of approach (c), adopted by the web-analytics community, has been studying user engagement through online behaviour metrics that assess users’ depth of engagement within a site, e.g.,~the time spent on a webpage. Studies on user engagement in the contexts of desktop-based systems \cite{van2007realism} and websites \cite{diaz2014predicting,yom2013measuring} have shown that simple metrics such as dwell time are meaningful and robust in modelling user engagement. For example, the time spent on a resource has been validated as an effective metric for measuring user engagement in the context of web search \cite{agichtein2006improving, bilenko2008mining}, and recommendation tasks \cite{yi2014beyond}. Kelly et al. \cite{kelly2004display, fox2005evaluating} consider dwelling times as an indicator of page relevance or user satisfaction during search engine interactions.  \citet{yi2014beyond} recommend designing dwell time based user engagement metrics and claim that this would enable them to extract better user engagement signals for training recommendation systems thereby optimizing for long term user satisfaction. 

Therefore, we also propose to use the metric of time spent on mobile apps (dwell time) as our user engagement metrics within a large-scale dataset. For now, minimal research has been done for modelling mobile app dwell time from a large-scale dataset; only the basic aggregated statics on app usage time was reported. Falaki et al. \cite{falaki2010diversity} found that 90\% of app usage sessions would be less than 6 minutes and Xu et al. \cite{xu2011identifying} reported that the majority of total network access time for all apps is from 10 seconds to 1 hour for each subscriber in one week. Li et al. \cite{li2015characterizing} reported usage time for different app categories in total and found that communication apps account for 49\% cellar time against all apps. B{\"o}hmer \cite{bohmer2011falling} found that the Libraries \& Demos apps (default Updater, Google Services Framework, etc.) have the longest average usage time from opening to closing. However, these summarized basic statistics of app usage time can not provide an in-depth understanding of what factors could influence the dwell time user spend on an app and whether we could predict how long user would stay with an app. %Properly monitoring user engagement is one of the key ingredients to success and increasing engagement. 
In our work, we conduct the first empirical analysis of dwell time during app usage based on a large-scale data set collected from thousands of users; and we are the first to assess how different kinds of features (e.g., demographics, device type, hour of day, and last used app) impact the mobile app dwell time.

\subsubsection{\textbf{Influential Factors of App Usage}}
Lots of previous research has been conducted to uncover the factors that influence app usage behaviour, ranging from time, demographics, device types, last used app, to periodic patterns. \citet{xu2011identifying} and  \citet{li2015characterizing} discovered the temporal pattern of app usage, e.g., news apps were more frequently used in early mornings, whereas sports apps were more frequently used in evenings. They also found that, in general, the app usage frequency changed during the day, which grew from 6 am and reached its first peak around 11 am, and were most active during the evening (7 pm to 9 pm). Additionally, demographics have been identified as an important factor that would impact the way users engage with their apps \cite{seneviratne2015your, zhao2016discovering, kooti2017iphone, van2017describing, tian2020cohort}, e.g., how users select apps or make in-app purchases on their smartphones. \citet{li2017mining} ever analyzed the device-specific apps usage patterns from large-scale android users and they found that users rely less on the cellular network as the price of device model increases. For the correlations between apps usage,  Li et al. \cite{li2015characterizing, xu2011identifying, bohmer2011falling} all showed that the last used app would probably impact the next app the user is going to use. However, minimal research has been done for analyzing if any of these features could impact the dwell time of an app. In our work, we conduct the first empirical analysis of how long users will stay with an app on a large data set. We are the first to systematically model how user characteristics (e.g., demographics) and context features (e.g., time, last used app) impact the time user stay with an app. 

\subsection{App Usage and Engagement Prediction}
\label{app_usage_engagement_prediction}
\subsubsection{\textbf{Next App and App Engagement Prediction}}
Much research work has been conducted on predicting which app user will use (a.k.a.~next app prediction). \rebuttal{Based on the survey \cite{cao2017mining}, we could know that before 2017, most of the app prediction/recommendation works are conducted during 2012-2013.} Tan et al. \cite{tan2012prediction} tried to treat users' app usage patterns as periodic time series cycles and predict the app will be used based on a prediction algorithm with fixed cycle length. Liao et al. \cite{liao2012mining} also predicted the next app based on mined temporal profiles for each app. Additionally, Huang et al \cite{huang2012predicting}, Shin et al. \cite{shin2012understanding} and Zou et al. \cite{zou2013prophet} all pointed out that the latest used app and time are more effective than location and other context information in the next app prediction. After 2017, neural approaches have become increasingly popular. Some researchers proposed different neural models for predicting the next app, including CNN \cite{schmidhuber2015deep} and LSTM \cite{xu2020predicting}, etc. Xu et al. \cite{xu2020predicting} proposed a generic prediction model based on Long Short-term Memory (LSTM), to covert the temporal-sequence dependency and contextual information into a unified feature representation for next app prediction and stated that it outperforms other models. Additionally, Tian et al.~\cite{tian2020identifying} explored the prediction to identify if the pair of two app usage logs belong to the same task. As far as we know, no research has been done for predicting how long users will stay with an app. Only Mathur et al. \cite{mathur2016engagement} tried to model and predict users' attention-based engagements (two levels: focused attention and felt involvement) in the context of smartphones. They conducted the work based on physiological measures (headset readings) within 10 participants, and their research has no works regarding which app users are currently using. In our work, by leveraging a large-scale data set of users' app usage logs, we could be able to model users' engagement (dwell time) within specific apps by exploiting user characteristics, temporal context, and short/long-term behavioural patterns. Most importantly, we investigate the challenges of simultaneously predicting which app user will use and how long the user will stay with this app.

\subsubsection{\textbf{Engagement (Dwell Time) Prediction in Other Areas}}
For the novel prediction problem proposed in our work, although no research has been conducted on the app dwell time prediction, some researchers investigated the engagement (dwell time) modelling and prediction in other areas, e.g., session-based recommendation (SBR) systems \cite{bogina2017incorporating, zhou2018jump, wangcapturing}, videos watching \cite{wu2018beyond}, news and non-news pages reading \cite{seki2018analysis,homma2018analysis}, and media streaming \cite{vasiloudis2017predicting}. 

The SBR tasks aim to predict the next click/buying/dwell time based on users' interactions in a session, e.g., buying one item after viewing several products (within a commerce site). They stated that dwell time should be used as a proxy to user satisfaction of the clicked result since clicked through or not is not enough to identify the satisfaction of the user. In their scenario, researchers aim to predict which link/product user would click among a list of recommended similar results (under specific search query). Additionally, they assumed that the longer user stays with the clicked result,  the more satisfied the user will be with the result. However, in our dwell time prediction problem, a user would use an app occasionally with no recommendation context, and the longer user stays with this app does not directly mean the user is satisfied with it or not. The dwell time may be affected by the specific app (e.g.,~ a weather app or game app), or whether the user accessed it during commuting or at night while they have more leisure time. Therefore, predicting how long a user will stay with an app is to predict the usage pattern while using the app. It would allow the service provider to optimize the user experience along with its business goals, the apps can be tuned to be more exploratory or exploitative based on the expected length of the usage. The SBR tasks can be solved by item-to-item and matrix factorization methods \cite{hidasi2016general,musto2015word}, Markov Decision Process (MDP) based technique \cite{tavakol2014factored}. Recently, deep learning methods, Recurrent Neural Networks (RNN) have emerged as powerful methods of modelling sequential data in SBR \cite{bogina2017incorporating, zhou2018jump, wangcapturing}.

The engagement pattern of the other online services, like video watching \cite{wu2018beyond}, news/non-news page reading  \cite{seki2018analysis,homma2018analysis}, and media streaming \cite{vasiloudis2017predicting} have more similar characteristics of apps usage, i.e. how long user would stay could be affected by the category of the content or the original length of the video or news. Additionally, the underlying motivation for engagement prediction of these services is also the same as our work. They consider popularity and engagement as different measures of online behaviour. Although popularity describes the human behaviour of choosing one particular item, it does not describe how users would engage with this item. i.e., popularity is characterized as the willingness to click a video, whereas engagement is the video watching pattern after clicking. By knowing how long user would stay, we could be able to provide users with more satisfying content/services that increase long term user engagement and as a side-benefit. It also allows the service providers to optimize the user experience along with its business goals. Wu et al. \cite{wu2018beyond} conducted a large-scale measurement study of engagement on 5.3 million videos over a two-month period and measured a set of engagement metrics (e.g., watch time, watch percentage) for online videos. They predicted engagement from video context, topics, and channel reputation, etc. Seki et al. \cite{seki2018analysis} and Homma et al. \cite{homma2018analysis} all clarified the characteristics of relationships between dwell time on news/non-news pages reading in order to discover which features are effective for predicting the dwell time, including (1) Dwell time by Device: desktops and mobies; (2) Dwell time by access time; (3) Dwell time by if users visited from inside or outside the site; (4) Dwell time by click and non-click: if the user clicked links in the page; (5) Dwell time by scroll depth. Vasiloudis et al. \cite{vasiloudis2017predicting} explored the prediction of session length in a mobile-focused music streaming service. They predicted the length of a session using contextual and user-based features including gender, age, subscription status, device, network type, duration of the user's last session, and time elapsed since the last session.

Given the task similarity between these engagement prediction studies and our focus (app engagement prediction),  we selected two of them \cite{vasiloudis2017predicting, wu2018beyond} with the most features that could be extracted from our dataset as the baselines for comparing the performance of our proposed model regarding the app engagement prediction. We all have similar engagement characteristics and the motivation for engagement prediction, e.g., the engagement (dwell time) is originally correlated with the content category and more exploratory or exploitative service could be provided based on the expected length of the usage.

In summary, for the traditional next app prediction task, we selected four models from previous works that could be fitted to our dataset as baselines, include three works \cite{tan2012prediction, shin2012understanding,zou2013prophet} based on the temporal pattern and contextual features and one recent work based on neural approach (LSTM) \cite{xu2020predicting}. For the joint prediction problem of predicting the next app and engagement level together, since there is no existing similar joint prediction work (app dwell time has a high dependency on which app user is engaging), we added those two recent works \cite{vasiloudis2017predicting,wu2018beyond} for predicting dwell time as baselines. The only difference is they solely predict how long a user will stay based on the specific item, without predicting on which item the user will engage with. In order to make them comparable baselines to our work, we assumed the ground truth of the next app is known for those two prior works \cite{wu2018beyond, vasiloudis2017predicting} and leveraged them specifically for predicting engagement (dwell time) of the known next app. This demonstrates the upper bound of those approaches (oracle performance) regarding the joint prediction problem.

\section{Data and Descriptive Analysis}
We start by describing the data used in this study, followed by some fundamental analysis that demonstrates the characteristics of our data.   
\subsection{Dataset}
\label{sec:dataset}

\begin{table}
    \scriptsize
	\setlength{\tabcolsep}{3pt}
	\caption{Overall statistics of the dataset.}
	\label{tab:demog_info}
	\scalebox{1}{\begin{tabular}{lll|lll}
		\toprule
		\textbf{OS}&\textbf{\%Logs}&\textbf{\%Users}&\textbf{Device}&\textbf{\%Logs}&\textbf{\%Users}\\
		\midrule
	    android&78.0\%&57.5\%&Phone&94.3\%&90.0\%\\
		ios&22.0\%&42.5\%&Tablet&5.7\%&10.0\%\\
		\midrule
		\textbf{Age}&\textbf{\%Logs}&\textbf{\%Users}&	\textbf{Gender}&\textbf{\%Logs}&\textbf{\%Users}\\
		\midrule
		13-17&5.7\%&9.0\%&female&44.1\%&51.8\%\\
		18-24&13.9\%&18.4\%&male&55.9\%&48.2\%\\
		25-34&32.2\%&30.3\%\\
		35-54&43.2\%&36.5\%\\
		55+&5.0\%&5.8\%\\
		\bottomrule
	\end{tabular}}
\end{table}

The dataset used in our paper is collected from a mobile advertising company, a library that mobile developers integrate into their apps to measure app usage and allow in-app advertising. We collected a sample of logs from a week in March 2017 of more than 1.3 million logs with over 9K different apps and 12K users from the United States. Each log consists of the user’s general app usage information, such as demographics, timestamp, app category, app id and time spent. Each app belongs to one of 45 categories ranging from social, communication to business, etc. The definition of these app categories is consistent with the Google Play App taxonomy \cite{GooglePlayAppCat}.  Table~\ref{tab:demog_info} shows various statistics of our dataset. Either Android or iOS operates the devices. 51.8\% of app users are female, and most logs (over 70\%) are generated by users between 25 and 54 years old.\footnote{Users have been classified into five age ranges in our dataset: 13-17,18-24,25-34,35-54 and 55+.}

\begin{figure}
	 \centering
	   \includegraphics[width=0.8\textwidth]{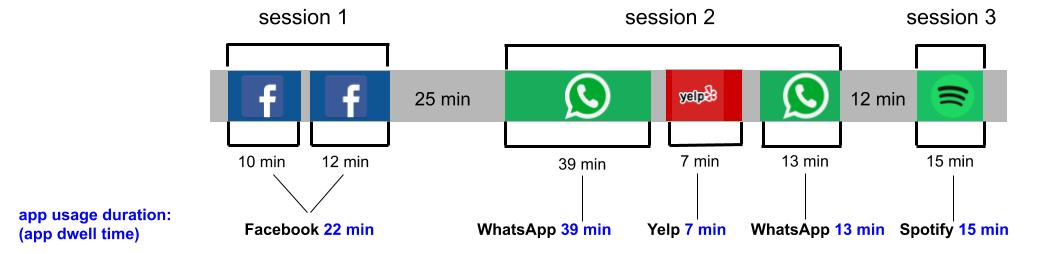}
	   \caption{Visualisation of app usage logs 
	   %(session: time threshold = 5 minutes) 
	   and the corresponding app usage duration. Five minutes of inactivity is used for segmenting mobile sessions.}
	   \label{fig:session}
\end{figure}

Following previous work \cite{carrascal2015situ}, we consider a five-minute range of inactivity as the signal of ending a session as shown in Figure~\ref{fig:session}. If the user leaves an app but revisits any app within 5 minutes the \emph{session} continues; otherwise, the \emph{session} ends. In our work, we aim to predict which app the user will use next and how long the user will stay with that app. The app usage duration we aim to predict is calculated as follows (as shown in Figure~\ref{fig:session}): 
we aggregate the consecutive app usage duration of the same app within each session. 
%whenever the user switches to a different app, the new app usage duration need to be recalculated; otherwise, the usage duration of the same app will be summed up. 
%For simplicity, we will often refer to \emph{app usage duration} as the time spent (dwell time) within each app usage in the rest of this paper, unless otherwise stated.
The app usage duration can be also referred to as app \emph{dwell time}. We adopt this definition as the unit of our analysis for the rest of the paper. 
Note that we only consider user-triggered events; i.e.~we do not include events that are triggered by background refresh when conducting our analysis. To reduce bias from users with a low level of engagement, we restricted our sample to those users who interacted with apps from at least five different categories. All the data was anonymized by removing all personally identifiable data prior to processing.

% As the dataset does not cover all the apps stored on a user's phone (i.e.~not all of the apps were registered with that mobile advertising company), our sample was restricted to users with at least five app categories.
%The dataset includes only timestamped activities and no personal information for the users covered by our study period. No individual user can be identified.

\subsection{Distributions of App Usage}
\begin{table}
    \scriptsize
	\centering
	\caption{Top 10 popular app categories.} 
	\label{tab:top_app_category}
   \scalebox{1}{\begin{tabular}{lll}
		\toprule
		\textbf{App Categories}&\textbf{\%Sessions}&\textbf{App Function Examples}\\
		\midrule
	    productivity&28.6\%&mail, calendar, notepad\\
		social&10.6\%&SNS, dating\\
		tools&8.3\%&caculator, screen lock, light\\
		communication&6.8\%&SMS, IM, free video calls\\
		entertainment&5.1\%&TV player, streaming video\\
		utilities&3.3\%&network, cleaner\\
		sports&3.1\%&live sports, sports news\\
		music&2.7\%&music player\\
		lifestyle&2.5\%&diary, discount, recipe\\
	    arcade&2.4\%&games\\
	\bottomrule
	\end{tabular}}
\end{table}

\label{engagment_level_definition}
\begin{figure}
	 \centering
	   \includegraphics[width = 0.75\textwidth]{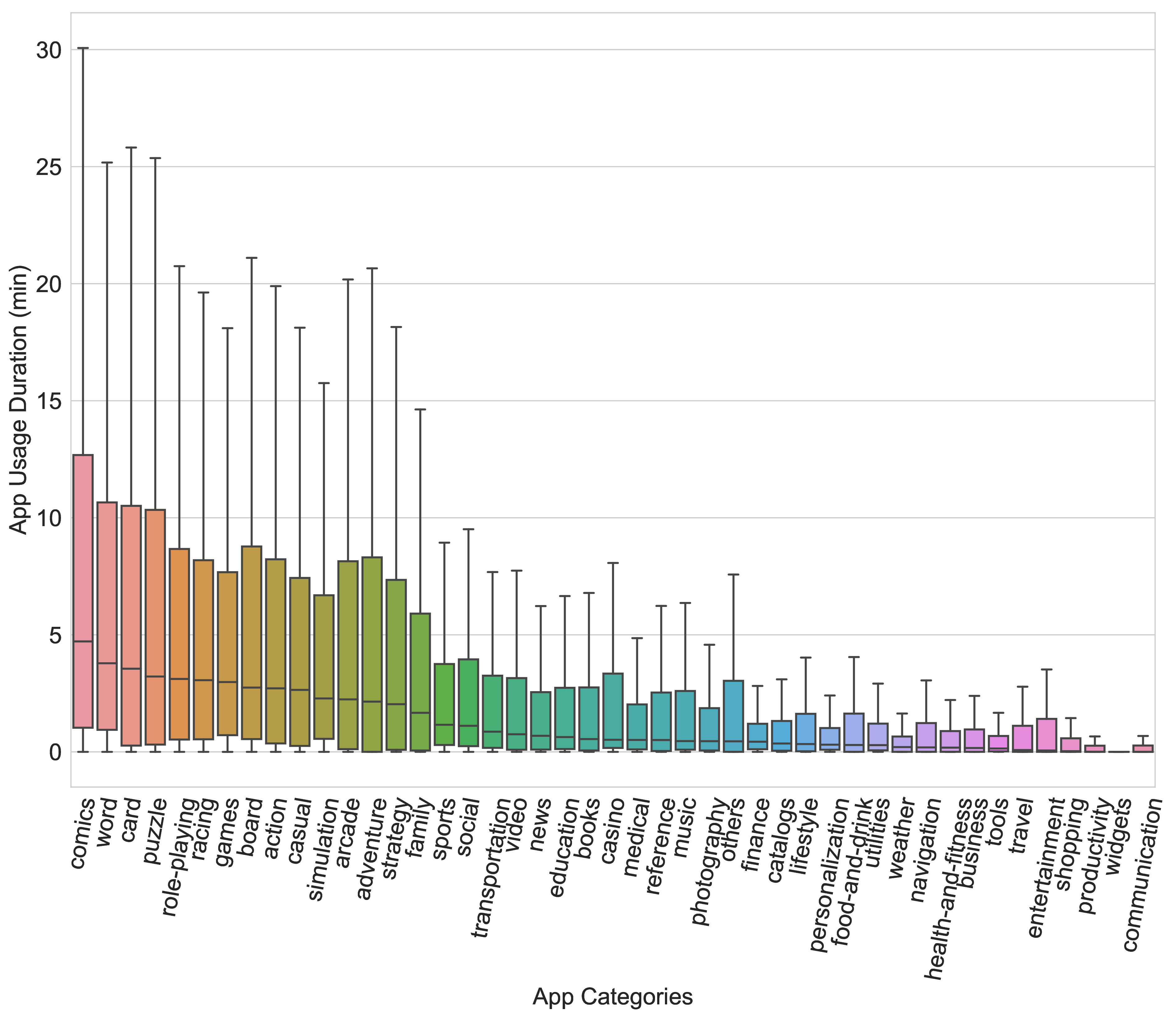}
	   \caption{Overall average app usage duration for different app categories.}
	   \label{fig:engagement_duration}
\end{figure}

\begin{figure}
\centering
%\captionsetup[subfigure]{oneside, skip= -50pt}
\subfloat[PDF (Probability Density Function)]{
			\label{fig:session_duration_pdf}
			\includegraphics[height = 1.9in]{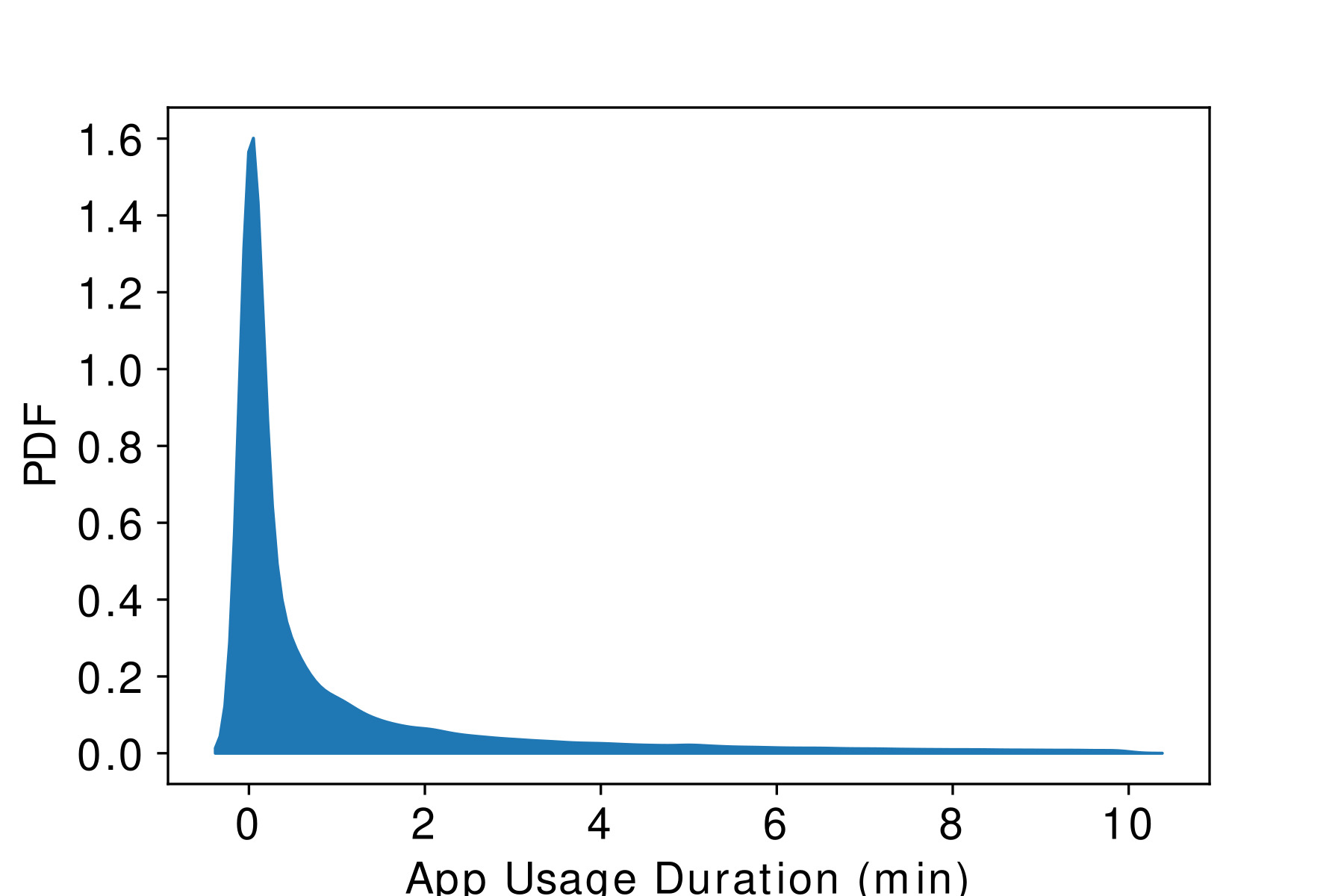} }
\subfloat[CDF (Cumulative Distribution Function)]{
		\label{fig:session_duration_cdf}	
		\includegraphics[height = 1.9in]{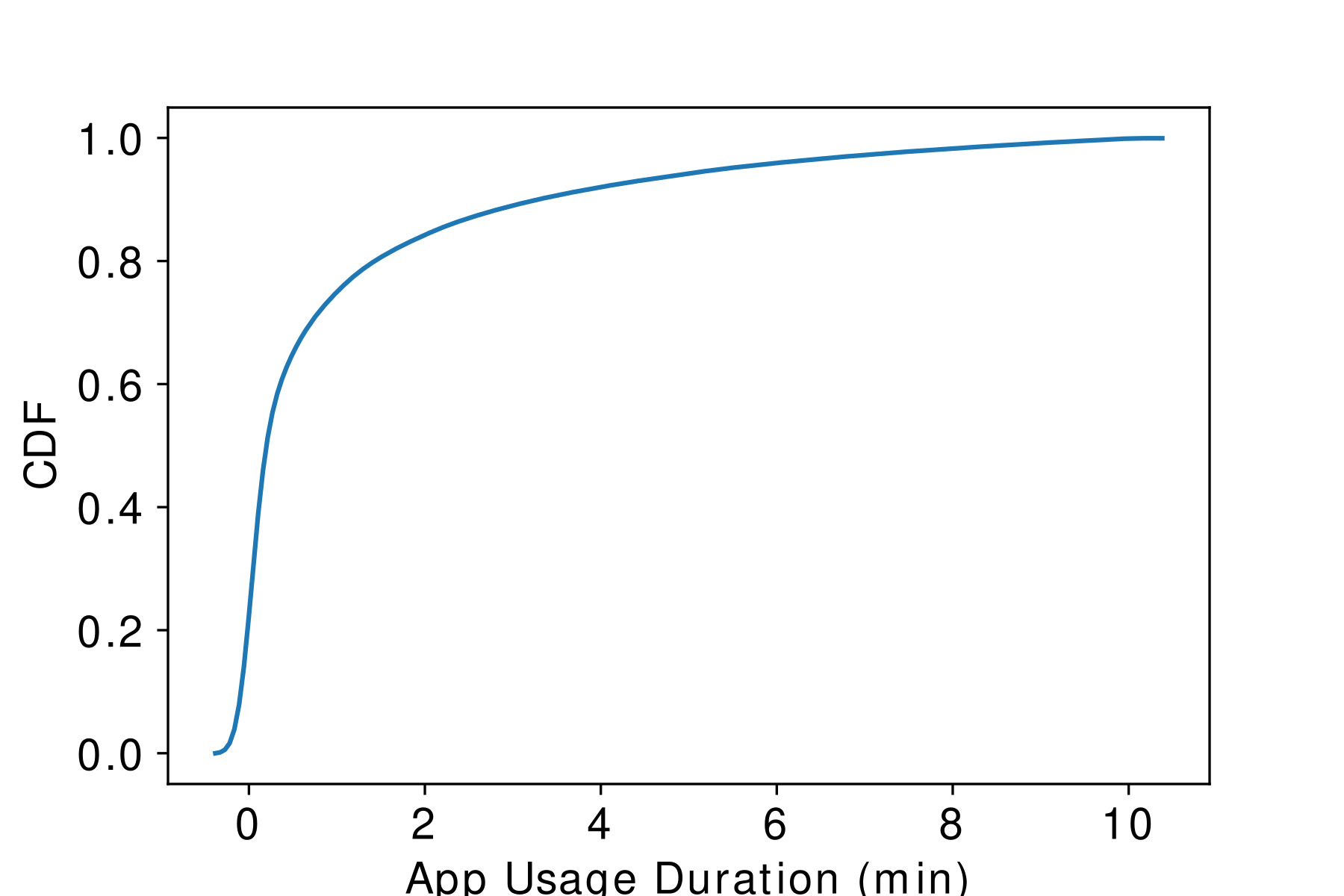} } 
	\caption{PDF and CDF of app usage duration across all apps.}
	\label{fig:session_duration}	
\end{figure}

We start by presenting the distributions of the popularity of apps. The popularity of an app category can be measured using the total number of sessions. Table~\ref{tab:top_app_category} shows the most popular app categories in our dataset. Generally, users are more likely to interact with app categories like productivity, social, tools, communication, and entertainment apps nowadays. We also illustrate several examples of the most popular functionalities of the corresponding apps within each category. Additionally, we show the average app usage duration for different app categories in Figure~\ref{fig:engagement_duration}, which ranges from less than 1 minute to over 10 minutes. We find that users always spend longer time when they are engaging with some app categories for relaxation, such as comics and games apps (i.e., card, word, puzzle, and board). It also states that different app categories initially have different lengths of app usage duration, e.g., game apps have a much longer duration than communication apps. On average for all the app categories, our users' app dwell time lasted about 2.5 minutes. In Figure~\ref{fig:session_duration}, we show the probability density function (PDF) and cumulative distribution function (CDF) of app usage duration. We can find that most of the app usage (93.7\%) is less than 10 minutes and 80\% of them only last less than 2 minutes.

\begin{figure}
	 \centering
	   \includegraphics[width=0.65\textwidth]{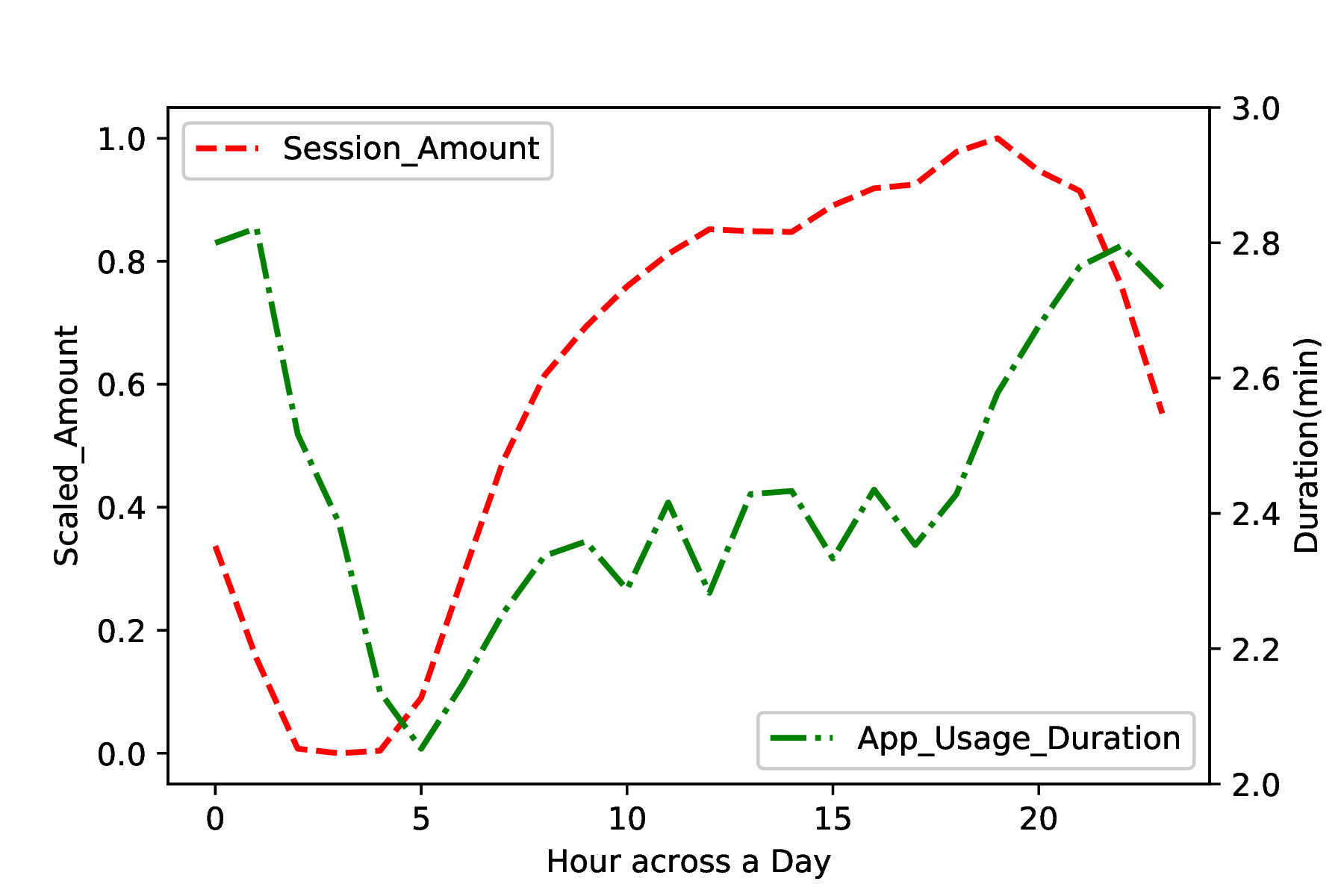}
	   \caption{App usage patterns across a day.}
	   \label{fig:hour_distribution}
\end{figure}

We then look at how users engage with their apps throughout the day. Figure~\ref{fig:hour_distribution} plots the scaled amount of sessions (using min-max normalization) and average app usage duration each hour. We can find that the total app usage (in terms of session amount) is at its maximum in the afternoon and evening, peaking at around 7 PM. This aligns with findings reported in~\cite{van2017describing,bohmer2011falling}. Figure~\ref{fig:hour_distribution} also shows the average app dwell time regarding the different hours of the day. Generally, the average app usage duration across the day is between 2 minutes and 3 minutes. The duration increases after 6 PM and drops after 1 AM. This might be due to people have more leisure time during the non-working hours. We indeed find that the app categories associated with longer duration at late night are game and tool (system cleaner/VPN) apps.

\subsection{Engagement Level Definition}
\label{engagment_level_definition}
To evaluate the performance of app usage duration prediction more intuitively, we propose to represent the time length of each duration as a discrete value, i.e., we classify the app usage duration into three engagement levels: light, medium, and intensive. Additionally, as we have found in Figure~\ref{fig:engagement_duration} that different app categories initially have different lengths of app usage duration, so it may not be reasonable to label the engagement levels without differentiating app categories. Specifically, to assign the corresponding engagement level of each app usage duration: (1) we first calculate the quantiles (33\% and 67\%) of the duration for different app categories respectively; (2) then we assign the level label to each duration based on their corresponding app category quantiles. For example, the 33\% and 67\% quantiles of weather apps are 6 seconds and 23 seconds respectively. If the current usage duration of the weather app is 10 seconds, then its engagement level is medium. For the game apps whose 33\% and 67\% quantiles are 1.3 minutes and 5.4 minutes respectively. When a game app usage duration is longer than 5.4 minutes, its engagement level is intensive.

\section{Inferring Users' App Dwell Time}
\label{infering_features}
To model the app engagement accurately, we need to uncover what information could be indicative factors for users' app dwell time. This is the main focus of this section.

\subsection{User and Device Characteristics}
\subsubsection{User Demographics}
\begin{figure}
%\vspace{-0.1in}
\centering	
\subfloat[Duration vs. Gender]{
			\label{fig:session_duration_gender}
			\includegraphics[height = 1.9in]{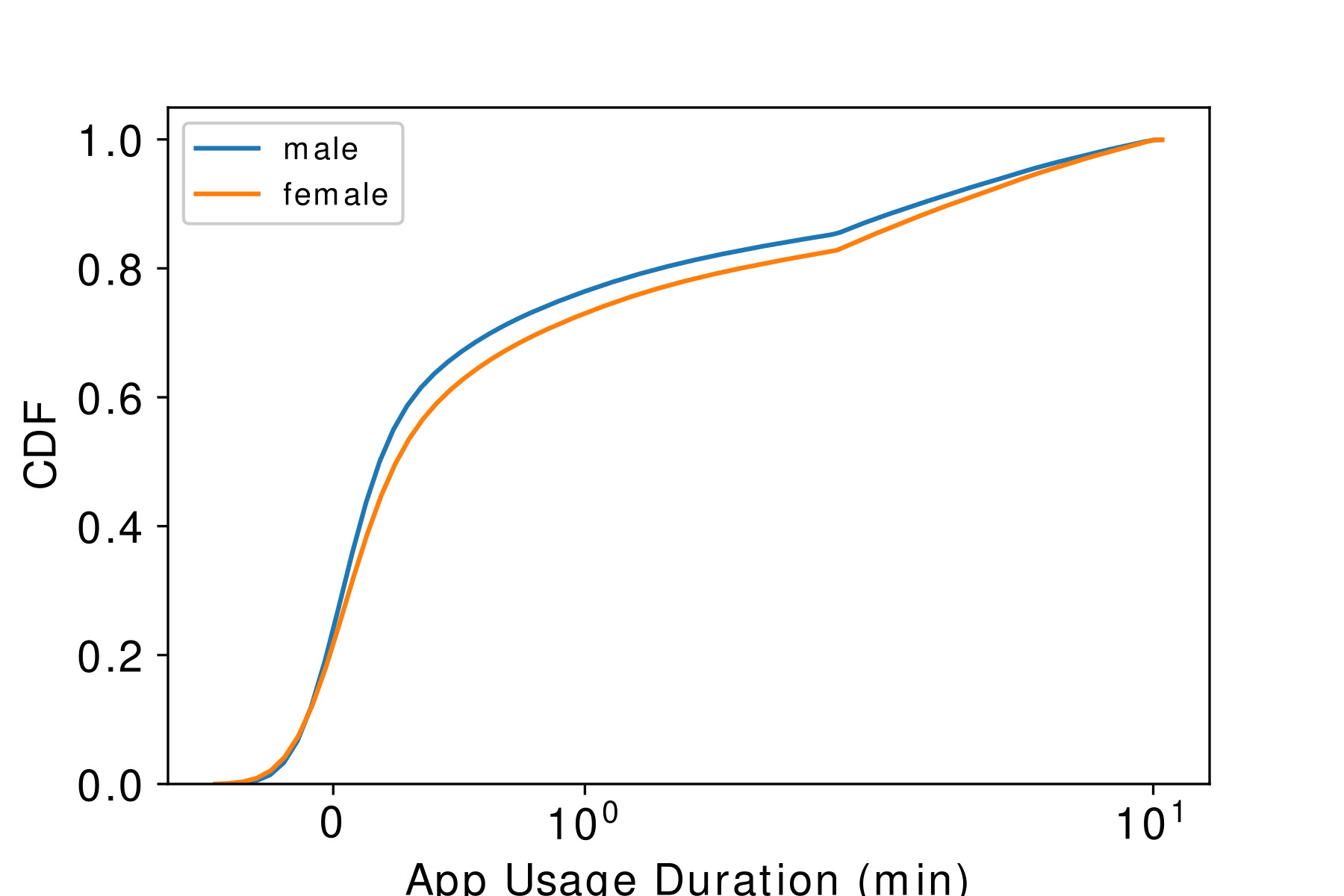} }
\subfloat[Duration vs. Age]{
		\label{fig:session_duration_age}		\includegraphics[height = 1.9in]{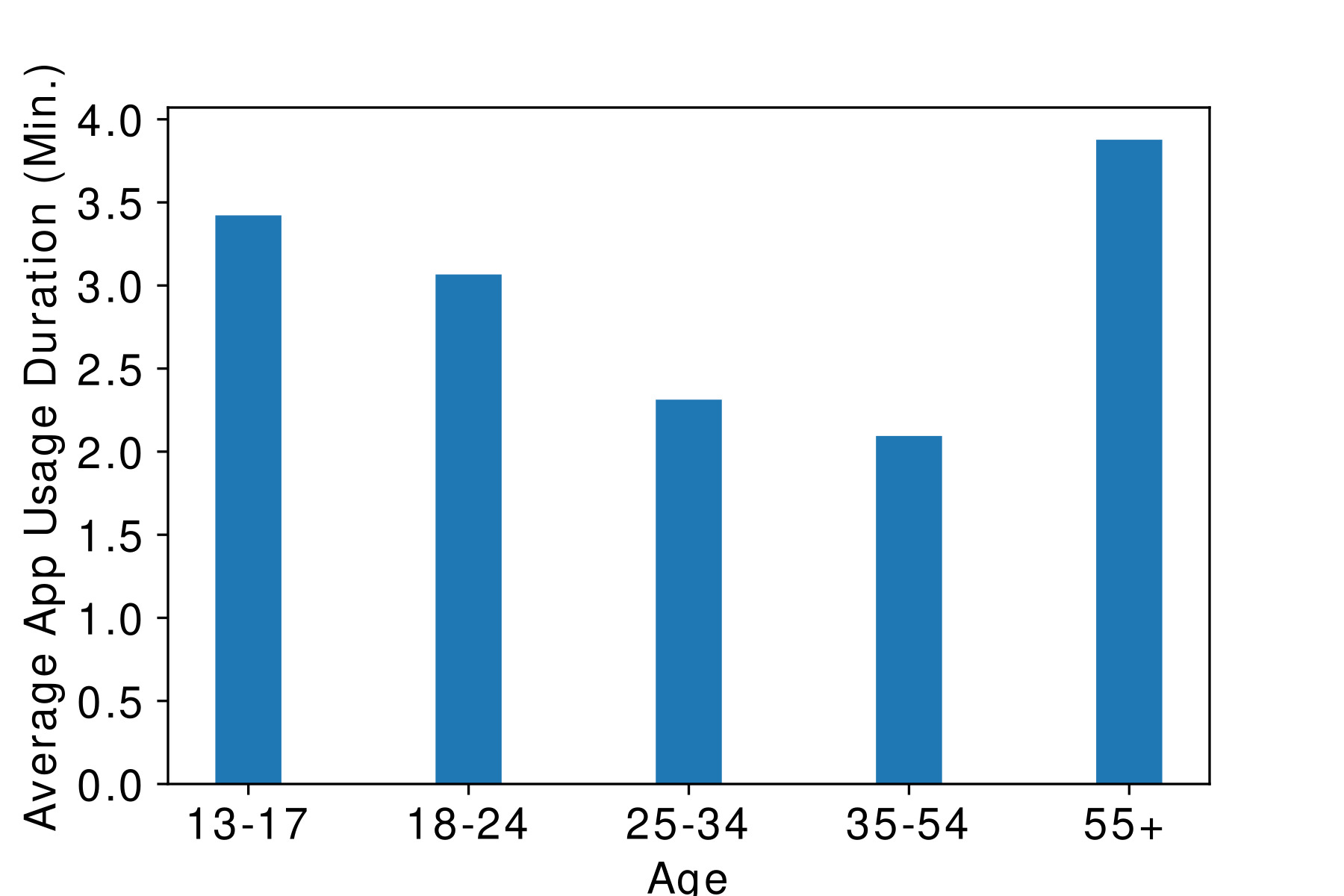} } 
	\caption{Effect of gender and age on app usage duration.}
	\label{fig:session_duration_user}	
\end{figure}

Prior studies \cite{zhao2016discovering, seneviratne2015your, kooti2017iphone} have shown that demographics (age and gender) have a significant impact on how users select apps or make in-app purchases on their smartphones. However, little is known about the correlation between users' demographic information and the time spent when using an app. Establishing such a relationship is the main focus of this subsection. By analyzing our user mobile app usage data, we demonstrate that on average, the app usage duration for female users is 2.8 minutes, which is longer than the male users who spent 2.2 minutes. To show the general pattern of usage duration across all apps, Figure~\ref{fig:session_duration_gender} presents the CDF of app usage duration for both the male and female users. We find that the duration distributions are similar between male and female users.

	\begin{table}
        \centering
        \scriptsize
		\caption{Top 5 app categories with the biggest gender effects -- the higher the gender effect is, the bigger difference exists in the app usage duration between male and female users. The gender group $G_1$ has longer app usage duration.}
		\label{tab:gender_bias}
	\scalebox{1}{\begin{tabular}{lcc|lcc}
			\toprule
			\textbf{App Category}&\textbf{Gender Group $G_1$}&\textbf{Gender Effect}&\textbf{App Category}&\textbf{Gender Group $G_1$}&\textbf{Gender Effect}\\
			\midrule	
			widgets&female&3.30&casino&male&1.81\\
			medical&female&2.54&video&male&1.64\\	
			entertainment&female&2.10&health-and-fitness&male&1.60\\	
			communication&female&1.64&finance&male&1.30\\
			shopping&female&1.63&navigation&male&1.27\\
			\bottomrule
		\end{tabular}}
	\end{table}

To investigate the effect of age, we demonstrate in Figure~\ref{fig:session_duration_age} that age affects app usage duration more significantly. Users between 25 and 54 years spend less time while engaging with apps than those who are teenagers or seniors. This scenario is consistent with results from the previous user studies \cite{ferreira2014contextual,pielot2014situ}. \citet{ferreira2014contextual} pointed out that micro-usage (less than 15 seconds) is popular across their participants aged between 22 and 40 years old. \citet{pielot2014situ} found that their participants  (aged between 24 and 43 years old) need to deal with the large volume of notifications coming from personal communication due to high social expectations and the exchange of time-critical information. 

Besides these findings of general patterns across all apps, we hypothesize that the impacts of demographics would vary across different app categories. Therefore, we further explore if users' app usage duration with each app category would be affected by their demographics. To measure the impacts brought by users with different genders in a more generalized manner, we calculate the \emph{gender effect} for each app category of the two genders. First, for each app category $C_a$, we calculate the average app usage duration of female and male users respectively, i.e., $D(C_a, Gi)$ where $Gi\in\{male, female\}$. Secondly we calculate the \emph{gender effect}, $GE(C_a)$, for each app category $C_a$ as:
    $$GE(C_a) = \frac{D(C_a, G_1)}{D(C_a,G_2)}$$
where $GE(C_a)$ measures the \emph{gender effect} of the app category $C_a$.
%and $G_1$ refers to \emph{male} users. 
The higher the effect is, the bigger difference exists in the app usage duration of the male and female users within this app category. The gender group in the numerator has a longer dwell time on this app category. Table~\ref{tab:gender_bias} shows the top 5 app categories with the highest gender effects. We can find the female users have a longer app usage duration than male users with apps like widgets, medical, entertainment, communication, and shopping. On the other side, male users stay longer with the casino, video, health-and-fitness, finance, and navigation apps.

 	\begin{table}
        \centering
        \scriptsize
		\setlength{\tabcolsep}{2pt}
		\caption{Top 5 app categories with the biggest age effects -- the higher the age effect is, the bigger difference exists in the app usage duration between users with different age. The age group $A_i$ has longer app usage duration when compared with other users.}
		\label{tab:age_bias}
		\scalebox{1}{\begin{tabular}{lcc|lcc|lcc}
			\toprule
			\textbf{App Category}&\textbf{Age Group $A_i$}&\textbf{Age Effects}&	\textbf{App Category}&\textbf{Age Group $A_i$}&\textbf{Age Effects}&	\textbf{App Category}&\textbf{Age Group $A_i$}&\textbf{Age Effects}\\
			\midrule	
			shopping&13-17&2.75&widgets&18-24&2.35&food-and-drink&25-34&1.59\\
			entertainment&13-17&1.81&entertainment&18-24&2.20&video&25-34&1.30\\
			social&13-17&1.28&business&18-24&1.65&medical&25-34&1.30 \\
			photography&13-17&1.27&tools&18-24&1.41&music&25-34&1.29\\	sports&13-17&1.24&adventure&18-24&1.41&transportation&25-34&1.27\\
			\midrule
			\textbf{App Category}&\textbf{Age Group $A_i$}&\textbf{Age Effects}&	\textbf{App Category}&\textbf{Age Group $A_i$}&\textbf{Age Effects}&&&\\
			\midrule
			lifestyle&35-54&1.29&productivity&55+&2.16&&&\\
			casino&35-54&1.27&entertainment&55+&2.11&&&\\
		    word&35-54&1.23&board&55+&2.00&&&\\
			travel&35-54&1.22&puzzle&55+&1.90&&&\\
			action&35-54&1.21&books&55+&1.88&&&\\
			\bottomrule
		\end{tabular}}
	\end{table}

Similarly, we calculate the \emph{age effect} brought by different users on the app dwell time.  Since there are five age groups in our dataset, we choose the average app usage duration of all users as the reference value to calculate the \emph{age effect}. Therefore, for each app category $C_a$, we calculate the average duration of users belonged to different age groups respectively, i.e., $D(C_a, Ai)$ where $Ai\in\{13$-$17$, $18$-$24$, $25$-$34$, $35$-$54$, $55+\}$. Then we calculate the \emph{age effect}, $AG(C_a)$, for each app category $C_a$ as:
    $$AG(C_a) = \frac{D(C_a, A_i)}{D(C_a,A)}$$
which measures the \emph{age effect} of the app category $C_a$. $A$ represents the users across all ages who have engaged with the app category $C_a$. The higher the effect is, the bigger difference exists in the app usage duration of the users with corresponding ages $A_i$. The age group in the numerator has a longer dwell time on this app category when compared with other users. Table~\ref{tab:age_bias} shows the app categories with top \emph{age effects}. 
It is interesting to find that teenage users have a longer duration in the shopping apps. %which could be due to the browsing behaviour since most products are not affordable for them. 
Users between 25 and 34 prefer to stay longer with the food-and-drink apps than other users. 
%Users in this age range may have more chances to dine together with their friends or colleagues and have higher requirements for the restaurants. 
It is not surprising to find that the older users over 55, will spend a much longer time when using the apps of entertainment, games (board/puzzle), and books apps since they have more spare time. For the longer time spent in productivity apps, this may result from that old users are not as efficient in the operations with the productivity apps (e.g., Microsoft Office, file managers). %the app developers may provide more simple designed functions for the older users to improve their user experience. 

\subsubsection{Device}
\citet{li2017mining} ever analyzed the influence brought by different mobile device models on users' online time and they found that users rely less on the cellular network as the price of the device model increases. However, there has been no empirical research conducted for the impacts on dwell time with different apps brought by device characteristics. In this section, we compare app usage duration across different device types. 
%If users holding specific devices stay longer on a few categories of apps each time, the developers, web content providers, and advertisers can leverage such knowledge to customize more accurate advertisements to the audience.  
There are mainly two different types of devices in our dataset: tablet and phone. We use the similar methodology of calculating the \emph{gender effect} for each app category to calculate the \emph{device effect} with the two device types. First, for each app category $C_a$, we calculate the average usage duration of phone users and tablet users respectively, i.e., $D(C_a, T_i)$ where $T_i \in \{phone, tablet\}$. Secondly, we calculate the \emph{device effect}, for each app category $C_a$ as:
$$DTE(C_a) = \frac{D(C_a, T_1)}{D(C_a, T_2)}$$
The higher the \emph{device effect} $DTE(C_a)$ is, the bigger difference exists in the app usage duration of the users with different device types.  Table~\ref{tab:device_effect} shows the app categories with the most significant difference across the different devices. We can find that the tablet users have a longer dwell time on productivity, books, business, board, and travel apps. On the other hand, phone users have a longer dwell time with navigation, shopping, weather, family, and personalization apps.
	\begin{table}
        \centering
        \scriptsize
		\caption{Top 5 app categories with the biggest device effects -- the higher the device effect is, the bigger difference exists in the app usage duration between the phone and tablet users. The users with device type $T_1$ have longer app usage duration.}
		\label{tab:device_effect}
		\scalebox{1}{\begin{tabular}{lcc|lcc}
			\toprule
			\textbf{App Category}&\textbf{Device Type $T_1$}&\textbf{Device Effects}&\textbf{App Category}&\textbf{Device Type $T_1$}&\textbf{Device Effects}\\
			\midrule	
			productivity&tablet&4.95&navigation&phone&8.04\\
		    books&tablet&3.29&shopping&phone&6.02\\
		    business&tablet&3.00&weather&phone&1.29\\
		    board&tablet&2.91&family&phone&1.22\\
		    travel&tablet&2.82&personalization&phone&1.19\\
		    \bottomrule
		\end{tabular}}
	\end{table}

\subsection{Temporal Patterns}
\begin{table}
\scriptsize
\centering
	\caption{Index of dispersion %(normalized variance)  = $\frac{\sigma^2}{\mu}$ 
	of average app usage duration distribution across a day.} 
	\label{tab:session_duration_variance}
        \scalebox{1}{\begin{tabular}{lc|lc}
		\toprule
		\textbf{App Category}&\textbf{Index of Dispersion}&\textbf{App Category}&\textbf{Index of Dispersion}\\
		\midrule    
        tools&0.02&comics&4.18\\
        productivity&0.02&adventure (game)&3.86\\
        utilities&0.02&strategy (game)&2.51\\
        music&0.04&racing (game)&2.05\\
        photography&0.05&family&1.59\\
        news&0.06&transportation&1.35\\
        widgets&0.07&medical&1.03\\
		\bottomrule
	\end{tabular}}
\end{table}

\begin{figure}
	 \centering
	   \includegraphics[width = 0.65\textwidth]{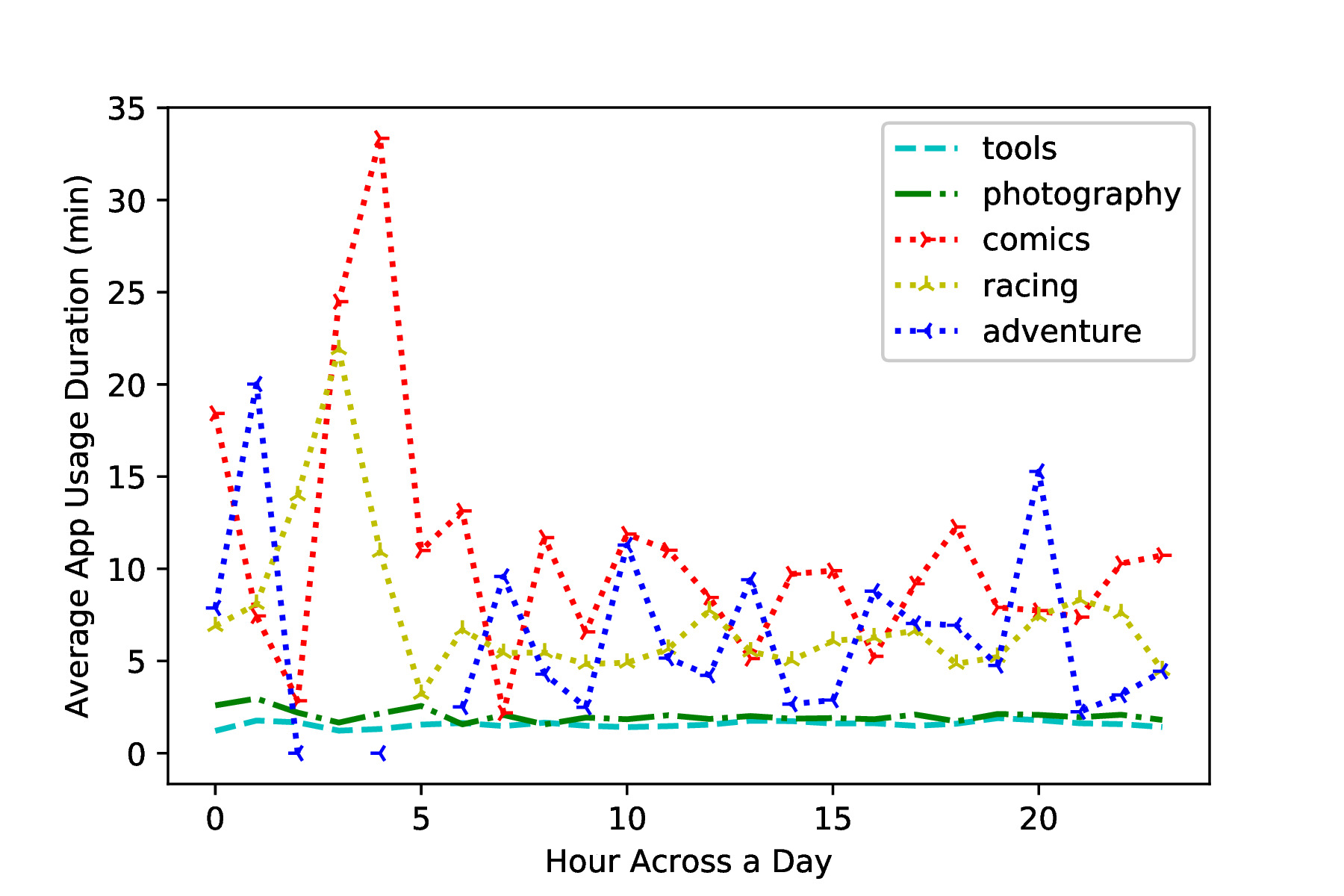}
	   \caption{Temporal patterns of average app usage duration for different app categories.}
	   \label{high_var_hour}
\end{figure}

The temporal usage pattern is always an essential factor for inferring which app user will use, where we usually model the possibility of using different apps at a specific time \cite{huang2012predicting, shin2012understanding, liao2012mining}. From our descriptive analysis in Figure~\ref{fig:hour_distribution}, we can find that the average app usage duration with all app categories across a day is about 2 to 3 minutes, and no sharp fluctuations exist. We use the index of dispersion  \cite{cox1966statistical} as a normalized measure of the dispersion for the app usage duration distribution across a day, which is defined as the ratio of the variance $\sigma^2$ to the mean $\mu$: $D =\frac{\sigma^2}{\mu}$. In general, the index of dispersion for all apps is 0.043. To validate that whether the temporal pattern varies on different app categories, we further explore the index of dispersion of the app usage duration across a day for different app categories respectively. Table~\ref{tab:session_duration_variance} shows the app categories with the smallest and biggest index of dispersion in app usage duration. We can find that the usage duration of most game apps will be significantly different across a day. However, for some functional apps like productivity, tools, utilities, and photography, whenever the user accesses them, the app usage duration does not change much. Figure~\ref{high_var_hour} illustrates the distribution of average usage duration across a day for several app categories, where we can find that besides the different variances, the temporal patterns could be significantly different with various app categories (each app category has its own specific temporal pattern). For example, comics apps have a longer usage duration around 4 AM; the usage duration of racing apps is peaking around 3 AM; longer time is spent in adventure apps around 8 PM and 1 AM.

\subsection{Short-term context}

\begin{figure}[t]
\subfloat[]{\label{fig_last:sub-first}
  \centering
  % include first image
  \includegraphics[height = 1.9in]{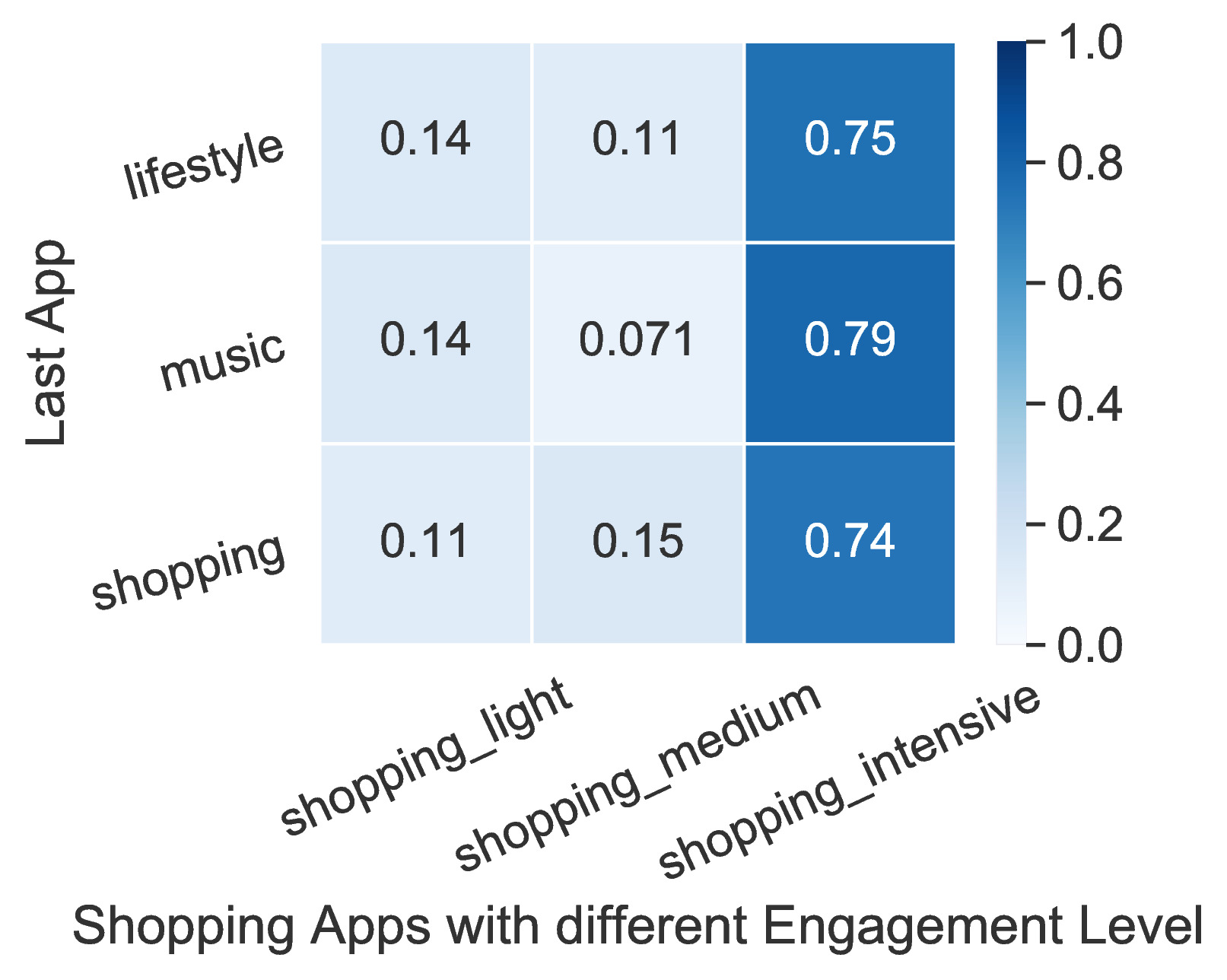}}
\subfloat[]{\label{fig_last:sub-second}
  \centering
  % include second image
  
  \includegraphics[height = 1.88in]{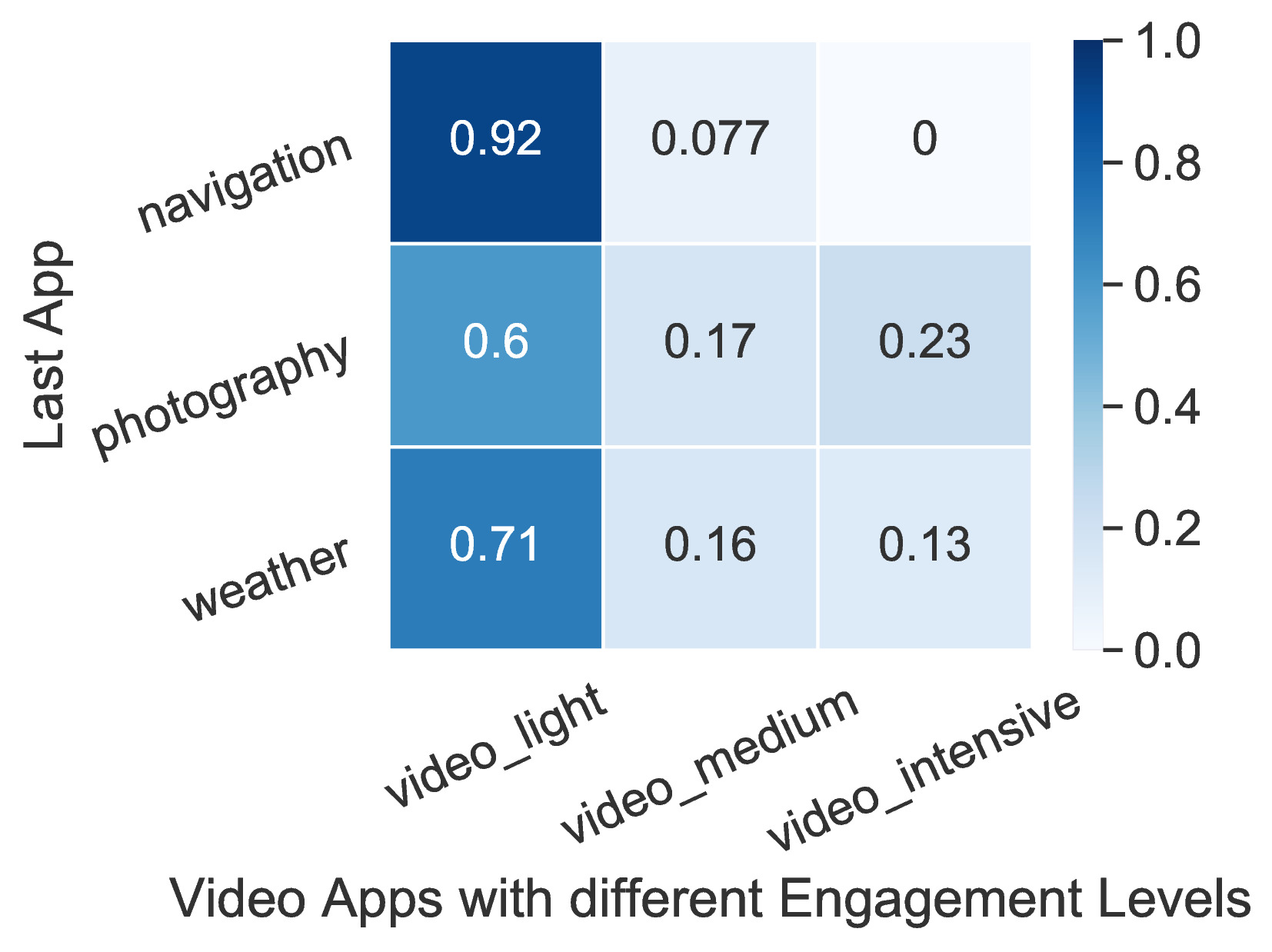}}\\
\subfloat[]{\label{fig_last:sub-third}
  \centering
  % include second image
  \includegraphics[height = 1.92in]{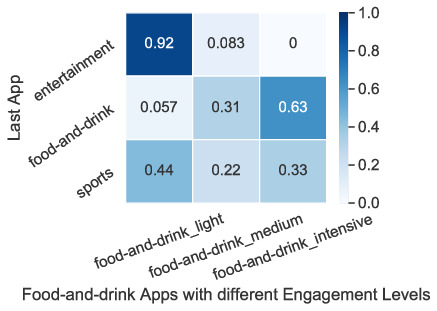}}
\subfloat[]{\label{fig_last:sub-fourth}
  \centering
  % include second image
  \includegraphics[height=1.88in]{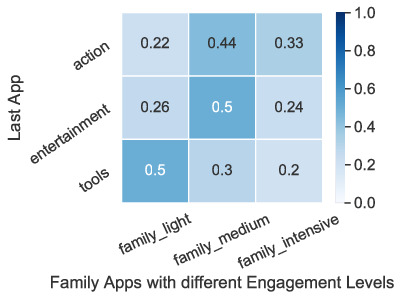}}
\caption{Correlation between last used app and the next app's engagement level (discrete representation of app usage duration: light, medium and intensive). The darker color means higher probabilities that the next app will be engaged with the corresponding engagement level.}
\label{fig:last_app_heatmap}
\end{figure}

\subsubsection{Last Used App}
Previous studies identified that some apps were often used together \cite{li2015characterizing,xu2011identifying, bohmer2011falling}. For example,  \citet{li2015characterizing} showed that some apps are installed together, whereas other studies \cite{xu2011identifying, bohmer2011falling} found that some genres of apps are highly likely to be accessed sequentially. Therefore, we aim to explore whether the last used app could also impact the next app usage duration. To avoid the biases of the original differences of app usage duration across different app categories and measure the impacts more intuitively, we quantify the app usage duration into the corresponding engagement level defined as in Sec~\ref{engagment_level_definition}. Figure~\ref{fig:last_app_heatmap} shows several examples for illustrating the different patterns between specific last apps and the engagement levels of next apps. Each cell in the table represents the transition probability $P_{ij}$ from the last used app $a_i$ to the next app $a_j$ at a given engagement level $e_{a_j}$. 
%To quantify the impacts brought by different last used apps to the next app engagement levels, 
By having $P_{ij}$ for all the three engagement levels, we calculate its standard deviation as $\sigma^i_j$. This represents the extent of how much the last app $a_i$ would have very different transition probabilities on different engagement levels of the next app $a_j$. The higher $\sigma^i_j$, the more likely the last app $a_i$ would result in a specific engagement level of the next app $a_j$.
%
%(bigger $\sigma_p$ means bigger differences exist in the $P_{ij}$). 
For a given app $a_j$, we average $\sigma^i_j$ across all the last used apps $a_i$ as $\sigma_j = \sum^{i=1..N} \sigma^i_j/N$, where $N$ is the amount of unique last used apps. %The metric $\sigma_j$ represents   
%So the higher $\sigma_a$ means the time spent on app category $a$ will be more significantly impacted by different last used apps. 
For all the apps, the $\sigma_j$ ranges between 0.11 and 0.32, whereas the median is 0.15. For the apps with higher $\sigma_j$, i.e., shopping apps ($\sigma_j$ $\approx$ 0.32) and video apps ($\sigma_j$ $\approx$ 0.29), the last used app could lead them to be used with specific engagement level (intensive for shopping and light for video). For example, for the video apps (as shown in Figure ~\ref{fig_last:sub-second}), if they are used after using navigation apps, the time spent on the video app is much more likely to be short. This could be explained by users who are on their commute to work. After checking the navigation, users may need to wait for the bus or subways, so they could have time for enjoying a short video. Additionally, we also show the app categories with average $\sigma_j$ and lower $\sigma_j$, i.e., food-and-drink apps ($\sigma_j$ $\approx$ 0.16) and family apps ($\sigma_j$ $\approx$ 0.11), whereas the last used app could not impact the engagement level of next app usage significantly. For example, no matter which app is used before the family apps (as shown in Figure~\ref{fig_last:sub-fourth}), the time spent on the family apps do not differ much.

\subsubsection{Last Engagement Level}
\begin{comment}
\begin{figure}
	 \centering
	   \includegraphics[width =0.9\textwidth]{pictures//last_engagement_heatmap.eps}
	   \caption{Transition probability of the last engagement level and next engagement level of the same app usage.}
	   \label{fig:last_engagement_heatmap}
\end{figure}
\end{comment}

\begin{figure}[t]
\subfloat[Books Apps]{\label{fig_last_en:sub-first}
  \centering
  % include first image
  \includegraphics[height =1.7in]{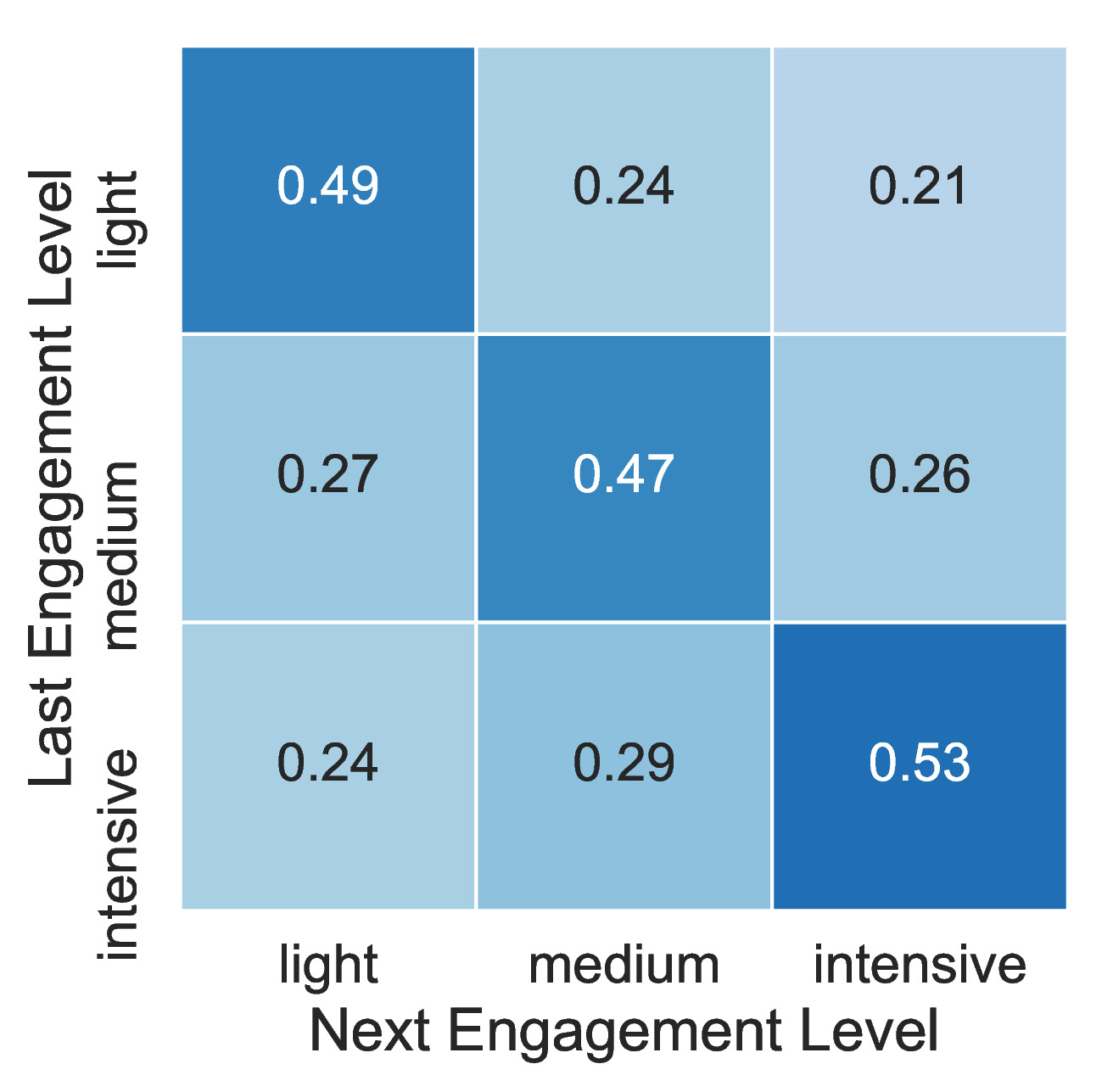}}
   \hspace{35pt}
\subfloat[Food-and-drink Apps]{\label{fig_last_en:sub-second}
  \centering
  % include second image
  \includegraphics[height = 1.7in]{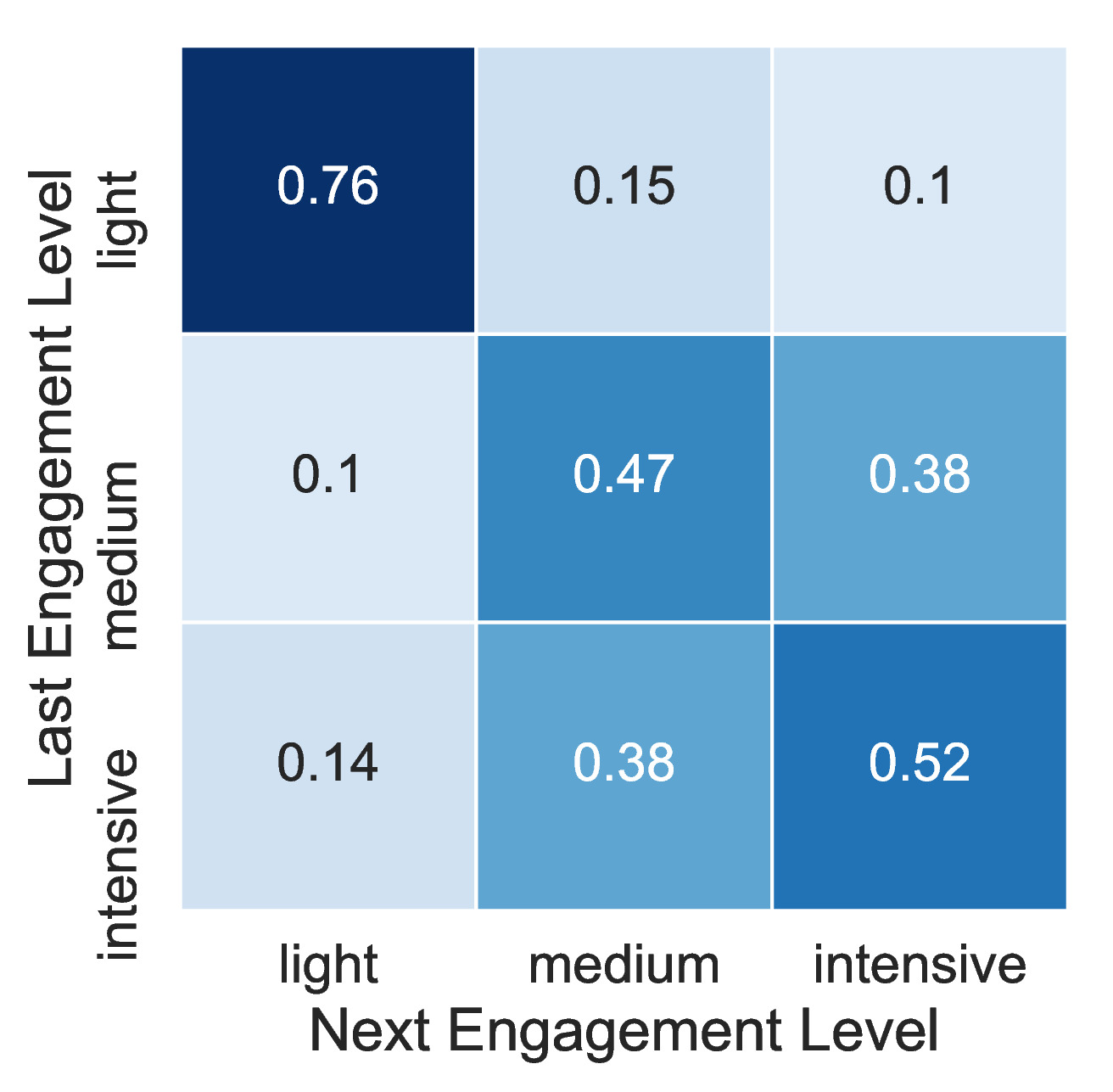}}\\
\subfloat[Comics Apps]{\label{fig_last_en:sub-third}
  \centering
  % include second image
  \includegraphics[height = 1.7in]{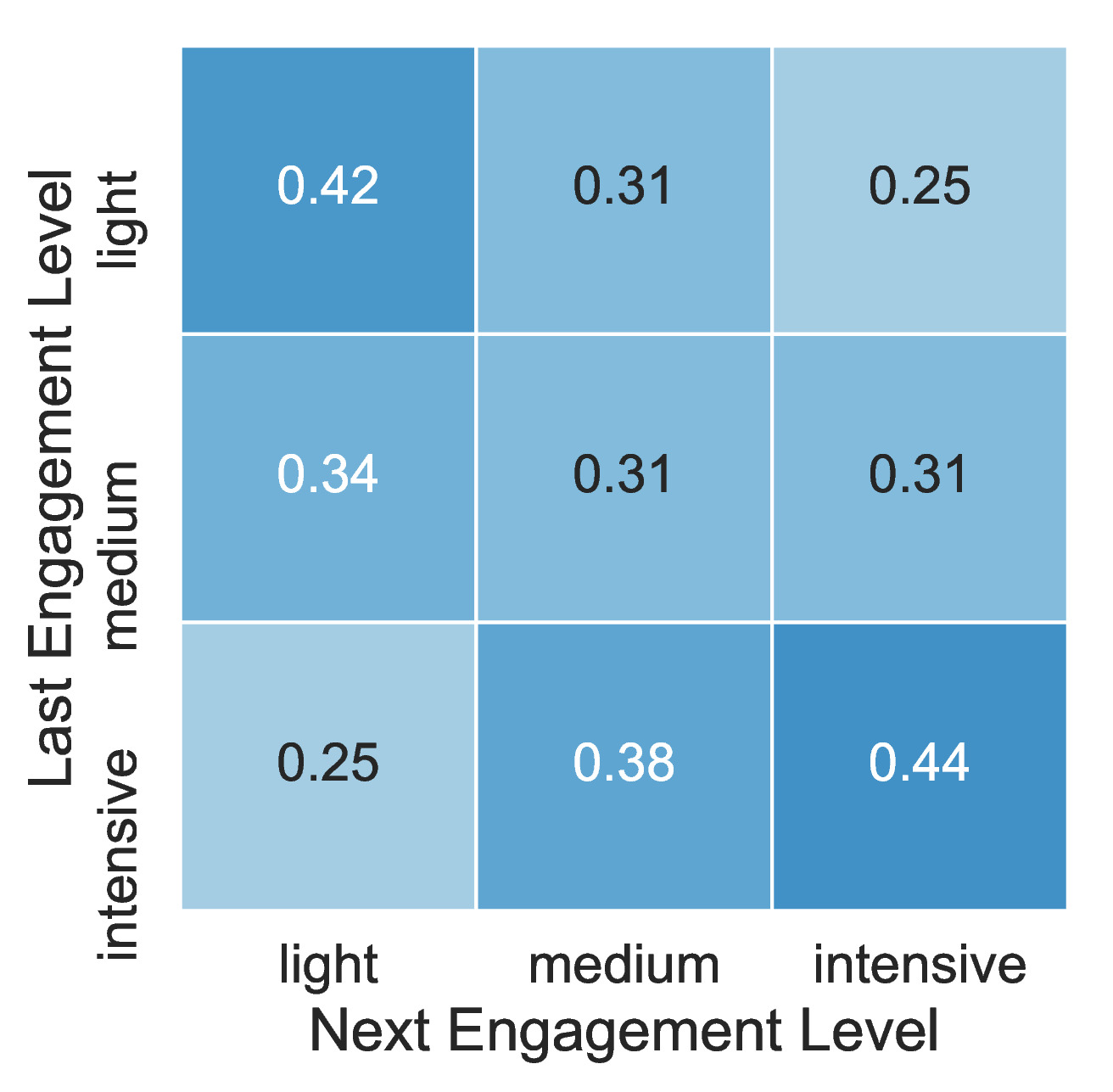}}
  \hspace{35pt}
\subfloat[Travel Apps]{\label{fig_last_en:sub-fourth}
  \centering
  % include second image
  \includegraphics[height=1.7in]{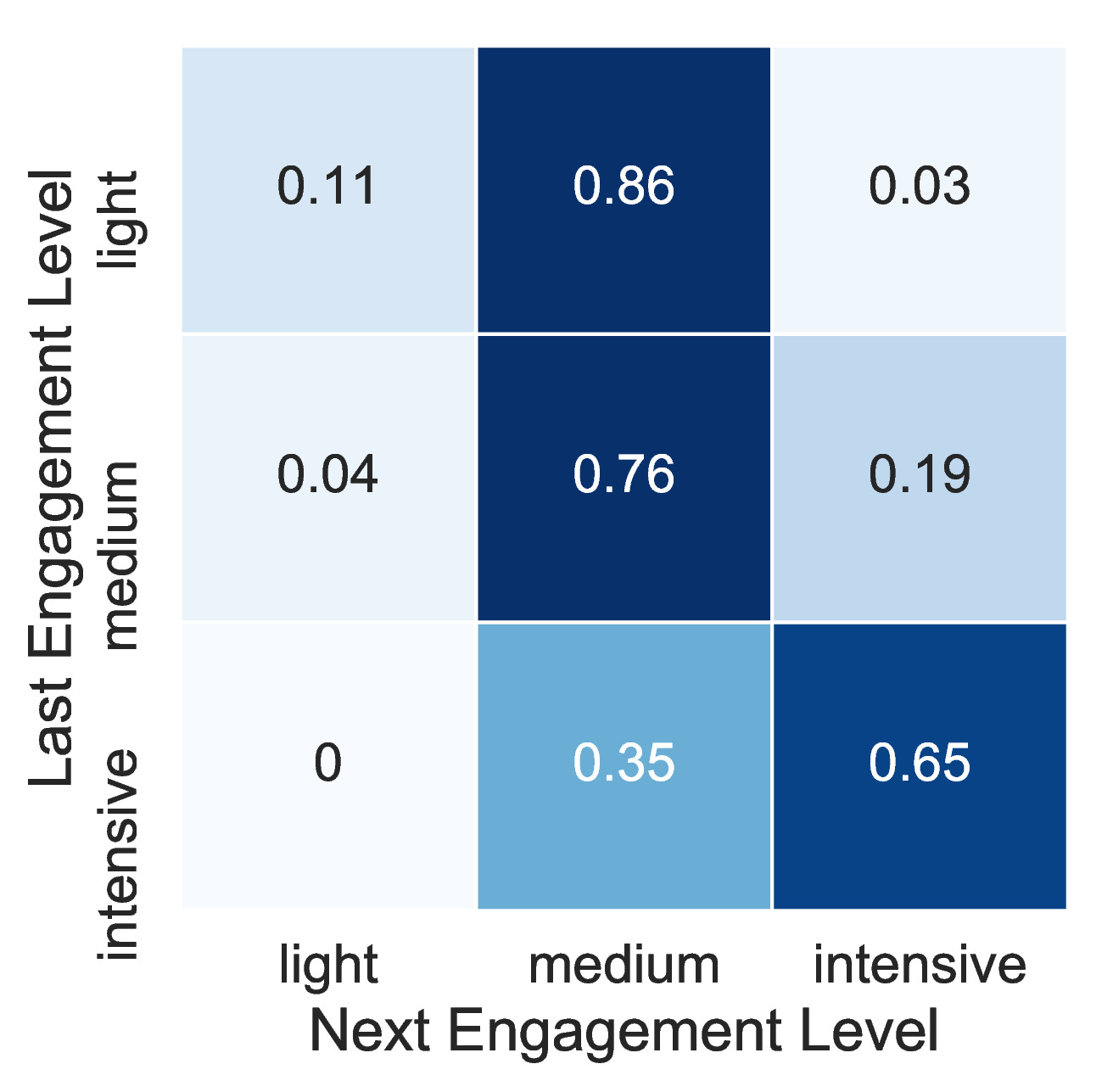}}
\caption{Transition probability of the last engagement level and next engagement level of the same app usage.}
\label{fig:last_engagement_heatmap}
\end{figure}

Besides the last used app, we hypothesize that the time spent on the last app (last engagement level) could also impact how long the user will stay with this app. 
To focus solely on the impact of the last engagement level, rather than the impact of last used app, we calculate the transition probability between the engagement levels of the last and next usage \emph{within the same app}.
We find four typical patterns across all different app categories.
%three common patterns and one specific pattern. 
We illustrate these correlation patterns between the last and next engagement levels in Figure~\ref{fig:last_engagement_heatmap}. Across all 45 app categories, 37.8\% (17/45) of them follow the pattern of maintaining a higher probability that the next engagement level is as same as the last engagement level (i.e.,~higher probability in the diagonal of the heatmap), such as the books apps shown in Figure~\ref{fig_last_en:sub-first}. 
Besides this common pattern, we also find that 44.4\% (20/45) of the app categories have a bidirectional (increasing/decreasing) trend with the closest level, like the food-and-drink apps (i.e., engagement level transition between medium and intensive). The last common pattern is illustrated by comics apps, 13.3\% (6/45) of app categories have a similar transition probability across all levels. The remaining pattern is illustrated by travel apps, where no other app categories have similar patterns with them. We can find that there is an apparent increasing trend between engagement level light and medium for travel apps, which could state that users may be more likely to get addicted to planning for a vacation within a travel app. %This also identifies that users could indulge themselves easily in reading comics. 

\subsection{Long-term Context}

\begin{figure}[t]
\subfloat[Weather Apps]{\label{fig_per:sub-first}
  \centering
  % include first image
  \includegraphics[height = 1.7in]{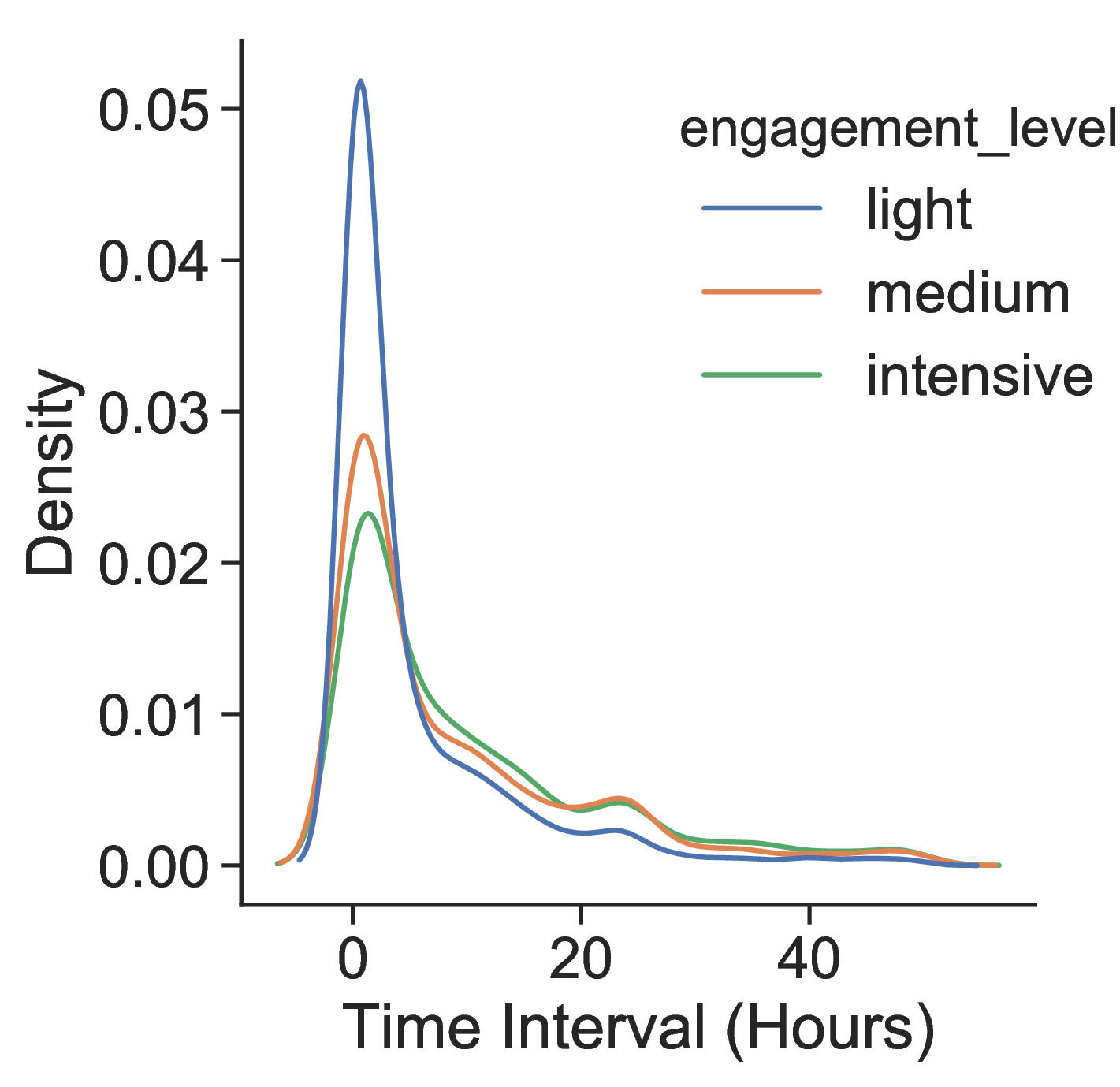}}
  \hspace{35pt}
\subfloat[Shopping Apps]{\label{fig_per:sub-second}
  \centering
  % include second image
  \includegraphics[height = 1.7in]{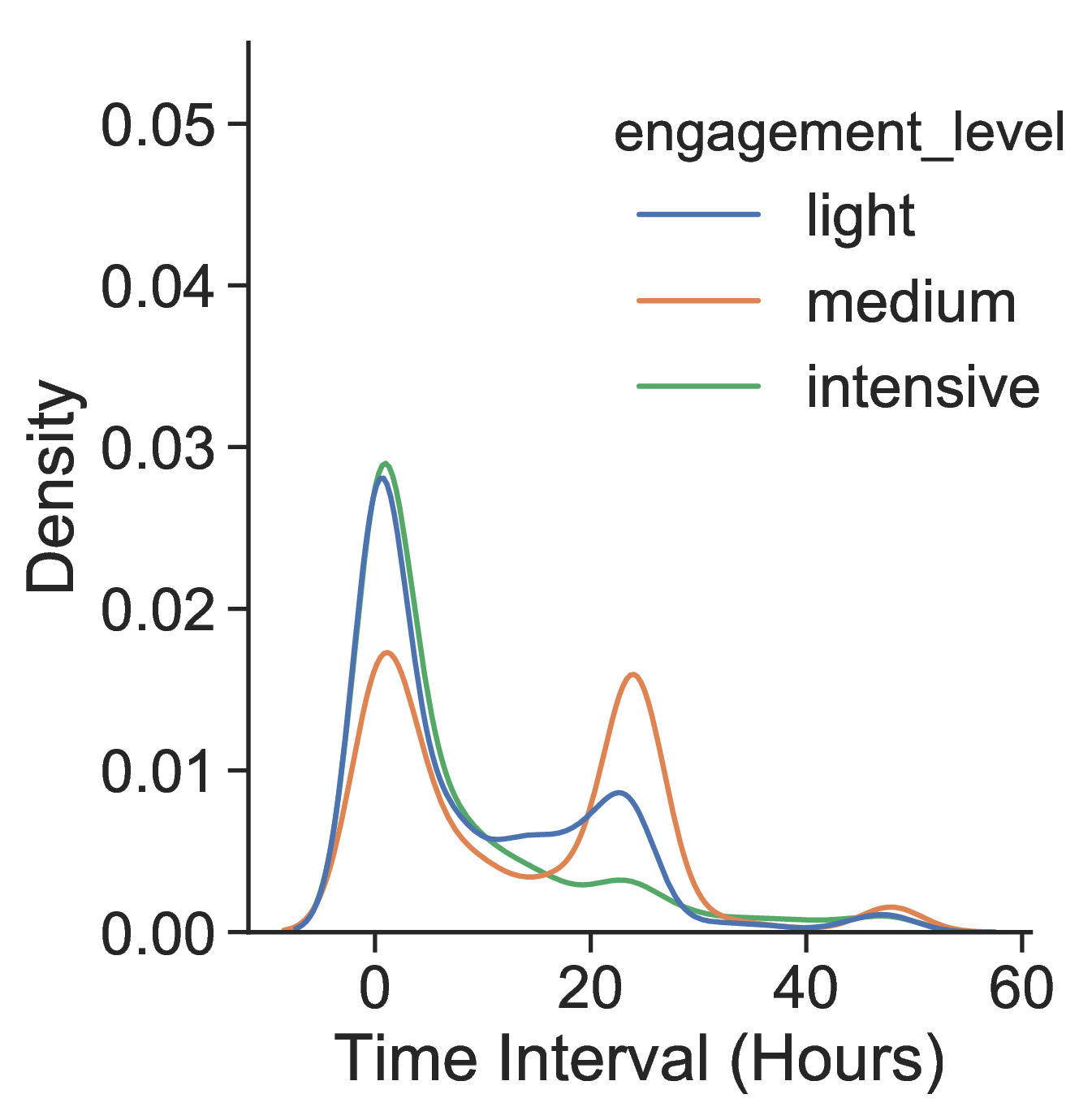}}
\caption{Two typical interval time distribution with different engagement levels (we limit the x-axis to 50 hours since over 98\% intervals are shorter than 50 hours): (1) General trend illustrated by weather apps: the shorter interval between two accesses, the less time is spent with the next usage; (2) Specific daily periodic pattern illustrated by shopping apps: similar length of time is spent on the same shopping app after a specific interval (i.e., 24-hour). }
\label{fig:engagement_periodic}
\end{figure}

\subsubsection{Periodic Pattern}
Some apps are used repeatedly after every specific period. For example, users may check the mail apps every hour or play with a game app every 3 hours \cite{liao2012mining, tan2012prediction}. To validate whether the periodic pattern also exists in the app dwell time (i.e.,~after a specific interval, users may stay with an app again for the similar length of time as before that interval), we first quantify the interval between the two accesses of the same app at the hour level,  e.g.,  28 min = 0 hour,   1.6 hours = 2 hours. We then generate the histogram of interval time length with different engagement levels respectively for each app category. Two typical patterns are found among all the app categories, which are shown in Figure~\ref{fig:engagement_periodic}. Figure~\ref{fig_per:sub-first} denotes the interval time length distribution of weather apps, which represents the general trend that a set of most other app categories will obey, i.e.,~the shorter interval between two accesses of the same app, the less time (engagement level: light) user will spend within the next usage of this app.

However, as shown in Figure~\ref{fig_per:sub-second} (i.e.,~ the shopping apps), the other typical pattern demonstrates that the less time between two accesses, the more likely the second app access will also result in a longer time stay (engagement level: intensive) on the app. This may be because when users are going to buy something, they will browse/revisit the shopping apps multiple times with short breaks (break for checking other information or chat with friends for asking advice) and finally place the order. 
Additionally, we find that there is a peak around the interval of 24 hours for the engagement level of light/medium for shopping apps. This states that users prefer to spend a similar length of time to regularly check the shopping apps every day (i.e., after every 24 hours), probably for checking the updated discount or product information. This pattern could also be observed with books apps.

\begin{comment}
\begin{figure}
	 \centering
	   \includegraphics[width = 0.65\textwidth]{pictures//Engagement_Periodic_2.eps}
	   \caption{Two typical interval time distribution with different engagement levels (we limit the x-axis to 50 hours since over 98\% intervals are shorter than 50 hours): (1) General trend illustrated by weather apps: shorter interval between two accesses, less time is spent with the next usage; (2) Specific daily periodic pattern illustrated by shopping apps: similar length of time is spent on the same shopping app after a specific interval (i.e., 24-hour). }
	   \label{fig:engagement_periodic}
\end{figure}
\end{comment}

\subsubsection{Historical Interest Pattern}
\begin{figure}
%\vspace{-0.1in}
\centering	
\subfloat[User A]{
			\label{fig:user_a_historical}
			\includegraphics[width=0.9\textwidth]{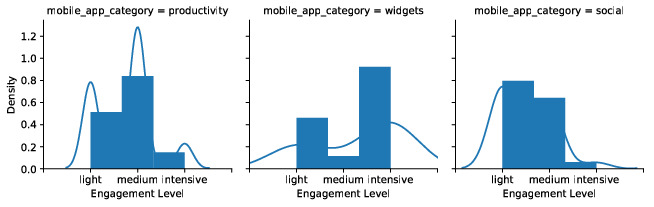} }\qquad
\subfloat[User B]{
		\label{fig:user_b_historical}		\includegraphics[width=0.9\textwidth]{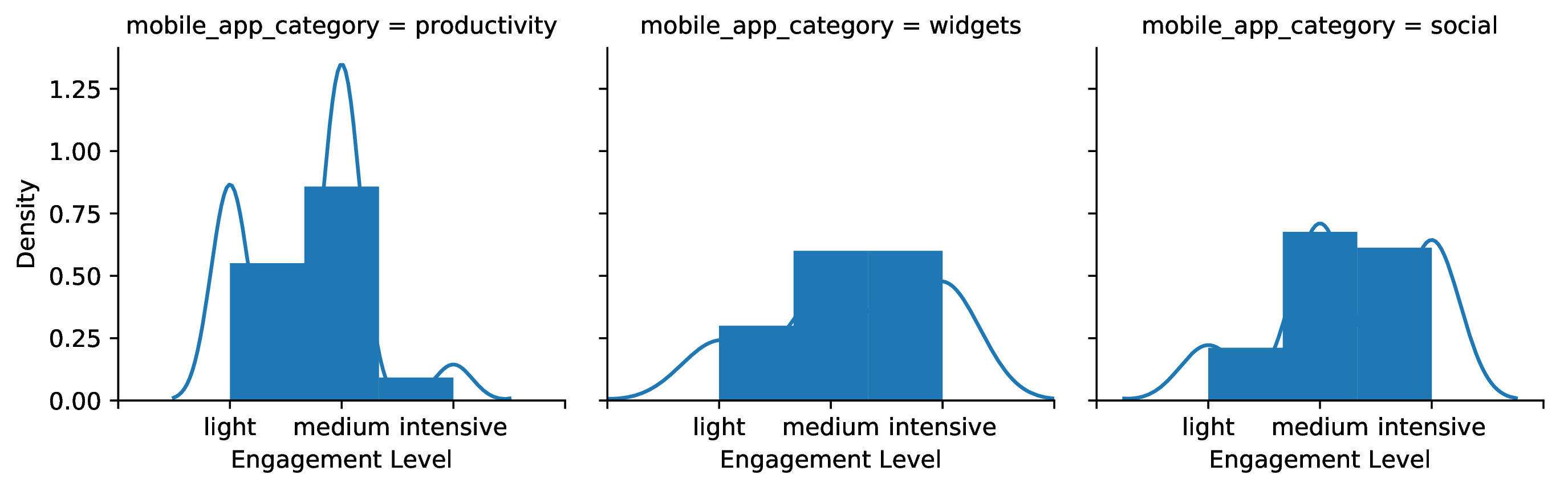} }\qquad 
\subfloat[User C]{
		\label{fig:user_c_historical}		\includegraphics[width=0.9\textwidth]{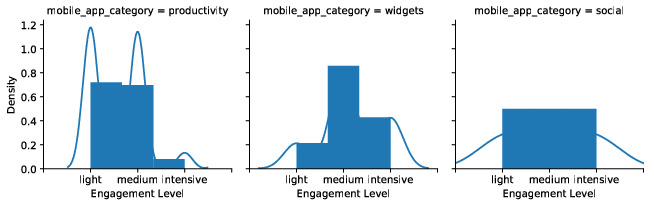} } 
	\caption{Historical pattern with app engagement levels.}
	\label{fig:Historical_Engagement}	
\end{figure}

Users' historical interests may be indicative of future intent. For example, users who love sports might potentially access sports apps and consume more sports-related content than others.  Therefore it is important to examine the long-term interest patterns of different users. To clarify that if users also have the historical habit for the app dwell time on different app categories, we select three users whose top three preferred (based on usage frequencies) app categories are all the same, which are productivity, widgets, and social apps. We then illustrate the histogram of their historical engagement levels within these three app categories in Figure~\ref{fig:Historical_Engagement}. We can find that \emph{User A} prefers to stay longer (engagement level: intensive) within the widgets apps but spend less time (engagement level: light) with social apps than others. Differently, \emph{User B} has a longer app usage duration (engagement level: intensive) with social apps. Therefore, even for the users with the same historical app preferences,  their engagement habits would be different. In a summary, different users may have their own specific historical engagement patterns on app dwell time for various apps. 

\section{App Usage and Engagement Prediction}
\label{sec:models}
Based on the statistical analysis in Section~\ref{infering_features} of app usage duration, we further focus on answering the important question: could we predict which app user will use and how long the user will stay on this app simultaneously? In this section, we start by formulating the next app usage and app dwell time prediction problem, followed by presenting several joint learning strategies for solving these two prediction problems simultaneously.

\subsection{Problem Formulation}
\label{sec:problem_formula}
For the next app prediction, many previous researchers have extracted the predictive features and proposed methods to solve it \cite{huang2012predicting,shin2012understanding, zou2013prophet,yan2012fast}. Similar to Ricardo et al. \cite{baeza2015predicting}, we model the prediction of the next app as a supervised classification problem. For the novel prediction problem proposed in our work, which is to infer users' app dwell time,  we also model it as a multi-class classification problem and the continuous dwell time (app usage duration) is represented as the discrete engagement levels illustrated in Section \ref{engagment_level_definition}. More specifically, given the current context $c$ and a user $u$, we need to predict which app $a$ the user $u$ will use next and the engagement level $e$ (light/medium/intensive) for measuring how long the user will stay with this app. We formulate our prediction problem as quaternion, i.e., $\{c,u,a,e\}$. In contrast to a traditional next app usage prediction formulation as $a^* = \operatorname*{argmax}_a \mathcal{F}(a|c,u)$, we formulate the joint learning task for predicting next app and engagement level as follows:
\begin{equation}
(a,e)^* = \operatorname*{argmax}_{a,e} \mathcal{F}((a,e)|c,u).
\end{equation}
Note that the $a$ is in the condition for $e$ since the engagement level is defined based on their corresponding app $a$ (\S~\ref{engagment_level_definition}).

\subsection{Joint Learning Prediction Model} %Building}
\label{sec:jointmodels}
We propose three methods to solve the prediction problem formulated in \S~\ref{sec:problem_formula}: sequential based model (\S~\ref{sequential}), stacking based model (\S~\ref{sec:stacking}) and boosting based model (\S~\ref{sec:boosting}). 
\begin{comment}
We decompose the joint learning task into subtasks:
\textbf{Subtask 1} (NEXT APP PREDICTION) \emph{Given a user $u$ and the current context $c$, we predict the app $a$ will be used.} undoubtedly, the next app usage prediction is a key issue to the joint learning task. If a predicted app is not the one user will use next, the entire task might become meaningless. Subtask 1 aims at modelling app usage. Here we could apply a supervised next app prediction model to measure which app user will use next.
\begin{equation}
    a^* = \operatorname*{argmax}_a P(a|c,u).
\end{equation}
\textbf{Subtask 2} (ENGAGEMENT LEVEL PREDICTION) \emph{Given a user $u$, the current context $c$, and a candidate app $a$, we predict which engagement level $e$ (light/medium/intensive) user will have within this app.}
This subtask characterizes another key issue for our task, which is to model how long a user will stay with an app. We can apply another supervised learning model to infer which engagement level the user will have.
\begin{equation}
    e^* = \operatorname*{argmax}_e P(e|a,c,u).
\end{equation}
Yet, these two subtasks should not be isolated. Since the choice of the app and the engagement level are provided as a pair, moreover, the result of the second subtask (engagement level prediction) will be influenced by the first subtask (next app prediction). So we propose three different strategies for solving these two subtasks jointly. 
\end{comment}

\subsubsection{\textbf{Sequential based Joint Prediction}}
\label{sequential} 
The most straightforward method for solving this joint learning problem is to perform these two prediction problems (next app and engagement level prediction) sequentially. It is undoubted that the next app usage prediction is a key issue to the joint learning task. If a predicted app is not the one user will use next, the engagement level prediction might become meaningless. Therefore, we first apply a supervised next app prediction model to measure which app user will use next:
\begin{equation}
    a^* = \operatorname*{argmax}_a P(a|c,u). 
\end{equation}
Since the engagement level is dependent on which app the user will use next, %where we need to capture the correlations. 
therefore, another supervised learning model is applied to infer which engagement level the user will have given the predicted app user will use:
\begin{equation}
    e^* = \operatorname*{argmax}_e P(e|a,c,u).
\end{equation}
Therefore, the sequential based joint prediction could be formulated as follows:
\begin{equation}
(a,e)^* = \operatorname*{argmax}_{a,e}P(a|c,u)P(e|a,c,u).
\end{equation}

\begin{figure}
	 \centering
	   \includegraphics[width =0.85\textwidth]{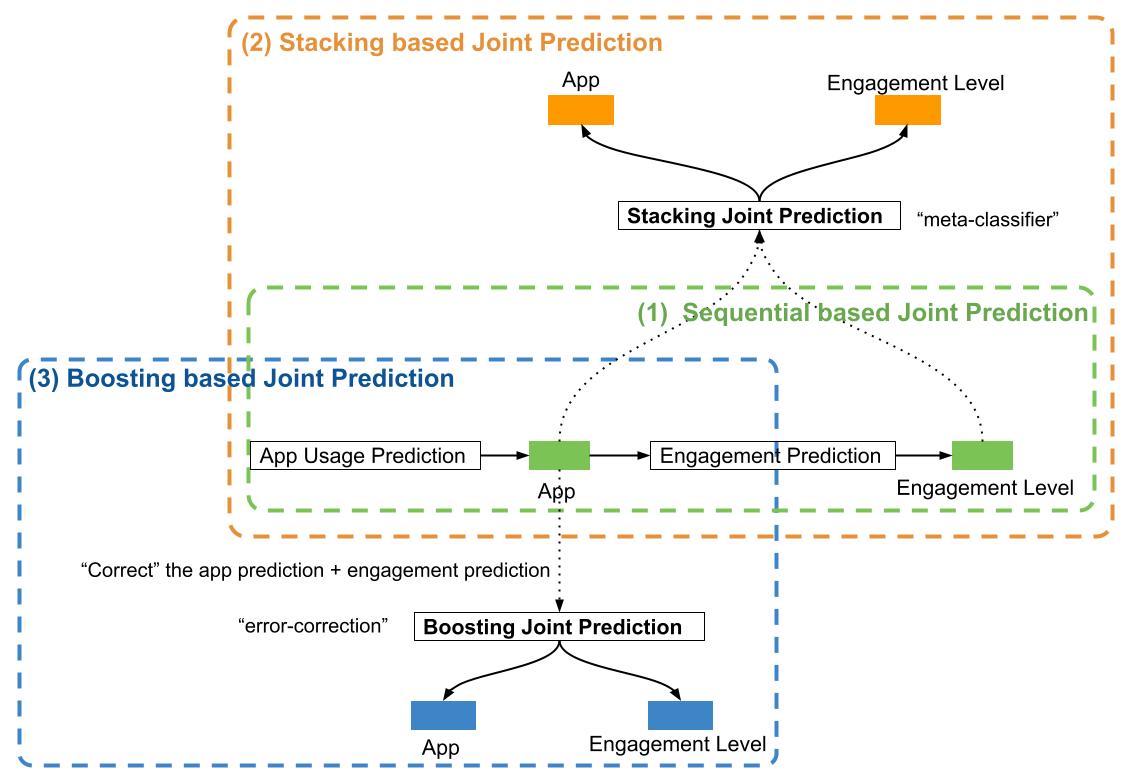}
	   \caption{Overview of three joint learning strategies: (1) Sequential based joint prediction; (2) Stacking based joint prediction adds a "meta"-classifier to the final stage after we have the prediction results of next app and engagement level sequentially; (3) Boosting based joint prediction has an "error-correction" of next app prediction in the second step for learning of engagement level prediction.}
	   \label{fig:sequential_stacking}
\end{figure}

As shown in Figure~\ref{fig:sequential_stacking} (in the green circle), in this sequential-based joint prediction strategy, a next app prediction model is needed to find the proper app $a^*$ user will use and then an engagement level prediction model is needed to infer the engagement level $e^*$ based on $a^*$. However, the predicted app $a^*$ is not guaranteed to be the right app; moreover, if the app is not the user will use next, the inferred engagement level based on this app would become meaningless. To this end, we propose two other joint learning methods for predicting these two problems more effectively.

\subsubsection{\textbf{Stacking based Joint Prediction}}
\label{sec:stacking}
Stacking \cite{syarif2012application} is an ensemble learning technique that combines multiple classification or regression models via a meta-classifier or a meta-regressor. The base-level models are trained based on a complete training set, then the meta-model is trained on the outputs of the base level models as features. So the main idea for our stacking based joint prediction model is to add a "meta"-classifier to the final stage after we got the prediction results on app and engagement level respectively from the sequential based model (\S~\ref{sequential}). Then we may improve the performance by adding this "meta-correction" step at the end of the prediction. In our scenario, the stacking consists of two levels which are base learner as level-0 and stacking model leaner as level-1, as shown in Figure~\ref{fig:sequential_stacking} (in the orange circle). So the base learners (level-0) are composed of the next app usage prediction model (Eq. (2)) and engagement level prediction model (Eq. (3)), which are the same as in sequential based joint prediction shown in Figure~\ref{fig:sequential_stacking}. The outputs of each of the sequential classifiers ($a'$ and $e'$) are collected to create a new dataset. Then the new dataset is used for stacking model learner (level-1) to provide the final output ($a^*$ and $e^*$). In this way, the predicted classifications from the two base classifiers at level-0 can be used as input variables into a meta-classifier as a stacking model learner, which will attempt to learn from the data on how to combine the predictions from the base models to achieve the best classification accuracy. The stacking based joint prediction in our scenario could be formulated as:

\begin{align}
    (a,e)' &= \operatorname*{argmax}_{a,e}P(a|c,u)P(e|a,c,u),\\
    (a,e)^*&= \operatorname*{argmax}_{a,e}P((a,e)|(a,e)').
\end{align}

\subsubsection{\textbf{Boosting based Joint Prediction Model}}
\label{sec:boosting}
\begin{figure}
	 \centering
	   \includegraphics[width = \textwidth]{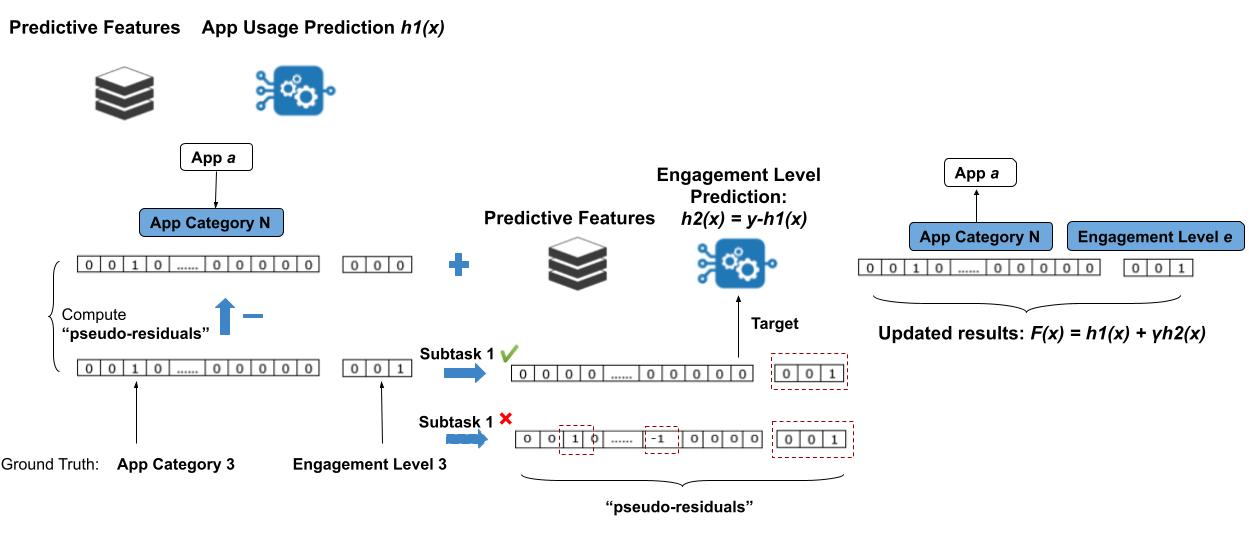}
	   \caption{Boosting based Joint Prediction}
	   \label{fig:Boosting}
	   %\vspace{-0.5cm}
\end{figure}

Boosting \cite{drucker1994boosting} is another ensemble method for improving the prediction model of any given learning algorithms. The idea of boosting is to train weak learners sequentially, each trying to correct its predecessor.  According to this idea, we aim to fit our two learners (next app and engagement level) iteratively such that the training of the model at a given step depends on the models fitted at the previous steps. Then we can improve the predictions from our first learner of app prediction by adding the "error-correction" prediction in the second learner of engagement level prediction. This is shown in Figure~\ref{fig:sequential_stacking} (in the blue circle).

We firstly introduce how boosting method works in the supervised learning problem. For a given data set with $n$ samples and $m$ features $D = \{(x, y)\}(|D| = n,  x \in R^m)$, the goal is to find an approximation $\hat{\mathcal{F}}(x)$ to a function $\mathcal{F}(x)$ that minimizes the expected value of some specified loss function $L(y, \mathcal{F}(x))$. The boosting method assumes a real-valued $y$ and seeks an approximation in the form of a weighted sum of $K$ additive functions (called base/weak learners) $h_k(x)$:

\begin{equation}
\hat{\mathcal{F}}(x) = \sum^K_{k = 1}\gamma_kh_k(x).
\end{equation}
So the model tries to find an approximation $\mathcal{F}(x)$ that minimizes the loss function $L$, and the model is updated as follows. 
\begin{equation}
\gamma_k = \operatorname*{argmin}_{\gamma}\sum^{n}_{i = 1} L(y_i, \mathcal{F}_{k-1}(x_i) + \gamma h_k(x_i)).
\end{equation}

\begin{algorithm}[t]
\SetAlgoLined{
\textbf{Input:} Training data  $D=\{x_i = (c_i, u_i), y_i = (a_i, e_i) \}^n_{i=1}$.\\
\textbf{Output:} Boosting based joint classifier $ \mathcal{F}(x) = \mathcal{F}((a,e)|c,u) =  \sum^K_{k = 1}\gamma_kh_k(x)$.\\
\emph{\textbf{Step 1:} Initialize the model with next app prediction}\\
Learn $h_1(x_i) = P(a_i|c_i, u_i)$ based on $D$.\\
\emph{\textbf{Step 2:} Improve the predictions from next app prediction by adding "error-correction" prediction in the second learner of engagement level prediction.}\\
 1. Compute so-called pseudo-residuals: $r_i = y_i-h_1(x_i)$.\\
 2. Fit the second learner $h_2(x_i)$ to pseudo-residuals. Train it using the training set $\{(x_i, r_i)\}^n_{i=1}$.\\
3. Compute multiplier $\gamma$ by solving the following one-dimensional optimization problem:
$\gamma = \operatorname*{argmin}\sum^n_{i=1}L(y_i, \mathcal{F}_1(x_i)+\gamma h_2(x_i))$.\\
\emph{\textbf{Step 3:} }Update the model:\\
$\mathcal{F}(x_i)= \mathcal{F}_1(x_i)+\gamma h_2(x_i)$.

 \caption{Boosting based Joint Training}
 \label{alg:boosting}
}
\end{algorithm}

To be specific, in our scenario, the dataset $D$ include features $x = (c, u)$ and label $y = (a, e)$. We have two ($K = 2$) base learners in our prediction, where  $h_1(x)$ is the next app usage learner and $h_2(x)$ is the engagement level leaner. The steps for applying the boosting method in our scenario are as shown in Figure~\ref{fig:Boosting} in details: 
\begin{enumerate}
    \item Fit a model to the data, $h_1(x_i) = P(a_i|c_i,u_i)$ for predicting the next app $a_i$.
    \item Compute the pseudo-residuals between the current predicted app $a_i'$ and the ground truth of finalized results: app with the corresponding engagement level together ($a_i, e_i$). Since the engagement level is defined based on different app categories, additionally, to avoid the sparsity issue caused by using a specific app in comparison, we calculate the "residual" of app prediction results within the app category level. The app category and engagement level are all represented by one-hot vectors and concatenated together during the "residual" calculation as shown in Figure~\ref{fig:Boosting}. 
    \item Then we fit a model to the residuals, $h_2(x_i) = y_i - h_1(x_i) = r_i$. As the boosting based joint prediction process shown in Figure~\ref{fig:Boosting}, if the first learner output the right predicted app, then the "residual" for the second learner to learn is still only the engagement level ($r_i = e_i$). On the other side, if the first learner output a wrong predicted app, then the residual for the second learner to learn is not only about the engagement level, it also needs to correct the previous results on app prediction. 
    \item Update the new model $\mathcal{F}(x_i) = \mathcal{F}_1(x_i) + \gamma h_2(x_i)$. Since the "residual" is calculated based on app category level, the finalized output of app prediction results is also on the category level. We will introduce how to infer the specific app from the predicted app category based on our next app prediction model in \S~\ref{sec:estimating}.
\end{enumerate}   
The boosting joint prediction algorithm is shown in Alg. \ref{alg:boosting}. Finally, the boosting based joint prediction could be formulated as:

\begin{equation}
\begin{split}
(a, e)^*& = \gamma_1h_1(x) + \gamma_2h_2(x)\\
        & = a' + r \\
        & = a' + [a_r, e]\\
        & = \operatorname*{argmax}_{a} P(a|c,u) + \operatorname*{argmax}_{r} P([a_r, e]|a',c, u),
\end{split}
\end{equation}
where $r$ is the "pseudo-residuals" when comparing the current predicted app with the ground truth (next app and engagement level), and $a_r$ is the difference between the first time predicted app $a'$ and the target app $a$.

\subsection{Estimating Components within Joint Prediction Strategies}
\label{sec:estimating}
After introducing the three strategies for building the joint prediction model, we can find that the next app prediction and engagement level prediction are two main components for all three strategies. Furthermore, some of these components could be reusable across different strategies, e.g., the next app prediction of sequential-based strategy could also be used by stacking and boosting based strategies at the first stage of app prediction as shown in Figure~\ref{fig:sequential_stacking}. In the following sections, we discuss how to construct these two components, i.e., the next app prediction (Eq. (2)) and engagement level prediction (Eq. (3)), adaptively for each strategy with our proposed predictive features.

\subsubsection{Next App Usage Prediction}
\label{app_prediction}
%In this section, we discuss the methodology for next app usage prediction. 
Many previous researchers \cite{zou2013prophet, huang2012predicting, baeza2015predicting} have proposed different methodologies to solve the next app usage prediction problem based on personalized mechanism. Additionally, some researchers \cite{do2014and,zhao2019appusage2vec} indicated that the generic (user-independent) model can improve the predictive performance of personalized models when the data is not sufficient. A generic model is trained using data from all available users. Inspired by the work of \citet{do2014and}, which achieved the best performance by combining the generic model with the personalized model together, we also propose a hybrid next app prediction model with generic and personalized models combined. 
When building a generic model, the main challenge lies in the fact that the dimensionality of context and output varies depending on all of the users. To learn a generic model and apply it for a given user, generic features and output are needed. In our work, the engagement level is defined based on different app categories. Additionally, the "residual" calculating in boosting based joint prediction strategy (Figure.~\ref{fig:Boosting}) is also based on app category. Therefore, to ensure the generalization of the generic model and to benefit the further prediction of engagement level, the output of the generic model corresponds to the app category. We infer the specific app user will use based on the user's personalized logs given the predicted app category. Specifically, our proposed hybrid next app prediction model could be formulated as the following function:
\begin{equation}
\begin{split}
a^* &= \operatorname*{argmax}_{a}{P(a|c,u)} \\
    &= \operatorname*{argmax}_{a}\{P_{g}(a_c|c,u)P_{p}(a|c,u), a \in a_c\},
\end{split}
\end{equation}
where $P_{g}(a_c|c,u)$ is the probability that user $u$ will use app category $a_c$ based on our generic app category prediction model; and $P_{p}(a|c,u)$ is the probability that user $u$ will use the app $a$ based on their own personalized app prediction model and we will limit the $a$ to the apps belonged to the predicted app category $a_c$. It means we firstly get the app category $a_c$ user will use based on all users' logs and then further rank the apps belonging to this app category based on the personalized logs of user $u$. Through this way, we can have the intermediate output of the app category, which could be used for benefiting the later engagement level prediction. What is more, this generic category-level prediction model could alleviate the cold-start problems with new apps/users resulted from user-specific prediction models, which may be trained on a limited quantity of logs.

For the generic app category prediction model, we extracted all the features that have correlation with app usage patterns, which have been validated by previous works \cite{do2014and,zhao2019appusage2vec} and also available in our dataset (\S~\ref{sec:dataset}): User characteristics (age, gender, country, device type, operation system) \cite{zhao2016discovering,kooti2017iphone,li2017mining}, temporal features (hour of day, day of week) \cite{huang2012predicting, shin2012understanding, liao2012mining}, historical preferences \cite{do2014and},  last one/two apps used \cite{zou2013prophet} and periodic features \cite{tan2012prediction,liao2012mining,liao2013mining}. To enrich our predictive features, we further expand users' characteristics by adding their total app usage duration, total app usage frequency, and unique app amount as user characteristic features. To ensure the generalization of the generic model, users' historical preferences are represented based on the total access frequency of each app category. Moreover, we also add users' app preferences on the last day, the last hour and the last session to expand the context features. In summary, we extracted 14 features related to app usage patterns from 4 aspects: user characteristics, temporal features,  short-term context features, and long-term context features. All the features used for next app prediction are described in Table \ref{tab:app_general_features}. Then the personalized next app prediction model is generated based on the predictive features proposed in previous works \cite{shin2012understanding,zou2013prophet,baeza2015predicting}: hour of day, day of weekday, most recently used apps (the last one/two apps),  and periodic feature. 

\begin{table}[ht]
\centering
\scriptsize
\caption{\rebuttal{All the features used in our next app prediction model related to users characteristics and context}}
\begin{threeparttable}
\scalebox{1}{\begin{tabular}[t]{|p{2cm}|lp{7cm}|}
\hline
\textbf{Feature Type}&\textbf{Feature}&\textbf{Description}\\
\hline
\multirow{6}{*}{\textbf{User and Device}}&Age*&Users' age group: 13-17, 18-24, 25-34, 35-54 and 55+\\
&Gender*&Users' gender: male and female\\
&Device Type*&Users' device type: phone and tablet\\
&Total App Usage Duration*&Users' total app usage duration for all apps\\
&Total App Usage Frequency*&Users' total access frequency for all apps\\
&Total Unique App Amount*&The amount of unique apps the user has accessed\\
\hline
\multirow{12}{*}{\textbf{Context}}&Temporal Features&\\
\cline{2-3}
&Hour of the Day*&Different hours of a day: 0 - 23\\
&Day of the Week*&Different days of a week: Monday to Sunday\\
\cline{2-3}
&Short-term Context Features&\\
\cline{2-3}
&App Preference in the Last Day&Access frequency of each app category in the last day\\ 
&App Preference in the Last Hour&Access frequency of each app category in the last hour\\ 
&App Preference in the Last Session&Access frequency of each app category in the last session\\
&Last Used App*&The last used app in the same session\\
&Last Used Two Apps&The last used two apps in the same session\\
\cline{2-3}
&Long-term context Features&\\
\cline{2-3}
&Periodic Feature&Time intervals between the current time and the last usage of each app category\\
&Historical App Preference&Total access frequency for each app category\\
\hline
\end{tabular}}
 \begin{tablenotes}
      \item *: The features marked with * are the common predictive features also used in app engagement level prediction.
    \end{tablenotes}
\end{threeparttable}
\label{tab:app_general_features}
\end{table}%

\subsubsection{App Engagement Level Prediction}
\label{sec:app_engagement_prediction}
In this subsection, we further explore the app engagement level prediction models that could fit in the different joint prediction strategies. As we mentioned before, we model the novel prediction problem of app engagement level as a multi-class classification problem, where the engagement level is defined based on different app categories. In the sequential based joint prediction strategy (\S~\ref{sequential}) and the level-0 classifiers of the stacking based joint prediction strategy (\S~\ref{sec:stacking}) (Figure~\ref{fig:sequential_stacking}), the engagement level prediction model could be trained for each app category respectively. For example, if we have predicted that the user will use the news app $a^*$ next, we just select the classifier which has been trained specifically based on the logs of using news apps. Then we use this classifier to predict the engagement level. So the engagement level prediction model for sequential-based and stacking-based joint learning strategies could be formulated as:
\begin{equation}
\begin{split}
      e^* & = \operatorname*{argmax}_e P(e|a,c,u)\\
          & = \operatorname*{argmax}_e \{P_{a_c}(e|c,u), a^* \in a_c\},
\end{split}
\end{equation}
where $a_c$ is the corresponding app category of the predicted app $a^*$ based on the next app prediction model. It states that we select the engagement level classifier exactly based on the prediction results coming from the next app prediction results. For each app category, we extract the predictive features based on the analysis in section~\ref{infering_features}: demographic features, device features, hour of day, day of week, last used app, last engagement level, last engagement level of the same app category, periodic feature and historical engagement level feature. Besides the common user characteristic features and some of the context features that have been listed in Table~\ref{tab:app_general_features}, we describe the additional features for the app engagement level prediction model in Table~\ref{tab:app_engagement_features}. These features are all established to have impacts on users' app dwell time in \S~\ref{infering_features}, and we will further analyze the feature importance in the following experimental results section \S~\ref{sec:feature_importance}.

Different from the engagement level prediction model $P(e|a,c,u)$ in sequential (Eq. (4)) and stacking (Eq. (5)) based strategies, the boosting strategy need to infer the "residual" with $P(r|a,c,u)$ (Eq. (11)) and then sum it to the first app prediction result for getting the finalized app and engagement level together. During this process, the next app user will use would be re-inferred during the engagement level prediction, we cannot select a specific classifier based on the first predicted app. Therefore the engagement level prediction model of the boosting-based strategy can only be constructed as a generic model which could be applied to all apps instead of a specific app category. Specifically, the first predicted app $a'$ from the next app prediction model would be treated as an input feature in the following engagement level prediction, which is represented as a one-hot vector as shown in Figure~\ref{fig:Boosting}. Additionally, to ensure the generalization of the input in the engagement level prediction model within boosting-based joint prediction, all the predictive features will also need to be expanded for representing the behaviour pattern coming from all the different app categories (e.g., the feature of historical engagement level for predicted app category need to be extended to historical engagement level for all different app categories). In this model, no matter which app category is predicted to be used next, the same engagement level prediction model will be applied. Since the finalized app and engagement level would be inferred together within the boosting strategy, the formulation of engagement level prediction could not be decomposed, which has been shown in \S~\ref{sec:boosting}.

\begin{table}[ht]
\centering
\scriptsize
\caption{\rebuttal{Additional features used in our app engagement level prediction models related to user characteristics and context}}
\begin{threeparttable}
\scalebox{1}{\begin{tabular}[t]{|p{1.5cm}|p{5cm}p{6cm}|}
\hline
\textbf{Feature Type}&\textbf{Feature}&\textbf{Description}\\
\hline
\multirow{12}{*}{\textbf{Context}}&Short-term Context Features&\\
\cline{2-3}
&Last Engagement Level&The last engagement level of all app categories\\
&Last Engagement Level of Predicted App Category& The last engagement level of the usage on predicted app category\\
\cline{2-3}
&Long-term Context Features&\\
\cline{2-3}
&Periodic Feature&Time intervals since the last use of all app categories\\
&Periodic Feature of Predicted App Category&Time intervals since the last use of predicted app category\\
&Historical Engagement Levels&Historical sum of engagement levels for all app categories\\
&Historical Engagement Levels of Predicted App Category&Historical count of each engagement level for predicted app categories\\
\hline
\end{tabular}}
\end{threeparttable}
\label{tab:app_engagement_features}
\end{table}%

\section{Experimental Results}
In this section, we measure the performance of our proposed prediction models (\S\ref{sec:models}). Experiments are conducted on a real-world app usage log data (\S\ref{sec:dataset}). We use 70\% of the data for each user as training data and the remaining 30\% as test data. We first evaluate the performance of our proposed prediction model on the classic prediction problem, predicting the next app (\S\ref{app_prediction}). Then the three joint learning strategies for predicting the next app and engagement level simultaneously (\S\ref{sec:jointmodels}) are thoroughly evaluated from different perspectives. 

\subsection{Evaluation}
In our proposed prediction models, we measure the performance of all prediction problems based on four metrics: accuracy, precision, recall and F1 score: The accuracy in our problem is defined as the fraction of correctly classified samples; The precision is defined as the ratio: 
\begin{equation}
\frac{tp}{(tp + fp)} , 
\end{equation} 
where $tp$ is the number of true positives and $fp$ the number of false positives. The precision is intuitively the ability of the classifier not to label as positive a sample that is negative. The recall is the ratio: \begin{equation}
\frac{tp}{tp+fn} , 
\end{equation} 
where $fn$ is the number of false negatives. The recall is intuitively the ability of the classifier to find all the positive samples; The F1 score can be interpreted as a weighted average of the precision and recall, where an F1 score reaches its best value at 1 and worst score at 0. The relative contribution of precision and recall to the F1 score is equal. The formula for the F1 score is:
\begin{equation}
 F1 = 2*\frac{precision *accuracy}{precision + accuracy}.
 \end{equation}
 Please note that in our multi-class case, all the precision, recall, and f1 scores are the weighted averages of scores for each class. In order to account for label imbalance,  metrics are calculated for each label, and their averages are weighted by support (the number of true instances for each label). 
 
For the next app prediction results, we will count the item as correctly predicted when the predicted app is exactly the same as the ground truth app. While evaluating the joint prediction problem (both next app and engagement level), a correctly classified sample indicates that both the predicted app and predicted engagement level are correct. If either of the prediction is wrong, it will not be counted as a correct classification.

\subsection{Baselines}
\subsubsection{Next App Usage Prediction}
In order to comprehensively measure the performance of our proposed hybrid next app prediction model (\S\ref{app_prediction}), we compare it with state-of-the-art counterparts. %Although some methods described in Section 2 are related to our work, some of them are not comparable with our model because they use some additional information. 
\rebuttal{Based on the available sources of evidence in our dataset, we first select the two common baseline methodologies: MFU (Most Frequently Used) and MRU (Most Recently Used) \cite{shin2012understanding, xu2013preference, zou2013prophet}. We then tested additional three methodologies from previous works which conducted the next app prediction by combining app usage history and contexts in a unified manner: SVM+Context \cite{shin2012understanding}, CPD \cite{tan2012prediction} and BN \cite{zou2013prophet}. Recently, the neural approaches are popular in solving the time sequence problems, we also investigated the performance of app usage prediction based on LSTM \cite{xu2020predicting} as the baseline:}

\begin{itemize}
   \item \textit{MFU}: the predicted app is the most frequently used app.
   \item \textit{MRU}: the predicted app is the most recently used app. 
   \item \textit{SVM+Context}: Shin et al. \cite{shin2012understanding} used a SVM classifier \cite{bishop2006pattern} and context information (i.e., day of week, hour of day, last used app and time since last app usage) to predict the next app user will use. 
   \item \textit{CPD}:  Tan, et al. \cite{tan2012prediction} proposed a prediction framework: Prediction Algorithm with Fixed Cycle Length (PAFCL). They hypothesized that each application has different usage probabilities in the different time slots of a cycle. CPD (Cumulative Probability Distribution) is the method used to choose applications with higher probability as the candidates based on computing the probabilities of each used app in the specific time slot.
   \item \textit{BN}: Zou et al. \cite{zou2013prophet} proposed a Bayes Network (BN) model which is a linear combination of $p(a_n = A|a_{n-1} = A_{n-1})$ (based on last used app $A_{n-1}$) and $p(a_n = A|a_{n-2} = A_{n-2})$ (second last used app $A_{n-2}$).
   \rebuttal{\item \textit{LSTM}: Xu et al. \cite{xu2020predicting} proposed a generic prediction model based on Long Short-term Memory (LSTM), which is an enhancement of the recurrent neural network (RNN) model. The proposed model converts the temporal-sequence dependency and contextual information into a unified feature representation for the next app prediction.} 
\end{itemize}

\subsubsection{Next App and Engagement Level Joint Prediction} 
Since we are the first work proposed to predict which app user will use and how long the user will stay on that app simultaneously, where the time spent has a high dependency with which app user is engaging, no previous related work could be identified as our baseline methodologies. Therefore, we propose the baselines from two kinds of approaches. Firstly, by following the classic baselines (MFU and MRU) in the next app prediction problem, we present two naive prediction methods as the baselines for our proposed joint prediction problem. Secondly, two recent works on dwell time prediction of online services (e.g., video \cite{wu2018beyond} and media streaming \cite{vasiloudis2017predicting}) are selected as baselines. Similar to our work, how long a user stays with an item in these services originally have associations with the content category, and could also be affected by user characteristics and contextual features. The only difference is that they solely predict how long a user will stay based on the specific item, whereas no prediction on which item the user will engage is required. In order to make them comparable baselines, we assumed the ground truth of the next app is known and leveraged those two prior works \cite{vasiloudis2017predicting,wu2018beyond} for predicting engagement (dwell time) of the oracle app. This demonstrates the upper bound of those approaches (oracle performance) within the joint prediction task.

\begin{itemize}
    \item \textit{MFU}: In our joint prediction problem, we discuss the usage frequency of the two fields together: the next app and engagement level. Then the MFU baseline states that every time we recommend the tuple $\left(a, e\right)$ as the app $a$ and engagement level $e$ based on the popularity of the tuples.
    \item \textit{MRU}: Similar to the MRU in next app prediction, we hypothesize that a correlation between two sequential tuples of app and engagement level may exist. For example, a user may always like to access the weather app for 10 seconds and then access the news app for 5 minutes. Therefore we generate a correlation-based baseline approach for our joint prediction problem, which aims to predict the next app and engagement level based on the correlation between two sequential tuples $(a', e')$ and $(a, e)$. We first calculate all the transition probabilities from one tuple to another. When we know the last usage tuple is $(a', e')$, we could recommend the tuple $(a, e)$ that has the highest probability to be used next. 
    \rebuttal{\item \textit{CSP}: Wu et al. \cite{wu2018beyond} conducted a large-scale measurement study of engagement on videos. They predicted engagement from video context, topics, and channel reputation, etc. They used linear regression with L2-regularization to predict engagement metrics and state that the Channel Specific Predictor (CSP) performs best, which is to train a separate predictor for each channel instead of using the shared predictor. To fit into our scenario, we generate the same CSP for each app category with their proposed features available in our dataset (e.g., the one-hot encoding of category, mean number of daily usage, mean, std and five points summary of past engagement (dwell time), etc.). The performance of the joint prediction problem is reported based on the oracle app and the predicted dwell time (which would be transformed to engagement levels according to our definition in Sec. \ref{engagment_level_definition}.)}
    \rebuttal{\item \textit{Aggregated}: Vasiloudis et al. \cite{vasiloudis2017predicting} presented the first analysis of session length in a mobile-focused online service (i.e. music streaming service). They showed that the time length of sessions can differ significantly between users. They used gradient boosted trees with appropriate objectives to predict the length of a session using contextual and user-based features. Their experiment results showed that the aggregated model trained with all the data performed better than the personalized models trained on each user's data. To fit into our prediction problem, we also trained the aggregated model based on all data with their proposed features available in our dataset (e.g., gender, age, device, duration of the user's last session and time elapsed since the last session, etc.). As we mentioned before, the performance of the joint prediction problem is reported based on the oracle app and the predicted dwell time (which would be transformed to engagement levels according to our definition in Sec. \ref{engagment_level_definition}).}
\end{itemize}

\subsection{Hybrid Next App Prediction Model}
%\todo{significance labels should always be on the numbers, not on the method}
\label{app_usage_prediction_evaluate}

Our analysis starts with the hybrid next app prediction results given it is the unified component leveraged by all the three different joint learning strategies. It is used at the first step for inferring the next app user will use before predicting its engagement level (Figure~\ref{fig:sequential_stacking}). %preliminarily while solving our proposed joint prediction problem (Figure~\ref{fig:sequential_stacking}). 
Table~\ref{tab:personalized_classifier} shows the performance of baseline methodologies and our proposed hybrid next app prediction model. We test our model with a set of state-of-the-art classification models, including Random Forests \cite{breiman2001random}, L2-regularized Logistic Regression \cite{hosmer2013applied}, K Nearest Neighbours \cite{cover1967nearest} and Support Vector Machines \cite{burges1998tutorial}. These models construct different prediction functions for the data from different aspects, and they can provide more robust results for our prediction.
Since our proposed hybrid next app prediction model (\S\ref{app_prediction}) involves the prediction results of two predictive components (Eq. (13)), we select the best classifiers for both of them within the hybrid model, which combines the results of generic app category prediction with Random Forest classifier and the results of personalized next app prediction with SVM classifier.

\begin{table}[t]
    \centering
    \scriptsize
 	\caption{Performance comparison of next app prediction models ($\ast$ indicates statistical significant (p$\leq$0.01) using two-tailed T-test when compared our hybrid next app prediction model to the best baseline model (LSTM).}
 	\label{tab:personalized_classifier}
    \scalebox{1}{\begin{tabular}{ l  c  c  c c}
\toprule
  \multirow{2}{*}{\textbf{Method}}&\multicolumn{4}{c}{\textbf{Measurement}}\\
 \cline{2-5}
 &\textbf{Accuracy}&\textbf{Precision}&\textbf{Recall}&\textbf{F1}\\
\midrule
MFU &0.486&0.489&0.486&0.486\\
MRU &0.518&0.521&0.518&0.518\\
CPD \cite{tan2012prediction}&0.424&0.419&0.424&0.369\\
BN \cite{zou2013prophet}&0.453&0.483&0.453&0.467\\
SVM+Context \cite{shin2012understanding}&0.606&0.563&0.606&0.572\\
\rebuttal{LSTM \cite{xu2020predicting}} &\rebuttal{0.525}&\rebuttal{0.660}&\rebuttal{0.525}&\rebuttal{0.576}\\
\midrule
Our Hybrid Model $^\ast$&\textbf{0.640}&\textbf{0.607}&\textbf{0.640}&\textbf{0.613}\\
\bottomrule
 \end{tabular}}
\end{table}

From Table~\ref{tab:personalized_classifier}, we can find that our proposed hybrid next app prediction model could significantly improve the performance of the best baseline model (LSTM) by 6.4\% on the F1 measure. The worse performance of LSTM is expected since it could not incorporate any user characteristics and contexts (e.g., access time, device type, etc.) during the prediction. While our proposed hybrid next app prediction model not only takes the user characteristics and contextual information into consideration but also learn from users’ common patterns to overcome important sources of prediction errors resulting from insufficient training data. It also states that through the use of community similarity and common usage patterns learned based on different app categories, we can improve the next app usage prediction by identifying the generic usage patterns present in similar users - rather than relying solely on the specific app usage patterns. This is also consistent with the findings from previous studies \cite{do2014and, zhao2019appusage2vec} which claim that the generic model can improve the predictive performance of models solely based on individual's logs. Therefore, our proposed hybrid next app prediction model is adopted to apply to further engagement level prediction for different joint learning strategies.

\subsection{Next App Usage and App Engagement Level Joint Prediction Strategies}
\label{joint_prediction_evaluate}
We first evaluate how different models perform for solving our proposed joint prediction problem: which app user will use next and how long the user will stay with this app? Then we further investigate the predictive ability of features in the two prediction problems respectively and the prediction effectiveness of our proposed different joint prediction strategies. 

As mentioned, our hybrid next app prediction model is applied to the first stage prediction on the next app for all the joint learning strategies (Figure~\ref{fig:sequential_stacking}). For the prediction of engagement level, the predicted app will be represented in different ways based on different joint prediction strategies. For the sequential and stacking based joint model, the predicted app will be used to select the specific engagement level classifier with the corresponding app category (Eq. (14)). For the boosting based joint model, the predicted app coming from the first stage of next app prediction could only be treated as the input feature for the next step prediction of engagement level (Eq. (12)). Table~\ref{tab:ensembling} shows the ultimate performance of different joint prediction strategies and all baselines.

\begin{table}[t]
    \centering
    \scriptsize
 	\caption{Performance comparison of joint learning prediction models ($\ast$ indicates statistical significant (p$\leq$0.01) using two-tailed T-test compared to the best baseline (CSP)).}
 	\label{tab:ensembling}
     \scalebox{1}{\begin{threeparttable}
     \begin{tabular}{l  c  c  c  c }
 				\hline
 \multirow{2}{*}\textbf{Model}&\multicolumn{4}{c}{\textbf{Measurement}}\\
 			\cline{2-5}
&\textbf{Accuracy}&\textbf{Precision}&\textbf{Recall}&\textbf{F1}\\
   \midrule
   MFU &0.286&0.286&0.286&0.286\\
  
   MRU &0.308&0.307&0.308&0.308\\
   \rebuttal{Aggregated \cite{vasiloudis2017predicting} $^\triangleright$} &\rebuttal{0.347}&\rebuttal{0.742}&\rebuttal{0.347}&\rebuttal{0.448}\\
    \rebuttal{CSP \cite{wu2018beyond} $^\triangleright$}&\rebuttal{0.339}&\rebuttal{0.729}&\rebuttal{0.339}&\rebuttal{0.467}\\
   \midrule
   Sequential$^\ast$&0.375&0.382&0.375&0.369\\

   Stacking$^\ast$&0.374&0.381&0.374&0.365\\

   Boosting$^\ast$&\textbf{0.485}&\textbf{0.483}&\textbf{0.485}&\textbf{0.483}\\
\bottomrule
 \end{tabular} 
 \begin{tablenotes}
 \item $\triangleright$: We assumed the ground truth of the next app is known and leveraged these baselines for predicting engagement (dwell time) of the oracle app. The performance reported in table demonstrates the upper bound of those approaches (oracle performance) regarding the joint prediction problem.
	\end{tablenotes}
 \end{threeparttable}}
\end{table}

We can find that all our proposed joint prediction models are better than the \rebuttal{two classic baselines: MFU and MRU. For another two baselines, we can observe that even we assumed the ground truth of the next app is known, the upper bound performance of these approaches could not beat our proposed best joint prediction model based on boosting strategy. To be specific, it states that we assume the predicted app is exactly the same as the ground truth and only evaluate the engagement level prediction performance, these baselines have performed worse than our boosting-based joint prediction model. It mostly because they didn't model the engagement with comprehensive characteristics as our proposed model, where we take the user characteristics, short/long-term usage patterns all into consideration. Therefore, if the next app prediction task is added, the worse performance of the joint prediction problem should also be expected for these baselines.} Among all the three proposed joint prediction strategies, stacking and boosting based strategies are two advanced approaches originally proposed to improve the performance with the most straightforward strategy, sequential based strategy. However, we find that the stacking based strategy does not improve the performance when compared with the benchmark sequential based strategy. This might be due to that,
%the stacking principle involves training a learning algorithm to combine the predictions of several other learning algorithms for the same prediction task and then it could yield performance better than any single one of the trained models. However, 
in our scenario, the base learners within stacking are not trained for the same target (we have two different prediction tasks: app and engagement level) where the stacking principles do not apply.
%which may be the reason why the stacking based strategy could not improve the performance in our joint learning prediction task. 
On the other hand, the boosting strategy works best compared to all other joint models. It respectively outperforms the baseline models over 56\% and the sequential/stacking models about 31\% on F1 measure. We mainly focus on investigating the boosting and sequential strategies in the later sections about how boosting helps in the joint prediction. Firstly, besides the overall performance reported in Table~\ref{tab:ensembling}, we look into the detailed prediction results. It demonstrates that the accuracy of next app prediction and engagement level prediction respectively are: 64\% (app) and 58.6\% (engagement level) for sequential strategy, 85.6\% (app) and 56.7\% (engagement level) for boosting strategy, where the accuracy of engagement level is calculated only based on the data with right predicted apps. So we can find that the improvement of overall performance for boosting strategy is mainly because of the "error-correction" step which corrects the app prediction results. 

In the following sections, we will first analyze the feature importance within the two prediction problems respectively, especially focus on exploring the predictive ability of the newly proposed features. Then we will dig into how is the effectiveness of boosting based strategy compared to the sequential based strategy.

\subsubsection{Feature Analysis}
\label{sec:feature_importance}
\begin{table}[t]
    \centering
    \scriptsize
 	\caption{Feature importance (MDI) of app category prediction. }
 	\label{tab:randome_forest_feature_importance}
     \scalebox{1}{\begin{tabular}{|l|c|c|}
 	\hline
     \textbf{Feature}&\textbf{Feature Type}&\textbf{Importance (MDI)}\\
 	\hline
   Total$\_$App$\_$Usage$\_$Frequency&User&0.023\\
   Hour&Temporal&0.011\\
   Periodic$\_$Feature&Long-term&0.011\\
   Total$\_$App$\_$Usage$\_$Duration&User&0.010\\
   Historical$\_$App$\_$Preference&Long-term&0.005\\
   Total$\_$Unique$\_$App$\_$Amount&User&0.005\\
   Age&User&0.004\\
   App$\_$Preference$\_$Last$\_$Day&Short-term&0.003\\
   Weekday&Temporal&0.002\\
   Gender&User&0.002\\
   App$\_$Preference$\_$Last$\_$Hour&Short-term&0.002\\
   Device$\_$Type&User&0.001\\
   App$\_$Preference$\_$Last$\_$Session&Short-term&0.001\\
   \hline
 \end{tabular}}
\end{table}

The sequential based strategy is implemented by conducting the app prediction and the engagement level prediction sequentially, where the app prediction is absolutely independent with the further engagement level prediction. Therefore, we discuss the analysis of the most impactful features for those two tasks respectively. The next app prediction within the sequential strategy is implemented as same as our proposed hybrid next app prediction model, which would infer what app category the user will use and then select the specific app given this predicted app category. To make the comparison more intuitively, we conduct the analysis within the app category level. As we mentioned above, the random forest is selected as the best classifier for the app category prediction problem. The random forest can be used to rank features by their importance in the classiﬁer, which provides useful insights about the discriminative power of the features in the considered problem setting. \textit{Mean Impurity Decrease (MDI)} is the most common way to obtain feature importance from random trees \cite{breiman2001random}. It is computed by averaging across all the trees in the forest the amount of impurity removed by each feature while traversing down the tree, weighted by the proportion of samples that reached that node during training. Using this method, we obtained the feature importance for all features in the app category prediction, as listed in Table~\ref{tab:randome_forest_feature_importance}. 

We find that besides the critical temporal feature \emph{hour}, the most important categories of features are mostly user and long-term context features. Compared to user characteristics features such as demographics and devices, the total usage (frequency, duration, and unique app amount) maintain higher impacts on the next app category prediction. The hour of day feature has been used to identify the salient pattern within app usage behaviour in many previous works \cite{huang2012predicting, shin2012understanding, liao2012mining}, e.g. the user usually set an alarm at around 23:00. The popularity of app usage (historical app preference) is also a famous feature that has been established by previous works \cite{liao2013mining, do2014and}. Additionally, our finding of the periodic pattern is consistent with the previous works on next app prediction, where the periodic patterns have been identified as the effective feature for inferring the app usage pattern \cite{tan2012prediction,liao2012mining,liao2013mining}. Liao et al. \cite{liao2013mining} also stated that the periodical usage feature is the most difﬁcult one to be substituted by other features.

\begin{table}[t]
    \centering
    \scriptsize
 	\caption{Top feature weights (standardized coefficients $\geq$ 0.001) for logistic regression model of engagement level prediction. $\ast$ indicates p-value $\leq$ 0.001 using Chi-Squared test.}
 	\label{tab:Feature_Importance_Specified}
     \scalebox{1}{\begin{tabular}{|l|c|c|}
 	\hline
     \textbf{Feature}&\textbf{Feature Type}&\textbf{Weight}\\
 	\hline
   Historical$\_$Level$\_$Light$^\ast$&Long-term&0.571\\
   Historical$\_$Level$\_$Medium$^\ast$&Long-term&0.197\\
   Total$\_$App$\_$Usage$\_$Duration$^\ast$&User&0.175\\
   Historical$\_$Level$\_$Intensive$^\ast$&Long-term&0.112\\
   Periodic$\_$Feature$^\ast$&Long-term&0.070\\
   Age$^\ast$&User&0.032\\
   Last$\_$Used$\_$App$^\ast$&Short-term&0.019\\
   Hour$^\ast$&Temporal&0.008\\
   Weekday&Temporal&0.002\\
   Last$\_$Engagement$\_$Level$^\ast$&Short-term&0.001\\
   \hline
 \end{tabular}}
\end{table}

The engagement level prediction is the novel problem proposed in our work and we extracted many different predictive features from user\&device characteristics, temporal pattern, and short/long-term context (\S\ref{infering_features}) to infer how long the user will stay with an app. Similarly, since the engagement level prediction within the sequential strategy can be studied separately, we discuss the impacts brought by different features for engagement level prediction within the sequential based strategy. We opt to build a classifier for predicting users' engagement with three levels: \emph{light}, \emph{Medium} and \emph{Intensive}. Hence, the problem of modelling user engagement turns into a multi-class classification problem. Similar to the next app prediction, we test and empirically compare the performance of a wide range of classification techniques, including Random Forests (RF), L2-regularized Logistic Regression (LR), K Nearest Neighbours (KNN), and Support Vector Machines (SVM), for predicting the app engagement level, where the Logistic Regression classifier performs best. 

Then we examine the contribution of each feature based on the feature coefficients in the logistic regression model. To compare the importance of different features, we divide each numeric variable by two times its standard deviation~\cite{gelman2008scaling}. Through this way, the resulting coefficients are directly comparable for both binary variables (e.g.,~categorical dummy variable) and numerical features. Table~\ref{tab:Feature_Importance_Specified} reports the top feature weights (standardized coefficients $\geq$ 0.001) of the Logistic Regression model for the engagement level prediction. We can find that most of the top features are originated from the long-term context features: historical engagement level preference and periodic pattern. It is not surprising that the historical pattern has more influence on how long a user will stay with an app. If the user always prefers to play games for a longer time, then he may still spend more time in the game app this time. The periodic feature has more impacts on users' engagement level than the temporal context, hour, and weekday. This demonstrates that no matter when the user uses this app, the time since the last use of this app is more important for inferring how long the user will engage with this app. Additionally, the last used app has more impacts on predicting the engagement level compared to all short-term context features. This could provide more insights for the app developers to recommend the contents of different time lengths according to the last used app to improve the user experience. Another finding is that we observe age is the most important signal among all demographics and device characteristics when inferring the app usage duration.

\subsubsection{How Effective are Boosting-based vs. Sequential-based Strategies?}
We have shown the boosting based strategy outperforms the benchmark sequential based strategy by the performance margin of 31\% on the F1 measure. Therefore we further conduct some error analysis to understand the underlying reasons.
%the effectiveness of the boosting based strategy, compared to the sequential based strategy.

\begin{figure}
	 \centering
	   \includegraphics[height = 1.8in]{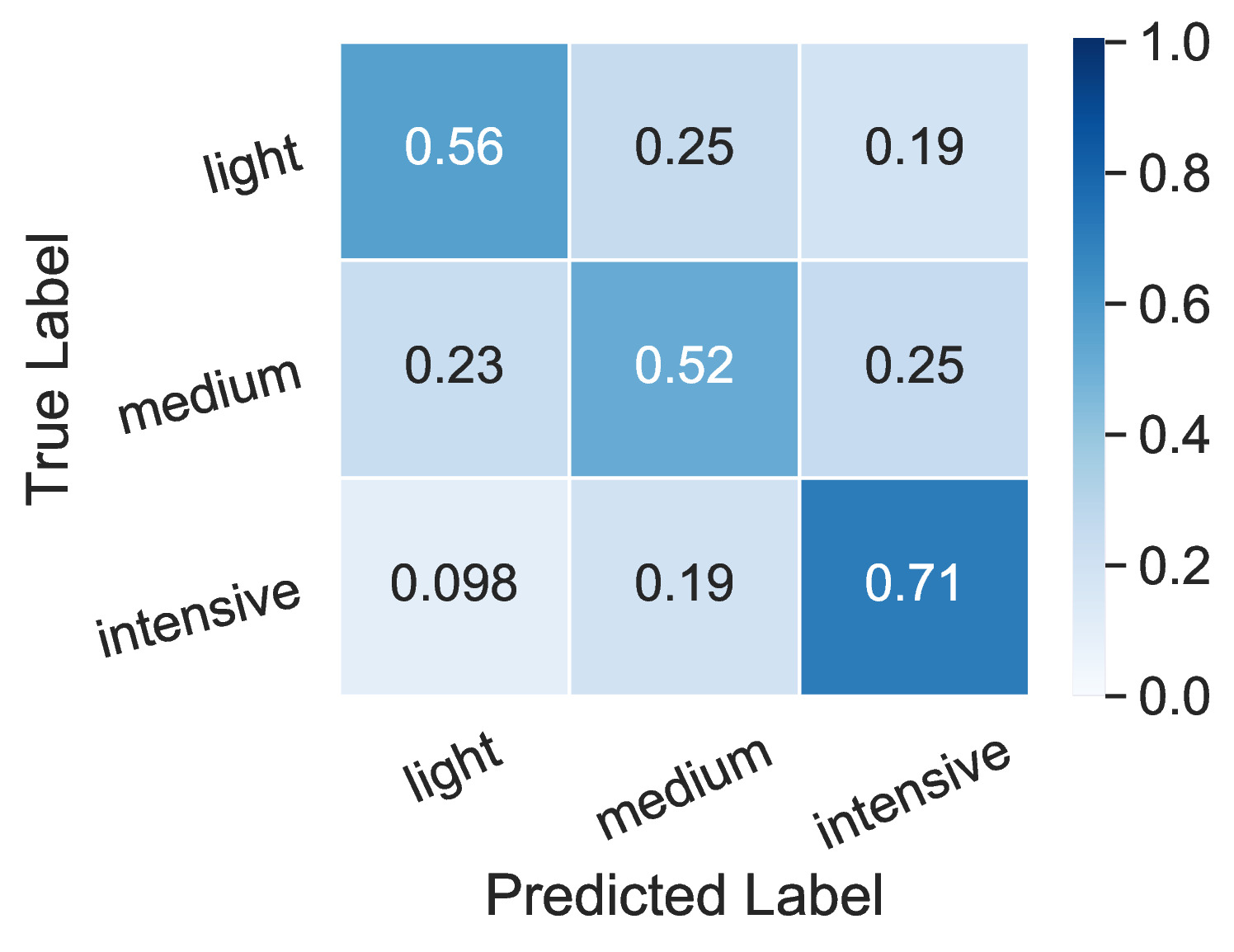}
	   \caption{The confusion matrix of our prediction model on the engagement level.} %Deeper color represents larger values.}
	   \label{fig:engagement_level_error_heatmap}
\end{figure}

Firstly, for the novel prediction problem on engagement level, the sequential and boosting based strategies achieve similar performance, which are 58.6\% and 56.7\%  respectively on accuracy. It demonstrates that even the boosting based strategy gets better overall performance in our joint prediction problem, it cannot improve the engagement level prediction results. We further analyze the prediction ability on app engagement level prediction with our proposed sequential strategy. We select only the engagement level prediction results when the next app is correctly predicted. Figure~\ref{fig:engagement_level_error_heatmap} shows the confusion matrix of the app engagement level prediction results. For the wrongly predicted results of all engagement levels, we can find that they have higher probabilities to be predicted into the adjacent level. For example, for the intensive engagement level, 19\% of them are misclassified as a medium level, which covers about 95\% of the wrongly predicted results. Similarly, for the light engagement level, 25\% of them are misclassified as a medium level, which is higher than those that are misclassified as intensive. The engagement level prediction results of boosting based strategy also get a similar confusion matrix as shown in Figure~\ref{fig:engagement_level_error_heatmap}. %Therefore we can conclude that even we represent users' time spent with the app into discrete values and model the app usage duration prediction problem as a three-level classification problem, our proposed model could still support the continuous numeric attribute in the prediction results. 
%The sequential and boosting strategies perform similarly in the engagement level prediction, so 
We now focus on exploring how boosting is more effective in the next app prediction as follows.

For the next app prediction, we have reported that the boosting based strategy improves the accuracy from 64\% to 85.6\%, which is resulted from the "error-correction" step. We then compare the app prediction results of sequential based and boosting based strategies. For the sequential based strategy, we found that the incorrect results are mainly generated from three reasons:

\begin{table}[t]
\centering
\scriptsize
 \caption{\rebuttal{Illustration of Case Studies in Next App Prediction}}
\label{tab:case_study}
\scalebox{0.85}{\begin{tabular}{p{2.2cm}|p{2cm}p{2cm}p{2cm}p{6cm}}
\toprule
\textbf{Error Reason Type}&\textbf{Predicted Result}&\textbf{Ground Truth}&\textbf{Correction while Boosting}&\textbf{Correction/Uncorrection Reason}\\
\midrule
Global Popularity&productivity&widgets, medical, transportation&widgets, medical, transportation&The "pesudo-residuals" are applied as ground truth in boosting makes the model less impacted by the global popularity of app categories within the dataset.\\
\midrule
Short Usage Interval&card&casino&casino&Additional information are provided by the engagement features added in the boosting strategy. Last engagement of casino is longer than the card app, so these cases are corrected from card to casino.\\
\cmidrule{2-5}
&arcade&adventure&No&No additional information could be provided based on engagement.\\
\midrule
Similar App Usage Frequency&lifestyle&food-and-drinks&food-and-drinks&Boosting helps when the engagement pattern provides additional information. Specifically, the results are corrected by boosting since the historical engagement pattern of food-and-drink apps usage of this user is longer than lifestyle apps. \\
\cmidrule{2-5}
&health-and-fitness&medical&No&None of them are corrected by boosting since no difference exists in the historical engagement pattern of these two apps of the user.\\
\bottomrule
\end{tabular}}
\end{table}

\noindent (a) \textit{\textbf{Global popularity}}. \rebuttal{We find that for 66.7\% (30/45) of the app categories, the top category they were misclassified as is \emph{productivity}. For example, 37.4\% of widgets apps are wrongly predicted as productivity, 31,7\% of medical apps are wrongly predicted as productivity, and 27.2\% of transportation apps are also wrongly predicted as productivity.} This could be easily explained since the productivity apps have the highest global popularity among all the app categories. It results in our generic app category classifier inferring that the user has a higher probability to use the productivity apps given all users' data. By analysing these cases, we find that the boosting strategy resolves most of the issues resulted from global popularity. For example, 25.4\% of widgets apps and 12.8\% of transportation apps are corrected by boosting from being misclassified into the productivity apps. This is because that the prediction model in the boosting step does not involve the global popularity of app categories. Since the "pseudo-residuals" (Figure~\ref{fig:Boosting}) are applied as the ground truth, this makes the model more sensitive for predicting users' app usage based on the context and user characteristics, rather than the global popularity of app categories within the dataset.

\noindent (b) \textit{\textbf{Short usage interval}}. Another important factor that accounts for the misclassification within the sequential based strategy is the time interval (periodic feature) between app usage. We observe that for those users whose historical app preference deviates from an average user (e.g., productivity is not the top preferred app), the periodic feature (time since the last usage) would have stronger impacts. \rebuttal{For example, 30.8\% of the adventure apps are wrongly predicted as arcade apps given the time from the last usage of arcade app is short.} Since the model learned that the shorter the time interval, the higher probability the same app would be used again. In these cases, even the adventure has a higher usage frequency in the user's historical app usage pattern, it would be wrongly predicted as arcade since arcade has a higher probability to be re-accessed in a short period of time. 

\noindent (c) \textit{\textbf{Similar app usage frequency}}. The third major reason for the misclassified cases is that when two app categories are all recently used with a similar frequency, the one with the higher personalized historical usage frequency would be selected. \rebuttal{For example, 11.4\% of medical apps are wrongly predicted as health-and-fitness apps because that the medical app and health-and-fitness app are always used in the same session,} the health-and-fitness app is predicted as the result since it has higher popularity within the user's historical app usage pattern. 

For the latter two error analysis in the sequential based strategy (described above as (b) and (c)), we find that the boosting strategy only helps some of the cases when the engagement information could provide additional insights. For example, among the wrongly predicted 30.8\% adventure apps,  only 5\% of adventure apps are corrected from arcade by boosting since the last engagement level of adventure is longer than arcade. For those cases without this additional information from engagement, they are still wrongly predicted.  However, for some specific cases, e.g., 7.8\% of casino apps are predicted as card apps for the same reason as adventure\&arcade apps, but all of these 7.8\% cases are corrected to casino with boosting. This is because for all the cases the last engagement of casino is longer than the card app. Similarly, for the cases which are wrongly predicted because they have similar recent usage patterns, the boosting only helps when the engagement pattern provides additional info. For example, 3.7\% of food-and-drink apps are corrected by boosting from lifestyle apps. This is because the historical engagement pattern of food-and-drink apps usage of this user is longer than lifestyle apps. However, for the 11.4\% wrongly predicted medical apps, none of them is corrected by boosting since no difference exists in the historical engagement level pattern of these two apps. 

To make the benefits brought by boosting strategy clearer, we also added a summary table of the case studies as Table~\ref{tab:case_study}, where we can find the corresponding reasons about why some of the cases are corrected by boosting but some others are not.

\section{Discussion}
In this study, we explored the factors that affect users' app dwell time from users' app usage logs and show evidence that the next app and how long the user will stay on this app could be predicted simultaneously. First, to answer our first research question (\textbf{RQ1}), we take a systematic approach to uncover the dependency of users' app usage duration on user characteristics and context features based on a large-scale dataset. We then showed that the features related to users' historical engagement pattern and periodic usage pattern are good predictors of how long users will stay with an app. To solve the second research question (\textbf{RQ2}), we propose three different joint prediction strategies and demonstrate that the boosting based strategy performs the best. We further conduct the error analysis on the boosting strategy compared to the benchmark sequential based strategy. Based on the analysis, we find that besides the benefits brought by the "pseudo-residuals" within the boosting principle, the engagement features also provide additional insights to help infer which app user will use. These findings inspire us that we should think of adding more engagement related features when predicting the next app. As below, We discuss how our findings can be applied to future mobile systems and applications, the limitations in our study \rebuttal{and the differences between personalized and general engagement prediction.}

\subsection{Implications}
Firstly, the model proposed in our work can be applied for more tailored engagement-aware recommendations on mobile phones. As the operating system has access to all the features used for training the next app and app engagement models in this work, it is uniquely suited to predict a user's likely next app usage and engagement level under the current context. By doing so, the operating system can manage the delivery of content and services to the end-user by matching their engagement demands with the predicted engagement levels of the user. For instance, an app that shows mobile advertisements require high engagement from the end-users and can ask the app provider to push its content to the user when he/she is likely to be highly engaged. The media apps, like video and news apps, could recommend more satisfactory content based on the predicted engagement level of users to improve user experience.

Recently, there is an active area of research for the timely delivery of notifications on mobile devices. Researchers have primarily focused on understanding the receptivity of mobile notifications \cite{mehrotra2016my} and predicting opportune moments to deliver notifications in order to optimize metrics such as response time \cite{pejovic2014interruptme}. While response time is indeed a useful metric to optimize for, they do not capture how much engagement the user will show towards the notification. The primary purpose of a notification is to attract user attention and increase the possibility of user engagement with the notification content. As such, we believe that the models and features we explored in this work can also be incorporated in designing an effective notification delivery mechanism. 

\subsection{Limitations} 
Although we have explored several meaningful features that could be applied in benefiting users' app engagement (dwell time) prediction, we must acknowledge the limitations of the dataset used in our study. Firstly, \rebuttal{similar to all other real-world app usage datasets, the app popularity distribution follows Zipf’s law \cite{li2015characterizing, petsas2017measurement}, which indicates that only a few apps have high installation/usage whereas many apps have low installation/usage. But doing a balanced prediction would not fit the realistic evaluation settings, which would bring the bias of the users/contexts to be selected. To handle this imbalance issue within app categories while evaluation, we measure the performance of all prediction problems based on four metrics: accuracy, precision, recall and f1. All the metrics are calculated for each label, and their averages are weighted by support (the number of true instances for each label). We also conducted case studies to explore the prediction results based on different app categories.} Additionally, our dataset might not be representative of the entire population of mobile users, as the users and apps are only coming from the apps registered in this library. This means that not all the apps usage behaviour of users could be tracked and there may be a selection bias in the subset of users being studied. While this might occur to some extent, given the scale of our dataset (with over 1.3 million logs), we believe our data would still provide useful insights, and our predictive models are the best models so far, effective for most users. 

Lastly, the predictive features we extracted can be further enriched. Other features that precisely characterize users' app engagement behaviours could be further explored, such as the location, wifi access status, battery, device mode setting (e.g., silent), illumination, screen, and blue-tooth. In this work, our main aim is to validate that how long a user will stay with an app could be modelled based on the user characteristics and context features. We would like to further explore additional features and models for improvements in our future work.

Despite these limitations, we believe our work is the first-of-its-kind study to examine, and model the mobile app engagement purely based on features derived from users' app usage logs. We also hope that the framework, models, and insights developed in this work can bring clarity and guidance to aid future mobile system developers in designing better, and engagement-aware user experience.

\subsection{Personalized V.S. General Engagement Prediction}
We know that the performance of the personalized model could be highly impacted by whether there is sufficient data for training or not. In this work, we are interested in relatively short-term user app engagement patterns. Therefore, we collect app usage data from all users for a week. Due to the nature of the data, it might be more difficult to acquire sufficient per-user patterns, which is also validated by our experiment results regarding the next app prediction (i.e. personalized model could not outperform our proposed hybrid model). When the long-term data is available, the personalized approach might be more suitable and achieve better performance. We leave the exploration of the personalized model for app usage prediction in our future work. In terms of the app engagement level (dwell time) prediction,  we introduce the engagement levels according to different app categories for handling users’ different consumption behaviour on different contents. Hence our engagement level prediction results do not directly translate to user-specific engagement. Additionally, we believe that defining and measuring aggregate engagement is also useful for content producers, e.g., video producers on Youtube. The content providers often do not target a specific user, but a large number of audience. Other than predicting how long user will stay on the app, we also want to provide content producers with a new set of tools to create engaging content and forecast user behaviour. For future work, we would measure users' engagement at the personalized level as complementary to the aggregated engagement study, especially when we have more sufficient training data for individuals. It would help the mobile apps to provide more fine-grained services to a specific user based on the more accurate expected time length.

\section{Conclusion}
In this paper, we propose to predict both users' next app and the engagement level at the same time. For the first time - to the best of our knowledge - a comprehensive analysis of users' app dwell time is conducted based on the large-scale commercial mobile logs, especially focusing on inferring the correlations between different predictive features and users' app usage duration. We find that the users' historical engagement pattern, periodic behaviour pattern, and the recent usage pattern have more impacts when inferring users' app dwell time. To solve our joint prediction problem on the next app and app engagement level, we propose three strategies, where the boosting based joint prediction model works best. Our experimental results demonstrate that users' next app and engagement level could be effectively predicted at the same time, and our proposed prediction method outperforms all baseline experiments by a large margin. Our work can help for providing more satisfying services to users for improving users' experience on mobile devices.

\bibliographystyle{ACM-Reference-Format}
\bibliography{sample-base}


\begin{thebibliography}{70}

%%% the Web address.

\ifx \showCODEN    \undefined \def \showCODEN     #1{\unskip}     \fi
\ifx \showDOI      \undefined \def \showDOI       #1{#1}\fi
\ifx \showISBNx    \undefined \def \showISBNx     #1{\unskip}     \fi
\ifx \showISBNxiii \undefined \def \showISBNxiii  #1{\unskip}     \fi
\ifx \showISSN     \undefined \def \showISSN      #1{\unskip}     \fi
\ifx \showLCCN     \undefined \def \showLCCN      #1{\unskip}     \fi
\ifx \shownote     \undefined \def \shownote      #1{#1}          \fi
\ifx \showarticletitle \undefined \def \showarticletitle #1{#1}   \fi
\ifx \showURL      \undefined \def \showURL       {\relax}        \fi
% invisible to TeX
\providecommand\bibfield[2]{#2}
\providecommand\bibinfo[2]{#2}
\providecommand\natexlab[1]{#1}
\providecommand\showeprint[2][]{arXiv:#2}

\bibitem[\protect\citeauthoryear{Agichtein, Brill, and Dumais}{Agichtein
  et~al\mbox{.}}{2006}]%
        {agichtein2006improving}
\bibfield{author}{\bibinfo{person}{Eugene Agichtein}, \bibinfo{person}{Eric
  Brill}, {and} \bibinfo{person}{Susan Dumais}.}
  \bibinfo{year}{2006}\natexlab{}.
\newblock \showarticletitle{Improving web search ranking by incorporating user
  behavior information}. In \bibinfo{booktitle}{\emph{Proceedings of the 29th
  annual international ACM SIGIR conference on Research and development in
  information retrieval}}. ACM, \bibinfo{pages}{19--26}.
\newblock


\bibitem[\protect\citeauthoryear{Baeza-Yates, Jiang, Silvestri, and
  Harrison}{Baeza-Yates et~al\mbox{.}}{2015}]%
        {baeza2015predicting}
\bibfield{author}{\bibinfo{person}{Ricardo Baeza-Yates}, \bibinfo{person}{Di
  Jiang}, \bibinfo{person}{Fabrizio Silvestri}, {and} \bibinfo{person}{Beverly
  Harrison}.} \bibinfo{year}{2015}\natexlab{}.
\newblock \showarticletitle{Predicting the next app that you are going to use}.
  In \bibinfo{booktitle}{\emph{Proceedings of the Eighth ACM International
  Conference on Web Search and Data Mining}}. ACM, \bibinfo{pages}{285--294}.
\newblock


\bibitem[\protect\citeauthoryear{Bilenko and White}{Bilenko and White}{2008}]%
        {bilenko2008mining}
\bibfield{author}{\bibinfo{person}{Mikhail Bilenko} {and}
  \bibinfo{person}{Ryen~W White}.} \bibinfo{year}{2008}\natexlab{}.
\newblock \showarticletitle{Mining the search trails of surfing crowds:
  identifying relevant websites from user activity}. In
  \bibinfo{booktitle}{\emph{Proceedings of the 17th international conference on
  World Wide Web}}. ACM, \bibinfo{pages}{51--60}.
\newblock


\bibitem[\protect\citeauthoryear{Bishop}{Bishop}{2006}]%
        {bishop2006pattern}
\bibfield{author}{\bibinfo{person}{Christopher~M Bishop}.}
  \bibinfo{year}{2006}\natexlab{}.
\newblock \bibinfo{booktitle}{\emph{Pattern recognition and machine learning}}.
\newblock \bibinfo{publisher}{springer}.
\newblock


\bibitem[\protect\citeauthoryear{Bogina and Kuflik}{Bogina and Kuflik}{2017}]%
        {bogina2017incorporating}
\bibfield{author}{\bibinfo{person}{Veronika Bogina} {and} \bibinfo{person}{Tsvi
  Kuflik}.} \bibinfo{year}{2017}\natexlab{}.
\newblock \showarticletitle{Incorporating Dwell Time in Session-Based
  Recommendations with Recurrent Neural Networks.}. In
  \bibinfo{booktitle}{\emph{RecTemp@ RecSys}}. \bibinfo{pages}{57--59}.
\newblock


\bibitem[\protect\citeauthoryear{B{\"o}hmer, Hecht, Sch{\"o}ning, Kr{\"u}ger,
  and Bauer}{B{\"o}hmer et~al\mbox{.}}{2011}]%
        {bohmer2011falling}
\bibfield{author}{\bibinfo{person}{Matthias B{\"o}hmer}, \bibinfo{person}{Brent
  Hecht}, \bibinfo{person}{Johannes Sch{\"o}ning}, \bibinfo{person}{Antonio
  Kr{\"u}ger}, {and} \bibinfo{person}{Gernot Bauer}.}
  \bibinfo{year}{2011}\natexlab{}.
\newblock \showarticletitle{Falling asleep with Angry Birds, Facebook and
  Kindle: a large scale study on mobile application usage}. In
  \bibinfo{booktitle}{\emph{Proceedings of the 13th international conference on
  Human computer interaction with mobile devices and services}}. ACM,
  \bibinfo{pages}{47--56}.
\newblock


\bibitem[\protect\citeauthoryear{Breiman}{Breiman}{2001}]%
        {breiman2001random}
\bibfield{author}{\bibinfo{person}{Leo Breiman}.}
  \bibinfo{year}{2001}\natexlab{}.
\newblock \showarticletitle{Random forests}.
\newblock \bibinfo{journal}{\emph{Machine learning}} \bibinfo{volume}{45},
  \bibinfo{number}{1} (\bibinfo{year}{2001}), \bibinfo{pages}{5--32}.
\newblock


\bibitem[\protect\citeauthoryear{Burges}{Burges}{1998}]%
        {burges1998tutorial}
\bibfield{author}{\bibinfo{person}{Christopher~JC Burges}.}
  \bibinfo{year}{1998}\natexlab{}.
\newblock \showarticletitle{A tutorial on support vector machines for pattern
  recognition}.
\newblock \bibinfo{journal}{\emph{Data mining and knowledge discovery}}
  \bibinfo{volume}{2}, \bibinfo{number}{2} (\bibinfo{year}{1998}),
  \bibinfo{pages}{121--167}.
\newblock


\bibitem[\protect\citeauthoryear{Cao and Lin}{Cao and Lin}{2017}]%
        {cao2017mining}
\bibfield{author}{\bibinfo{person}{Hong Cao} {and} \bibinfo{person}{Miao Lin}.}
  \bibinfo{year}{2017}\natexlab{}.
\newblock \showarticletitle{Mining smartphone data for app usage prediction and
  recommendations: A survey}.
\newblock \bibinfo{journal}{\emph{Pervasive and Mobile Computing}}
  \bibinfo{volume}{37} (\bibinfo{year}{2017}), \bibinfo{pages}{1--22}.
\newblock


\bibitem[\protect\citeauthoryear{Carrascal and Church}{Carrascal and
  Church}{2015}]%
        {carrascal2015situ}
\bibfield{author}{\bibinfo{person}{Juan~Pablo Carrascal} {and}
  \bibinfo{person}{Karen Church}.} \bibinfo{year}{2015}\natexlab{}.
\newblock \showarticletitle{An in-situ study of mobile app \& mobile search
  interactions}. In \bibinfo{booktitle}{\emph{Proceedings of the 33rd Annual
  ACM Conference on Human Factors in Computing Systems}}. ACM,
  \bibinfo{pages}{2739--2748}.
\newblock


\bibitem[\protect\citeauthoryear{Cover, Hart, et~al\mbox{.}}{Cover
  et~al\mbox{.}}{1967}]%
        {cover1967nearest}
\bibfield{author}{\bibinfo{person}{Thomas~M Cover}, \bibinfo{person}{Peter~E
  Hart}, {et~al\mbox{.}}} \bibinfo{year}{1967}\natexlab{}.
\newblock \showarticletitle{Nearest neighbor pattern classification}.
\newblock \bibinfo{journal}{\emph{IEEE transactions on information theory}}
  \bibinfo{volume}{13}, \bibinfo{number}{1} (\bibinfo{year}{1967}),
  \bibinfo{pages}{21--27}.
\newblock


\bibitem[\protect\citeauthoryear{Cox}{Cox}{1966}]%
        {cox1966statistical}
\bibfield{author}{\bibinfo{person}{David~Roxbee Cox}.}
  \bibinfo{year}{1966}\natexlab{}.
\newblock \showarticletitle{The statistical analysis of series of events}.
\newblock \bibinfo{journal}{\emph{Monographs on Applied Probability and
  Statistics}} (\bibinfo{year}{1966}).
\newblock


\bibitem[\protect\citeauthoryear{Diaz-Aviles, Lam, Pinelli, Braghin, Gkoufas,
  Berlingerio, and Calabrese}{Diaz-Aviles et~al\mbox{.}}{2014}]%
        {diaz2014predicting}
\bibfield{author}{\bibinfo{person}{Ernesto Diaz-Aviles},
  \bibinfo{person}{Hoang~Thanh Lam}, \bibinfo{person}{Fabio Pinelli},
  \bibinfo{person}{Stefano Braghin}, \bibinfo{person}{Yiannis Gkoufas},
  \bibinfo{person}{Michele Berlingerio}, {and} \bibinfo{person}{Francesco
  Calabrese}.} \bibinfo{year}{2014}\natexlab{}.
\newblock \showarticletitle{Predicting user engagement in twitter with
  collaborative ranking}. In \bibinfo{booktitle}{\emph{Proceedings of the 2014
  Recommender Systems Challenge}}. ACM, \bibinfo{pages}{41}.
\newblock


\bibitem[\protect\citeauthoryear{Do and Gatica-Perez}{Do and
  Gatica-Perez}{2014}]%
        {do2014and}
\bibfield{author}{\bibinfo{person}{Trinh Minh~Tri Do} {and}
  \bibinfo{person}{Daniel Gatica-Perez}.} \bibinfo{year}{2014}\natexlab{}.
\newblock \showarticletitle{Where and what: Using smartphones to predict next
  locations and applications in daily life}.
\newblock \bibinfo{journal}{\emph{Pervasive and Mobile Computing}}
  \bibinfo{volume}{12} (\bibinfo{year}{2014}), \bibinfo{pages}{79--91}.
\newblock


\bibitem[\protect\citeauthoryear{Drucker, Cortes, Jackel, LeCun, and
  Vapnik}{Drucker et~al\mbox{.}}{1994}]%
        {drucker1994boosting}
\bibfield{author}{\bibinfo{person}{Harris Drucker}, \bibinfo{person}{Corinna
  Cortes}, \bibinfo{person}{Lawrence~D Jackel}, \bibinfo{person}{Yann LeCun},
  {and} \bibinfo{person}{Vladimir Vapnik}.} \bibinfo{year}{1994}\natexlab{}.
\newblock \showarticletitle{Boosting and other ensemble methods}.
\newblock \bibinfo{journal}{\emph{Neural Computation}} \bibinfo{volume}{6},
  \bibinfo{number}{6} (\bibinfo{year}{1994}), \bibinfo{pages}{1289--1301}.
\newblock


\bibitem[\protect\citeauthoryear{Ehmke and Wilson}{Ehmke and Wilson}{2007}]%
        {ehmke2007identifying}
\bibfield{author}{\bibinfo{person}{Claudia Ehmke} {and}
  \bibinfo{person}{Stephanie Wilson}.} \bibinfo{year}{2007}\natexlab{}.
\newblock \showarticletitle{Identifying web usability problems from
  eye-tracking data}. In \bibinfo{booktitle}{\emph{Proceedings of the 21st
  British HCI Group Annual Conference on People and Computers: HCI... but not
  as we know it-Volume 1}}. British Computer Society,
  \bibinfo{pages}{119--128}.
\newblock


\bibitem[\protect\citeauthoryear{Falaki, Mahajan, Kandula, Lymberopoulos,
  Govindan, and Estrin}{Falaki et~al\mbox{.}}{2010}]%
        {falaki2010diversity}
\bibfield{author}{\bibinfo{person}{Hossein Falaki}, \bibinfo{person}{Ratul
  Mahajan}, \bibinfo{person}{Srikanth Kandula}, \bibinfo{person}{Dimitrios
  Lymberopoulos}, \bibinfo{person}{Ramesh Govindan}, {and}
  \bibinfo{person}{Deborah Estrin}.} \bibinfo{year}{2010}\natexlab{}.
\newblock \showarticletitle{Diversity in smartphone usage}. In
  \bibinfo{booktitle}{\emph{Proceedings of the 8th international conference on
  Mobile systems, applications, and services}}. ACM, \bibinfo{pages}{179--194}.
\newblock


\bibitem[\protect\citeauthoryear{Ferreira, Goncalves, Kostakos, Barkhuus, and
  Dey}{Ferreira et~al\mbox{.}}{2014}]%
        {ferreira2014contextual}
\bibfield{author}{\bibinfo{person}{Denzil Ferreira}, \bibinfo{person}{Jorge
  Goncalves}, \bibinfo{person}{Vassilis Kostakos}, \bibinfo{person}{Louise
  Barkhuus}, {and} \bibinfo{person}{Anind~K Dey}.}
  \bibinfo{year}{2014}\natexlab{}.
\newblock \showarticletitle{Contextual experience sampling of mobile
  application micro-usage}. In \bibinfo{booktitle}{\emph{Proceedings of the
  16th international conference on Human-computer interaction with mobile
  devices \& services}}. ACM, \bibinfo{pages}{91--100}.
\newblock


\bibitem[\protect\citeauthoryear{Fox, Karnawat, Mydland, Dumais, and White}{Fox
  et~al\mbox{.}}{2005}]%
        {fox2005evaluating}
\bibfield{author}{\bibinfo{person}{Steve Fox}, \bibinfo{person}{Kuldeep
  Karnawat}, \bibinfo{person}{Mark Mydland}, \bibinfo{person}{Susan Dumais},
  {and} \bibinfo{person}{Thomas White}.} \bibinfo{year}{2005}\natexlab{}.
\newblock \showarticletitle{Evaluating implicit measures to improve web
  search}.
\newblock \bibinfo{journal}{\emph{ACM Transactions on Information Systems
  (TOIS)}} \bibinfo{volume}{23}, \bibinfo{number}{2} (\bibinfo{year}{2005}),
  \bibinfo{pages}{147--168}.
\newblock


\bibitem[\protect\citeauthoryear{Gelman}{Gelman}{2008}]%
        {gelman2008scaling}
\bibfield{author}{\bibinfo{person}{Andrew Gelman}.}
  \bibinfo{year}{2008}\natexlab{}.
\newblock \showarticletitle{Scaling regression inputs by dividing by two
  standard deviations}.
\newblock \bibinfo{journal}{\emph{Statistics in medicine}}
  \bibinfo{volume}{27}, \bibinfo{number}{15} (\bibinfo{year}{2008}),
  \bibinfo{pages}{2865--2873}.
\newblock


\bibitem[\protect\citeauthoryear{Google}{Google}{2018}]%
        {GooglePlayAppCat}
\bibfield{author}{\bibinfo{person}{Google}.} \bibinfo{year}{2018}\natexlab{}.
\newblock \bibinfo{booktitle}{\emph{Select a category for your app or game}}.
\newblock
\urldef\tempurl%
\url{https://support.google.com/googleplay/android-developer/answer/113475?hl=en-GB}
\showURL{%
\tempurl}


\bibitem[\protect\citeauthoryear{Hidasi and Tikk}{Hidasi and Tikk}{2016}]%
        {hidasi2016general}
\bibfield{author}{\bibinfo{person}{Bal{\'a}zs Hidasi} {and}
  \bibinfo{person}{Domonkos Tikk}.} \bibinfo{year}{2016}\natexlab{}.
\newblock \showarticletitle{General factorization framework for context-aware
  recommendations}.
\newblock \bibinfo{journal}{\emph{Data Mining and Knowledge Discovery}}
  \bibinfo{volume}{30}, \bibinfo{number}{2} (\bibinfo{year}{2016}),
  \bibinfo{pages}{342--371}.
\newblock


\bibitem[\protect\citeauthoryear{Homma, Soejima, Yoshida, and Umemura}{Homma
  et~al\mbox{.}}{2018}]%
        {homma2018analysis}
\bibfield{author}{\bibinfo{person}{Ryosuke Homma}, \bibinfo{person}{Keiichi
  Soejima}, \bibinfo{person}{Mitsuo Yoshida}, {and} \bibinfo{person}{Kyoji
  Umemura}.} \bibinfo{year}{2018}\natexlab{}.
\newblock \showarticletitle{Analysis of User Dwell Time on Non-News Pages}. In
  \bibinfo{booktitle}{\emph{2018 IEEE International Conference on Big Data (Big
  Data)}}. IEEE, \bibinfo{pages}{4333--4338}.
\newblock


\bibitem[\protect\citeauthoryear{Hosmer~Jr, Lemeshow, and Sturdivant}{Hosmer~Jr
  et~al\mbox{.}}{2013}]%
        {hosmer2013applied}
\bibfield{author}{\bibinfo{person}{David~W Hosmer~Jr}, \bibinfo{person}{Stanley
  Lemeshow}, {and} \bibinfo{person}{Rodney~X Sturdivant}.}
  \bibinfo{year}{2013}\natexlab{}.
\newblock \bibinfo{booktitle}{\emph{Applied logistic regression}}.
  Vol.~\bibinfo{volume}{398}.
\newblock \bibinfo{publisher}{John Wiley \& Sons}.
\newblock


\bibitem[\protect\citeauthoryear{Huang, White, and Dumais}{Huang
  et~al\mbox{.}}{2011}]%
        {huang2011no}
\bibfield{author}{\bibinfo{person}{Jeff Huang}, \bibinfo{person}{Ryen~W White},
  {and} \bibinfo{person}{Susan Dumais}.} \bibinfo{year}{2011}\natexlab{}.
\newblock \showarticletitle{No clicks, no problem: using cursor movements to
  understand and improve search}. In \bibinfo{booktitle}{\emph{Proceedings of
  the SIGCHI conference on human factors in computing systems}}. ACM,
  \bibinfo{pages}{1225--1234}.
\newblock


\bibitem[\protect\citeauthoryear{Huang, Zhang, Ma, and Chen}{Huang
  et~al\mbox{.}}{2012}]%
        {huang2012predicting}
\bibfield{author}{\bibinfo{person}{Ke Huang}, \bibinfo{person}{Chunhui Zhang},
  \bibinfo{person}{Xiaoxiao Ma}, {and} \bibinfo{person}{Guanling Chen}.}
  \bibinfo{year}{2012}\natexlab{}.
\newblock \showarticletitle{Predicting mobile application usage using
  contextual information}. In \bibinfo{booktitle}{\emph{Proceedings of the 2012
  ACM Conference on Ubiquitous Computing}}. ACM, \bibinfo{pages}{1059--1065}.
\newblock


\bibitem[\protect\citeauthoryear{Kelly and Belkin}{Kelly and Belkin}{2004}]%
        {kelly2004display}
\bibfield{author}{\bibinfo{person}{Diane Kelly} {and}
  \bibinfo{person}{Nicholas~J Belkin}.} \bibinfo{year}{2004}\natexlab{}.
\newblock \showarticletitle{Display time as implicit feedback: understanding
  task effects}. In \bibinfo{booktitle}{\emph{Proceedings of the 27th annual
  international ACM SIGIR conference on Research and development in information
  retrieval}}. ACM, \bibinfo{pages}{377--384}.
\newblock


\bibitem[\protect\citeauthoryear{Kim, Kim, and Wachter}{Kim
  et~al\mbox{.}}{2013}]%
        {kim2013study}
\bibfield{author}{\bibinfo{person}{Young~Hoon Kim}, \bibinfo{person}{Dan~J
  Kim}, {and} \bibinfo{person}{Kathy Wachter}.}
  \bibinfo{year}{2013}\natexlab{}.
\newblock \showarticletitle{A study of mobile user engagement (MoEN):
  Engagement motivations, perceived value, satisfaction, and continued
  engagement intention}.
\newblock \bibinfo{journal}{\emph{Decision Support Systems}}
  \bibinfo{volume}{56} (\bibinfo{year}{2013}), \bibinfo{pages}{361--370}.
\newblock


\bibitem[\protect\citeauthoryear{Kooti, Grbovic, Aiello, Bax, and Lerman}{Kooti
  et~al\mbox{.}}{2017}]%
        {kooti2017iphone}
\bibfield{author}{\bibinfo{person}{Farshad Kooti}, \bibinfo{person}{Mihajlo
  Grbovic}, \bibinfo{person}{Luca~Maria Aiello}, \bibinfo{person}{Eric Bax},
  {and} \bibinfo{person}{Kristina Lerman}.} \bibinfo{year}{2017}\natexlab{}.
\newblock \showarticletitle{iPhone's Digital Marketplace: Characterizing the
  Big Spenders}. In \bibinfo{booktitle}{\emph{Proceedings of the Tenth ACM
  International Conference on Web Search and Data Mining}}. ACM,
  \bibinfo{pages}{13--21}.
\newblock


\bibitem[\protect\citeauthoryear{Lehmann, Lalmas, Yom-Tov, and Dupret}{Lehmann
  et~al\mbox{.}}{2012}]%
        {lehmann2012models}
\bibfield{author}{\bibinfo{person}{Janette Lehmann}, \bibinfo{person}{Mounia
  Lalmas}, \bibinfo{person}{Elad Yom-Tov}, {and} \bibinfo{person}{Georges
  Dupret}.} \bibinfo{year}{2012}\natexlab{}.
\newblock \showarticletitle{Models of user engagement}. In
  \bibinfo{booktitle}{\emph{International Conference on User Modeling,
  Adaptation, and Personalization}}. Springer, \bibinfo{pages}{164--175}.
\newblock


\bibitem[\protect\citeauthoryear{Li and Lu}{Li and Lu}{2017}]%
        {li2017mining}
\bibfield{author}{\bibinfo{person}{Huoran Li} {and} \bibinfo{person}{Xuan Lu}.}
  \bibinfo{year}{2017}\natexlab{}.
\newblock \showarticletitle{Mining device-specific apps usage patterns from
  large-scale android users}.
\newblock \bibinfo{journal}{\emph{arXiv preprint arXiv:1707.09252}}
  (\bibinfo{year}{2017}).
\newblock


\bibitem[\protect\citeauthoryear{Li, Lu, Liu, Xie, Bian, Lin, Mei, and Feng}{Li
  et~al\mbox{.}}{2015}]%
        {li2015characterizing}
\bibfield{author}{\bibinfo{person}{Huoran Li}, \bibinfo{person}{Xuan Lu},
  \bibinfo{person}{Xuanzhe Liu}, \bibinfo{person}{Tao Xie},
  \bibinfo{person}{Kaigui Bian}, \bibinfo{person}{Felix~Xiaozhu Lin},
  \bibinfo{person}{Qiaozhu Mei}, {and} \bibinfo{person}{Feng Feng}.}
  \bibinfo{year}{2015}\natexlab{}.
\newblock \showarticletitle{Characterizing smartphone usage patterns from
  millions of android users}. In \bibinfo{booktitle}{\emph{Proceedings of the
  2015 Internet Measurement Conference}}. ACM, \bibinfo{pages}{459--472}.
\newblock


\bibitem[\protect\citeauthoryear{Liao, Lei, Shen, Li, and Peng}{Liao
  et~al\mbox{.}}{2012}]%
        {liao2012mining}
\bibfield{author}{\bibinfo{person}{Zhung-Xun Liao}, \bibinfo{person}{Po-Ruey
  Lei}, \bibinfo{person}{Tsu-Jou Shen}, \bibinfo{person}{Shou-Chung Li}, {and}
  \bibinfo{person}{Wen-Chih Peng}.} \bibinfo{year}{2012}\natexlab{}.
\newblock \showarticletitle{Mining temporal profiles of mobile applications for
  usage prediction}. In \bibinfo{booktitle}{\emph{Data Mining Workshops
  (ICDMW), 2012 IEEE 12th International Conference on}}. IEEE,
  \bibinfo{pages}{890--893}.
\newblock


\bibitem[\protect\citeauthoryear{Liao, Li, Peng, Philip, and Liu}{Liao
  et~al\mbox{.}}{2013a}]%
        {liao2013feature}
\bibfield{author}{\bibinfo{person}{Zhung-Xun Liao}, \bibinfo{person}{Shou-Chung
  Li}, \bibinfo{person}{Wen-Chih Peng}, \bibinfo{person}{S~Yu Philip}, {and}
  \bibinfo{person}{Te-Chuan Liu}.} \bibinfo{year}{2013}\natexlab{a}.
\newblock \showarticletitle{On the feature discovery for app usage prediction
  in smartphones}. In \bibinfo{booktitle}{\emph{Data Mining (ICDM), 2013 IEEE
  13th International Conference on}}. IEEE, \bibinfo{pages}{1127--1132}.
\newblock


\bibitem[\protect\citeauthoryear{Liao, Pan, Peng, and Lei}{Liao
  et~al\mbox{.}}{2013b}]%
        {liao2013mining}
\bibfield{author}{\bibinfo{person}{Zhung-Xun Liao}, \bibinfo{person}{Yi-Chin
  Pan}, \bibinfo{person}{Wen-Chih Peng}, {and} \bibinfo{person}{Po-Ruey Lei}.}
  \bibinfo{year}{2013}\natexlab{b}.
\newblock \showarticletitle{On mining mobile apps usage behavior for predicting
  apps usage in smartphones}. In \bibinfo{booktitle}{\emph{Proceedings of the
  22nd ACM international conference on Information \& Knowledge Management}}.
  ACM, \bibinfo{pages}{609--618}.
\newblock


\bibitem[\protect\citeauthoryear{Mathur, Lane, and Kawsar}{Mathur
  et~al\mbox{.}}{2016}]%
        {mathur2016engagement}
\bibfield{author}{\bibinfo{person}{Akhil Mathur}, \bibinfo{person}{Nicholas~D
  Lane}, {and} \bibinfo{person}{Fahim Kawsar}.}
  \bibinfo{year}{2016}\natexlab{}.
\newblock \showarticletitle{Engagement-aware computing: Modelling user
  engagement from mobile contexts}. In \bibinfo{booktitle}{\emph{Proceedings of
  the 2016 ACM International Joint Conference on Pervasive and Ubiquitous
  Computing}}. ACM, \bibinfo{pages}{622--633}.
\newblock


\bibitem[\protect\citeauthoryear{Mehrotra, Pejovic, Vermeulen, Hendley, and
  Musolesi}{Mehrotra et~al\mbox{.}}{2016}]%
        {mehrotra2016my}
\bibfield{author}{\bibinfo{person}{Abhinav Mehrotra}, \bibinfo{person}{Veljko
  Pejovic}, \bibinfo{person}{Jo Vermeulen}, \bibinfo{person}{Robert Hendley},
  {and} \bibinfo{person}{Mirco Musolesi}.} \bibinfo{year}{2016}\natexlab{}.
\newblock \showarticletitle{My phone and me: understanding people's receptivity
  to mobile notifications}. In \bibinfo{booktitle}{\emph{Proceedings of the
  2016 CHI conference on human factors in computing systems}}. ACM,
  \bibinfo{pages}{1021--1032}.
\newblock


\bibitem[\protect\citeauthoryear{Musto, Semeraro, De~Gemmis, and Lops}{Musto
  et~al\mbox{.}}{2015}]%
        {musto2015word}
\bibfield{author}{\bibinfo{person}{Cataldo Musto}, \bibinfo{person}{Giovanni
  Semeraro}, \bibinfo{person}{Marco De~Gemmis}, {and} \bibinfo{person}{Pasquale
  Lops}.} \bibinfo{year}{2015}\natexlab{}.
\newblock \showarticletitle{Word Embedding Techniques for Content-based
  Recommender Systems: An Empirical Evaluation.}. In
  \bibinfo{booktitle}{\emph{Recsys posters}}.
\newblock


\bibitem[\protect\citeauthoryear{Nelissen, Snoeck, Broucke, and
  Baesens}{Nelissen et~al\mbox{.}}{2018}]%
        {nelissen2018swipe}
\bibfield{author}{\bibinfo{person}{Klaas Nelissen}, \bibinfo{person}{Monique
  Snoeck}, \bibinfo{person}{Seppe~Vanden Broucke}, {and} \bibinfo{person}{Bart
  Baesens}.} \bibinfo{year}{2018}\natexlab{}.
\newblock \showarticletitle{Swipe and tell: Using implicit feedback to predict
  user engagement on tablets}.
\newblock \bibinfo{journal}{\emph{ACM Transactions on Information Systems
  (TOIS)}} \bibinfo{volume}{36}, \bibinfo{number}{4} (\bibinfo{year}{2018}),
  \bibinfo{pages}{1--36}.
\newblock


\bibitem[\protect\citeauthoryear{Pan, Aharony, and Pentland}{Pan
  et~al\mbox{.}}{2011}]%
        {pan2011composite}
\bibfield{author}{\bibinfo{person}{Wei Pan}, \bibinfo{person}{Nadav Aharony},
  {and} \bibinfo{person}{Alex Pentland}.} \bibinfo{year}{2011}\natexlab{}.
\newblock \showarticletitle{Composite social network for predicting mobile apps
  installation}.
\newblock \bibinfo{journal}{\emph{arXiv preprint arXiv:1106.0359}}
  (\bibinfo{year}{2011}).
\newblock


\bibitem[\protect\citeauthoryear{Pejovic and Musolesi}{Pejovic and
  Musolesi}{2014}]%
        {pejovic2014interruptme}
\bibfield{author}{\bibinfo{person}{Veljko Pejovic} {and} \bibinfo{person}{Mirco
  Musolesi}.} \bibinfo{year}{2014}\natexlab{}.
\newblock \showarticletitle{InterruptMe: designing intelligent prompting
  mechanisms for pervasive applications}. In
  \bibinfo{booktitle}{\emph{Proceedings of the 2014 ACM International Joint
  Conference on Pervasive and Ubiquitous Computing}}. ACM,
  \bibinfo{pages}{897--908}.
\newblock


\bibitem[\protect\citeauthoryear{Petsas, Papadogiannakis, Polychronakis,
  Markatos, and Karagiannis}{Petsas et~al\mbox{.}}{2017}]%
        {petsas2017measurement}
\bibfield{author}{\bibinfo{person}{Thanasis Petsas}, \bibinfo{person}{Antonis
  Papadogiannakis}, \bibinfo{person}{Michalis Polychronakis},
  \bibinfo{person}{Evangelos~P Markatos}, {and} \bibinfo{person}{Thomas
  Karagiannis}.} \bibinfo{year}{2017}\natexlab{}.
\newblock \showarticletitle{Measurement, modeling, and analysis of the mobile
  app ecosystem}.
\newblock \bibinfo{journal}{\emph{ACM Transactions on Modeling and Performance
  Evaluation of Computing Systems (TOMPECS)}} \bibinfo{volume}{2},
  \bibinfo{number}{2} (\bibinfo{year}{2017}), \bibinfo{pages}{7}.
\newblock


\bibitem[\protect\citeauthoryear{Pielot, Church, and De~Oliveira}{Pielot
  et~al\mbox{.}}{2014}]%
        {pielot2014situ}
\bibfield{author}{\bibinfo{person}{Martin Pielot}, \bibinfo{person}{Karen
  Church}, {and} \bibinfo{person}{Rodrigo De~Oliveira}.}
  \bibinfo{year}{2014}\natexlab{}.
\newblock \showarticletitle{An in-situ study of mobile phone notifications}. In
  \bibinfo{booktitle}{\emph{Proceedings of the 16th international conference on
  Human-computer interaction with mobile devices \& services}}. ACM,
  \bibinfo{pages}{233--242}.
\newblock


\bibitem[\protect\citeauthoryear{Quinlan}{Quinlan}{2003}]%
        {quinlan2003just}
\bibfield{author}{\bibinfo{person}{Mary~Lou Quinlan}.}
  \bibinfo{year}{2003}\natexlab{}.
\newblock \bibinfo{booktitle}{\emph{Just ask a woman: Cracking the code of what
  women want and how they buy}}.
\newblock \bibinfo{publisher}{John Wiley \& Sons}.
\newblock


\bibitem[\protect\citeauthoryear{Revels, Tojib, and Tsarenko}{Revels
  et~al\mbox{.}}{2010}]%
        {revels2010understanding}
\bibfield{author}{\bibinfo{person}{Janeaya Revels}, \bibinfo{person}{Dewi
  Tojib}, {and} \bibinfo{person}{Yelena Tsarenko}.}
  \bibinfo{year}{2010}\natexlab{}.
\newblock \showarticletitle{Understanding consumer intention to use mobile
  services}.
\newblock \bibinfo{journal}{\emph{Australasian Marketing Journal (AMJ)}}
  \bibinfo{volume}{18}, \bibinfo{number}{2} (\bibinfo{year}{2010}),
  \bibinfo{pages}{74--80}.
\newblock


\bibitem[\protect\citeauthoryear{Schmidhuber}{Schmidhuber}{2015}]%
        {schmidhuber2015deep}
\bibfield{author}{\bibinfo{person}{J{\"u}rgen Schmidhuber}.}
  \bibinfo{year}{2015}\natexlab{}.
\newblock \showarticletitle{Deep learning in neural networks: An overview}.
\newblock \bibinfo{journal}{\emph{Neural networks}}  \bibinfo{volume}{61}
  (\bibinfo{year}{2015}), \bibinfo{pages}{85--117}.
\newblock


\bibitem[\protect\citeauthoryear{Seki and Yoshida}{Seki and Yoshida}{2018}]%
        {seki2018analysis}
\bibfield{author}{\bibinfo{person}{Yoshifumi Seki} {and}
  \bibinfo{person}{Mitsuo Yoshida}.} \bibinfo{year}{2018}\natexlab{}.
\newblock \showarticletitle{Analysis of user dwell time by category in news
  application}. In \bibinfo{booktitle}{\emph{2018 IEEE/WIC/ACM International
  Conference on Web Intelligence (WI)}}. IEEE, \bibinfo{pages}{732--735}.
\newblock


\bibitem[\protect\citeauthoryear{Seneviratne, Seneviratne, Mohapatra, and
  Mahanti}{Seneviratne et~al\mbox{.}}{2015}]%
        {seneviratne2015your}
\bibfield{author}{\bibinfo{person}{Suranga Seneviratne}, \bibinfo{person}{Aruna
  Seneviratne}, \bibinfo{person}{Prasant Mohapatra}, {and}
  \bibinfo{person}{Anirban Mahanti}.} \bibinfo{year}{2015}\natexlab{}.
\newblock \showarticletitle{Your installed apps reveal your gender and more!}
\newblock \bibinfo{journal}{\emph{ACM SIGMOBILE Mobile Computing and
  Communications Review}} \bibinfo{volume}{18}, \bibinfo{number}{3}
  (\bibinfo{year}{2015}), \bibinfo{pages}{55--61}.
\newblock


\bibitem[\protect\citeauthoryear{Shin, Hong, and Dey}{Shin
  et~al\mbox{.}}{2012}]%
        {shin2012understanding}
\bibfield{author}{\bibinfo{person}{Choonsung Shin}, \bibinfo{person}{Jin-Hyuk
  Hong}, {and} \bibinfo{person}{Anind~K Dey}.} \bibinfo{year}{2012}\natexlab{}.
\newblock \showarticletitle{Understanding and prediction of mobile application
  usage for smart phones}. In \bibinfo{booktitle}{\emph{Proceedings of the 2012
  ACM Conference on Ubiquitous Computing}}. ACM, \bibinfo{pages}{173--182}.
\newblock


\bibitem[\protect\citeauthoryear{Syarif, Zaluska, Prugel-Bennett, and
  Wills}{Syarif et~al\mbox{.}}{2012}]%
        {syarif2012application}
\bibfield{author}{\bibinfo{person}{Iwan Syarif}, \bibinfo{person}{Ed Zaluska},
  \bibinfo{person}{Adam Prugel-Bennett}, {and} \bibinfo{person}{Gary Wills}.}
  \bibinfo{year}{2012}\natexlab{}.
\newblock \showarticletitle{Application of bagging, boosting and stacking to
  intrusion detection}. In \bibinfo{booktitle}{\emph{International Workshop on
  Machine Learning and Data Mining in Pattern Recognition}}. Springer,
  \bibinfo{pages}{593--602}.
\newblock


\bibitem[\protect\citeauthoryear{Tan, Liu, Chen, and Xiong}{Tan
  et~al\mbox{.}}{2012}]%
        {tan2012prediction}
\bibfield{author}{\bibinfo{person}{Chang Tan}, \bibinfo{person}{Qi Liu},
  \bibinfo{person}{Enhong Chen}, {and} \bibinfo{person}{Hui Xiong}.}
  \bibinfo{year}{2012}\natexlab{}.
\newblock \showarticletitle{Prediction for mobile application usage patterns}.
  In \bibinfo{booktitle}{\emph{Nokia MDC Workshop}}, Vol.~\bibinfo{volume}{12}.
\newblock


\bibitem[\protect\citeauthoryear{Tavakol and Brefeld}{Tavakol and
  Brefeld}{2014}]%
        {tavakol2014factored}
\bibfield{author}{\bibinfo{person}{Maryam Tavakol} {and} \bibinfo{person}{Ulf
  Brefeld}.} \bibinfo{year}{2014}\natexlab{}.
\newblock \showarticletitle{Factored MDPs for detecting topics of user
  sessions}. In \bibinfo{booktitle}{\emph{Proceedings of the 8th ACM Conference
  on Recommender Systems}}. \bibinfo{pages}{33--40}.
\newblock


\bibitem[\protect\citeauthoryear{Tian, Zhou, Lalmas, Liu, and Pelleg}{Tian
  et~al\mbox{.}}{2020b}]%
        {tian2020cohort}
\bibfield{author}{\bibinfo{person}{Yuan Tian}, \bibinfo{person}{Ke Zhou},
  \bibinfo{person}{Mounia Lalmas}, \bibinfo{person}{Yiqun Liu}, {and}
  \bibinfo{person}{Dan Pelleg}.} \bibinfo{year}{2020}\natexlab{b}.
\newblock \showarticletitle{Cohort Modeling Based App Category Usage
  Prediction}. In \bibinfo{booktitle}{\emph{Proceedings of the 28th ACM
  Conference on User Modeling, Adaptation and Personalization}}.
  \bibinfo{pages}{248--256}.
\newblock


\bibitem[\protect\citeauthoryear{Tian, Zhou, Lalmas, and Pelleg}{Tian
  et~al\mbox{.}}{2020a}]%
        {tian2020identifying}
\bibfield{author}{\bibinfo{person}{Yuan Tian}, \bibinfo{person}{Ke Zhou},
  \bibinfo{person}{Mounia Lalmas}, {and} \bibinfo{person}{Dan Pelleg}.}
  \bibinfo{year}{2020}\natexlab{a}.
\newblock \showarticletitle{Identifying Tasks from Mobile App Usage Patterns}.
  In \bibinfo{booktitle}{\emph{Proceedings of the 43rd International ACM SIGIR
  Conference on Research and Development in Information Retrieval}}.
  \bibinfo{pages}{2357--2366}.
\newblock


\bibitem[\protect\citeauthoryear{Van~Canneyt, Bron, Haines, and
  Lalmas}{Van~Canneyt et~al\mbox{.}}{2017}]%
        {van2017describing}
\bibfield{author}{\bibinfo{person}{Steven Van~Canneyt}, \bibinfo{person}{Marc
  Bron}, \bibinfo{person}{Andy Haines}, {and} \bibinfo{person}{Mounia Lalmas}.}
  \bibinfo{year}{2017}\natexlab{}.
\newblock \showarticletitle{Describing Patterns and Disruptions in Large Scale
  Mobile App Usage Data}. In \bibinfo{booktitle}{\emph{Proceedings of the 26th
  International Conference on World Wide Web Companion}}. International World
  Wide Web Conferences Steering Committee, \bibinfo{pages}{1579--1584}.
\newblock


\bibitem[\protect\citeauthoryear{van Vugt, Konijn, Hoorn, Keur, and
  Eli{\'e}ns}{van Vugt et~al\mbox{.}}{2007}]%
        {van2007realism}
\bibfield{author}{\bibinfo{person}{Henriette~C van Vugt},
  \bibinfo{person}{Elly~A Konijn}, \bibinfo{person}{Johan~F Hoorn},
  \bibinfo{person}{I Keur}, {and} \bibinfo{person}{Anton Eli{\'e}ns}.}
  \bibinfo{year}{2007}\natexlab{}.
\newblock \showarticletitle{Realism is not all! User engagement with
  task-related interface characters}.
\newblock \bibinfo{journal}{\emph{Interacting with Computers}}
  \bibinfo{volume}{19}, \bibinfo{number}{2} (\bibinfo{year}{2007}),
  \bibinfo{pages}{267--280}.
\newblock


\bibitem[\protect\citeauthoryear{Vasiloudis, Vahabi, Kravitz, and
  Rashkov}{Vasiloudis et~al\mbox{.}}{2017}]%
        {vasiloudis2017predicting}
\bibfield{author}{\bibinfo{person}{Theodore Vasiloudis},
  \bibinfo{person}{Hossein Vahabi}, \bibinfo{person}{Ross Kravitz}, {and}
  \bibinfo{person}{Valery Rashkov}.} \bibinfo{year}{2017}\natexlab{}.
\newblock \showarticletitle{Predicting session length in media streaming}. In
  \bibinfo{booktitle}{\emph{Proceedings of the 40th International ACM SIGIR
  Conference on Research and Development in Information Retrieval}}.
  \bibinfo{pages}{977--980}.
\newblock


\bibitem[\protect\citeauthoryear{Wang, Chen, Zhuang, Lin, Xia, Du, and He}{Wang
  et~al\mbox{.}}{[n.d.]}]%
        {wangcapturing}
\bibfield{author}{\bibinfo{person}{Tianxin Wang}, \bibinfo{person}{Jingwu
  Chen}, \bibinfo{person}{Fuzhen Zhuang}, \bibinfo{person}{Leyu Lin},
  \bibinfo{person}{Feng Xia}, \bibinfo{person}{Lihuan Du}, {and}
  \bibinfo{person}{Qing He}.} \bibinfo{year}{[n.d.]}\natexlab{}.
\newblock \showarticletitle{Capturing Attraction Distribution: Sequential
  Attentive Network for Dwell Time Prediction}.
\newblock  (\bibinfo{year}{[n.\,d.]}).
\newblock


\bibitem[\protect\citeauthoryear{Wu, Rizoiu, and Xie}{Wu et~al\mbox{.}}{2018}]%
        {wu2018beyond}
\bibfield{author}{\bibinfo{person}{Siqi Wu}, \bibinfo{person}{Marian-Andrei
  Rizoiu}, {and} \bibinfo{person}{Lexing Xie}.}
  \bibinfo{year}{2018}\natexlab{}.
\newblock \showarticletitle{Beyond views: Measuring and predicting engagement
  in online videos}. In \bibinfo{booktitle}{\emph{Proceedings of the
  International AAAI Conference on Web and Social Media}},
  Vol.~\bibinfo{volume}{12}.
\newblock


\bibitem[\protect\citeauthoryear{Xu, Erman, Gerber, Mao, Pang, and
  Venkataraman}{Xu et~al\mbox{.}}{2011}]%
        {xu2011identifying}
\bibfield{author}{\bibinfo{person}{Qiang Xu}, \bibinfo{person}{Jeffrey Erman},
  \bibinfo{person}{Alexandre Gerber}, \bibinfo{person}{Zhuoqing Mao},
  \bibinfo{person}{Jeffrey Pang}, {and} \bibinfo{person}{Shobha Venkataraman}.}
  \bibinfo{year}{2011}\natexlab{}.
\newblock \showarticletitle{Identifying diverse usage behaviors of smartphone
  apps}. In \bibinfo{booktitle}{\emph{Proceedings of the 2011 ACM SIGCOMM
  conference on Internet measurement conference}}. ACM,
  \bibinfo{pages}{329--344}.
\newblock


\bibitem[\protect\citeauthoryear{Xu, Li, Zhang, Gao, Zhan, and Lu}{Xu
  et~al\mbox{.}}{2020}]%
        {xu2020predicting}
\bibfield{author}{\bibinfo{person}{Shijian Xu}, \bibinfo{person}{Wenzhong Li},
  \bibinfo{person}{Xiao Zhang}, \bibinfo{person}{Songcheng Gao},
  \bibinfo{person}{Tong Zhan}, {and} \bibinfo{person}{Sanglu Lu}.}
  \bibinfo{year}{2020}\natexlab{}.
\newblock \showarticletitle{Predicting and Recommending the next Smartphone
  Apps based on Recurrent Neural Network}.
\newblock \bibinfo{journal}{\emph{CCF Transactions on Pervasive Computing and
  Interaction}} \bibinfo{volume}{2}, \bibinfo{number}{4}
  (\bibinfo{year}{2020}), \bibinfo{pages}{314--328}.
\newblock


\bibitem[\protect\citeauthoryear{Xu, Lin, Lu, Cardone, Lane, Chen, Campbell,
  and Choudhury}{Xu et~al\mbox{.}}{2013}]%
        {xu2013preference}
\bibfield{author}{\bibinfo{person}{Ye Xu}, \bibinfo{person}{Mu Lin},
  \bibinfo{person}{Hong Lu}, \bibinfo{person}{Giuseppe Cardone},
  \bibinfo{person}{Nicholas Lane}, \bibinfo{person}{Zhenyu Chen},
  \bibinfo{person}{Andrew Campbell}, {and} \bibinfo{person}{Tanzeem
  Choudhury}.} \bibinfo{year}{2013}\natexlab{}.
\newblock \showarticletitle{Preference, context and communities: a
  multi-faceted approach to predicting smartphone app usage patterns}. In
  \bibinfo{booktitle}{\emph{Proceedings of the 2013 International Symposium on
  Wearable Computers}}. ACM, \bibinfo{pages}{69--76}.
\newblock


\bibitem[\protect\citeauthoryear{Yan, Chu, Ganesan, Kansal, and Liu}{Yan
  et~al\mbox{.}}{2012}]%
        {yan2012fast}
\bibfield{author}{\bibinfo{person}{Tingxin Yan}, \bibinfo{person}{David Chu},
  \bibinfo{person}{Deepak Ganesan}, \bibinfo{person}{Aman Kansal}, {and}
  \bibinfo{person}{Jie Liu}.} \bibinfo{year}{2012}\natexlab{}.
\newblock \showarticletitle{Fast app launching for mobile devices using
  predictive user context}. In \bibinfo{booktitle}{\emph{Proceedings of the
  10th international conference on Mobile systems, applications, and
  services}}. ACM, \bibinfo{pages}{113--126}.
\newblock


\bibitem[\protect\citeauthoryear{Yi, Hong, Zhong, Liu, and Rajan}{Yi
  et~al\mbox{.}}{2014}]%
        {yi2014beyond}
\bibfield{author}{\bibinfo{person}{Xing Yi}, \bibinfo{person}{Liangjie Hong},
  \bibinfo{person}{Erheng Zhong}, \bibinfo{person}{Nanthan~Nan Liu}, {and}
  \bibinfo{person}{Suju Rajan}.} \bibinfo{year}{2014}\natexlab{}.
\newblock \showarticletitle{Beyond clicks: dwell time for personalization}. In
  \bibinfo{booktitle}{\emph{Proceedings of the 8th ACM Conference on
  Recommender systems}}. ACM, \bibinfo{pages}{113--120}.
\newblock


\bibitem[\protect\citeauthoryear{Yom-Tov, Lalmas, Baeza-Yates, Dupret, Lehmann,
  and Donmez}{Yom-Tov et~al\mbox{.}}{2013}]%
        {yom2013measuring}
\bibfield{author}{\bibinfo{person}{Elad Yom-Tov}, \bibinfo{person}{Mounia
  Lalmas}, \bibinfo{person}{Ricardo Baeza-Yates}, \bibinfo{person}{Georges
  Dupret}, \bibinfo{person}{Janette Lehmann}, {and} \bibinfo{person}{Pinar
  Donmez}.} \bibinfo{year}{2013}\natexlab{}.
\newblock \showarticletitle{Measuring inter-site engagement}. In
  \bibinfo{booktitle}{\emph{Big Data, 2013 IEEE International Conference on}}.
  IEEE, \bibinfo{pages}{228--236}.
\newblock


\bibitem[\protect\citeauthoryear{Zhang, Ding, Chen, Huang, Ma, and Yan}{Zhang
  et~al\mbox{.}}{2012}]%
        {zhang2012nihao}
\bibfield{author}{\bibinfo{person}{Chunhui Zhang}, \bibinfo{person}{Xiang
  Ding}, \bibinfo{person}{Guanling Chen}, \bibinfo{person}{Ke Huang},
  \bibinfo{person}{Xiaoxiao Ma}, {and} \bibinfo{person}{Bo Yan}.}
  \bibinfo{year}{2012}\natexlab{}.
\newblock \showarticletitle{Nihao: A predictive smartphone application
  launcher}. In \bibinfo{booktitle}{\emph{International Conference on Mobile
  Computing, Applications, and Services}}. Springer, \bibinfo{pages}{294--313}.
\newblock


\bibitem[\protect\citeauthoryear{Zhao, Luo, Jiang, Wang, Xu, Li, Yin, and
  Pan}{Zhao et~al\mbox{.}}{2019}]%
        {zhao2019appusage2vec}
\bibfield{author}{\bibinfo{person}{Sha Zhao}, \bibinfo{person}{Zhiling Luo},
  \bibinfo{person}{Ziwen Jiang}, \bibinfo{person}{Haiyan Wang},
  \bibinfo{person}{Feng Xu}, \bibinfo{person}{Shijian Li},
  \bibinfo{person}{Jianwei Yin}, {and} \bibinfo{person}{Gang Pan}.}
  \bibinfo{year}{2019}\natexlab{}.
\newblock \showarticletitle{AppUsage2Vec: Modeling smartphone app usage for
  prediction}. In \bibinfo{booktitle}{\emph{2019 IEEE 35th International
  Conference on Data Engineering (ICDE)}}. IEEE, \bibinfo{pages}{1322--1333}.
\newblock


\bibitem[\protect\citeauthoryear{Zhao, Ramos, Tao, Jiang, Li, Wu, Pan, and
  Dey}{Zhao et~al\mbox{.}}{2016}]%
        {zhao2016discovering}
\bibfield{author}{\bibinfo{person}{Sha Zhao}, \bibinfo{person}{Julian Ramos},
  \bibinfo{person}{Jianrong Tao}, \bibinfo{person}{Ziwen Jiang},
  \bibinfo{person}{Shijian Li}, \bibinfo{person}{Zhaohui Wu},
  \bibinfo{person}{Gang Pan}, {and} \bibinfo{person}{Anind~K Dey}.}
  \bibinfo{year}{2016}\natexlab{}.
\newblock \showarticletitle{Discovering different kinds of smartphone users
  through their application usage behaviors}. In
  \bibinfo{booktitle}{\emph{Proceedings of the 2016 ACM International Joint
  Conference on Pervasive and Ubiquitous Computing}}. ACM,
  \bibinfo{pages}{498--509}.
\newblock


\bibitem[\protect\citeauthoryear{Zhou, Qian, Shen, Zhang, Wang, Liu, and
  Ou}{Zhou et~al\mbox{.}}{2018}]%
        {zhou2018jump}
\bibfield{author}{\bibinfo{person}{Tengfei Zhou}, \bibinfo{person}{Hui Qian},
  \bibinfo{person}{Zebang Shen}, \bibinfo{person}{Chao Zhang},
  \bibinfo{person}{Chengwei Wang}, \bibinfo{person}{Shichen Liu}, {and}
  \bibinfo{person}{Wenwu Ou}.} \bibinfo{year}{2018}\natexlab{}.
\newblock \showarticletitle{Jump: A joint predictor for user click and dwell
  time}. In \bibinfo{booktitle}{\emph{Proceedings of the 27th International
  Joint Conference on Artificial Intelligence. AAAI Press}}.
  \bibinfo{pages}{3704--3710}.
\newblock


\bibitem[\protect\citeauthoryear{Zou, Zhang, Li, and Pan}{Zou
  et~al\mbox{.}}{2013}]%
        {zou2013prophet}
\bibfield{author}{\bibinfo{person}{Xun Zou}, \bibinfo{person}{Wangsheng Zhang},
  \bibinfo{person}{Shijian Li}, {and} \bibinfo{person}{Gang Pan}.}
  \bibinfo{year}{2013}\natexlab{}.
\newblock \showarticletitle{Prophet: What app you wish to use next}. In
  \bibinfo{booktitle}{\emph{Proceedings of the 2013 ACM conference on Pervasive
  and ubiquitous computing adjunct publication}}. ACM,
  \bibinfo{pages}{167--170}.
\newblock


\end{thebibliography}

\end{document}